\documentclass[aps,twocolumn,amsmath,superscriptaddress,amssymb,longbibliography,prb,10pt]{revtex4-2}
\usepackage[colorlinks=true,allcolors=blue,hypertexnames=false]{hyperref}
\usepackage{amsmath}
\usepackage{dsfont}
\usepackage{textcomp}
\usepackage{tabularx}
\usepackage{graphicx}
\usepackage{soul}
\usepackage{feynmp-auto}
\usepackage{footmisc}
\usepackage[export]{adjustbox}
\usepackage{wrapfig}
\usepackage{bbm}

\makeatletter
\AtBeginDocument{\let\NAT@sort\z@}
\def\@bibdataout@aps{%
 \immediate\write\@bibdataout{%
  @CONTROL{%
   apsrev42Control%
   \longbibliography@sw{%
    ,author="48",editor="1",pages="0",title="0",year="1"%
   }{%
    ,author="48",editor="1",pages="0",title="",year="1"%
   }%
  }%
 }%
 \if@filesw
  \immediate\write\@auxout{\string\citation{apsrev42Control}}%
 \fi
}%
\makeatother

\newcommand{\beg}{\begin{equation}}
\newcommand{\en}{\end{equation}}
\newcommand{\bp}{\mathbf p}
\newcommand{\bq}{\mathbf q}

\newcommand{\bk}{\mathbf k}
\newcommand{\br}{\mathbf r}
\newcommand{\bR}{\mathbf R}
\newcommand{\bn}{\mathbf n}

\newcommand \bel  {\begin{align}}
\newcommand \enl  {\end{align}}

\newcommand{\eps}{\epsilon}
\newcommand{\veps}{\varepsilon}

\newcommand{\up}{\uparrow}
\newcommand{\dn}{\downarrow}
\newcommand{\dg}{^\dagger}

\renewcommand{\Re}{\mathrm{Re}}
\renewcommand{\Im}{\mathrm{Im}}

\newcommand{\darkred}[1]{{\color[rgb]{0.6,0,0}{#1}}}

\def\Xint#1{\mathchoice
   {\XXint\displaystyle\textstyle{#1}}%
   {\XXint\textstyle\scriptstyle{#1}}%
   {\XXint\scriptstyle\scriptscriptstyle{#1}}%
   {\XXint\scriptscriptstyle\scriptscriptstyle{#1}}%
   \!\int}
\def\XXint#1#2#3{{\setbox0=\hbox{$#1{#2#3}{\int}$}
     \vcenter{\hbox{$#2#3$}}\kern-.5\wd0}}

\def\dashint{\Xint-}

\begin{document}

\title{Amplitude mode in Eliashberg superconductors}

\author{Maxim Dzero}
\affiliation{Department of Physics, Kent State University, Kent, Ohio 44242, USA}

\author{Emil A. Yuzbashyan}
\affiliation{Department of Physics and Astronomy, Center for Materials Theory, Rutgers University, Piscataway, New Jersey 08854, USA}

\date{\today}

\begin{abstract}
In the BCS theory of superconductivity the amplitude (Higgs) mode is degenerate with the pair-breaking threshold: its frequency coincides with $2\Delta$, so it is an edge singularity rather than a true collective mode. This is why after a sudden weak perturbation at $t=0$ its oscillations decay algebraically as $1/\sqrt{t}$. We show that the retardation effects due to the phonon dynamics in the
strong-coupling theory lift this degeneracy: the Higgs mode detaches from the pair-breaking edge $2\Delta_g$, set by the spectral gap, and becomes a sharp subgap excitation. We determine its frequency from the longitudinal pair susceptibility evaluated directly on the real frequency axis, using the Eilenberger equations for Migdal--Eliashberg theory with
the Einstein phonon spectrum. The response of the mass renormalization supplies at least two fifths of
the displacement, and possibly all of it: with it switched off, the binding
falls below the resolution of the calculation. Its
consequences, however, are significant because a 
percent-level shift transfers about a quarter of the spectral weight of the amplitude
channel out of the two-quasiparticle continuum into a discrete energy level. What
experiment sees is therefore not the displacement itself but its consequence:
amplitude oscillations that persist instead of decaying as $1/\sqrt{t}$.

\end{abstract}

\maketitle
\section{Introduction}
States of matter with spontaneously broken symmetry are characterized by an order parameter and exhibit rigidity. A superconductor is described by a complex order parameter, and its rigidity has two distinct components: a stiffness with respect to spatial variations of the phase, and a restoring force — the curvature of the free energy at its minimum — with respect to variations of the amplitude. The phase (transverse) channel acquires the Bogoliubov–Anderson–Goldstone mode: gapless with linear dispersion in a neutral superfluid, lifted to the plasma frequency by the long-range Coulomb interaction in three dimensions, and surviving as a gapless plasmon in two dimensions \cite{Anderson1958,CarlsonGoldman73,Volkov1975,SchmidSchon1975,SchmidSchon1979,Volkov1979,Kulik1981,Anderson2015}. The amplitude (longitudinal) channel hosts the massive (Higgs) mode.  The amplitude of the
order parameter is probed directly only by this channel, which is why the Higgs
mode has attracted considerable attention from both the theoretical and the
experimental
communities \cite{ASchmid1968,VolkovKogan1974,SooryakumarKlein1980,LittlewoodVarma1981,
LittlewoodVarma1982,Spivak2004,Andreev2004,Amin2004,
Enolski2005a,Yuzbashyan2006,YuzbashyanDzero2006,BarankovLevitov2006,
Dzero2007,Levitov2007,Papenkort2007,Gurarie2009,
Shimano2013,Foster2013pwave,Shimano2014,Measson2014,
Volovik2014,Foster2014Floquet,PekkerVarma2015,TsujiAoki2015,
QReview2015,Kemper2015,Peronaci2015,Dzero2015Competing,
Cea2016,Werner2016,Murakami2016damping,Tsuji2016retarded,
Krull2016,Rufus2017,Matsunaga2017,MoorVolkovEfetov2017,Chou2017,
Katsumi2018,Joerg1,Nakamura2019,Udina2019,
Scaramazza2019,Vadimov2019,Smale2019,ShimanoTsuji2020,
Schwarz2020-Classification,Buzzi2021,Gao2021,Marino2022,
Wang2022,Shankar2022,ChubukovHiggs2023,Lyu2023,
Duraki2024,LiDzero2024,Young2024,Dyke2024,
Eremin2024,Uhrig2025,Nosov2025,Kamenev2025,
Breyer2025,Chu2025Topological,Uhrig2026,Kazi2026,
Awelewa2026}.

\begin{figure*}
\includegraphics[width=0.95\linewidth]{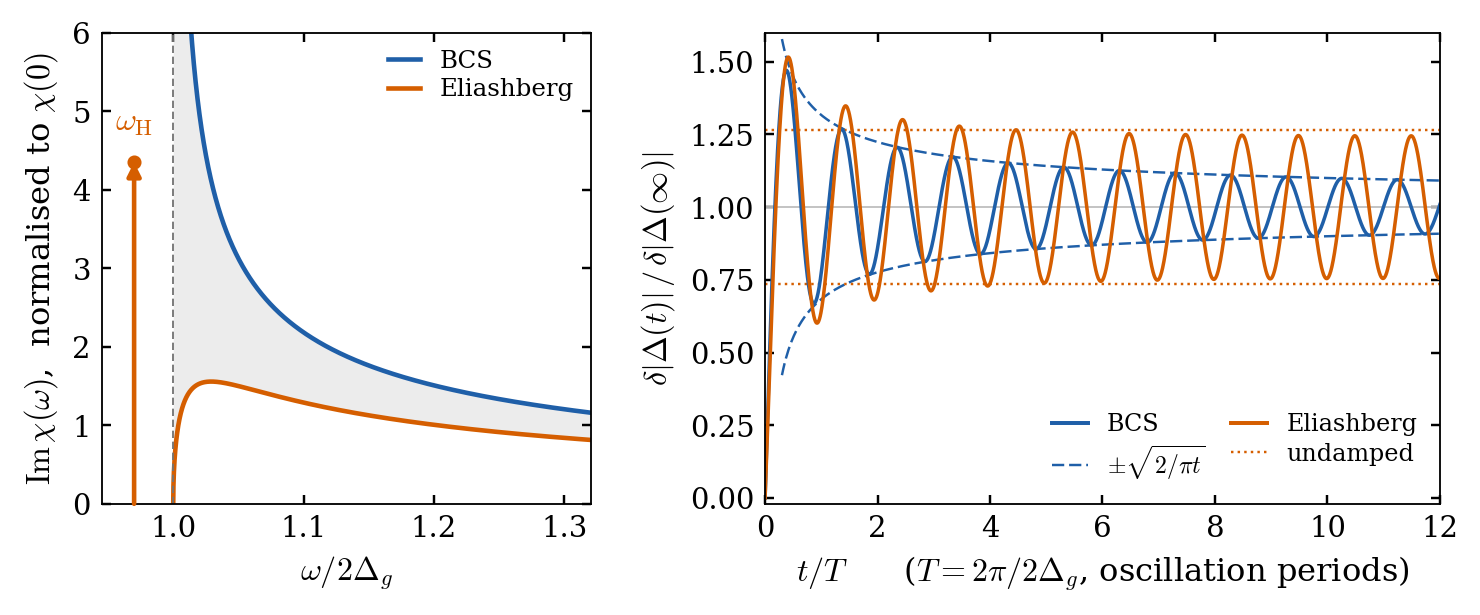}
\caption{\footnotesize Time-domain signature of the amplitude mode. {\bf Left panel:}
spectral function of the longitudinal pair susceptibility
$\chi(\omega)/\chi(0)$.
The pole corresponding to amplitude Higgs mode then holds approximately $25\%$ of
it. Shading marks the continuum weight that retardation removes from the
neighborhood of the edge. We depict the weight transferred from the edge into
the pole by the shaded area. 
{\bf Right panel:} order-parameter amplitude after a sudden weak pair field applied at
$t=0$, in units of its new equilibrium value, computed at that binding. The
BCS oscillation dephases across the continuum with envelope $\propto t^{-1/2}$
(dashed). The pole gives an undamped oscillation of relative amplitude $0.265$
(dotted), which exceeds the BCS envelope after $1.4$ periods.
$T=2\pi/2\Delta_g$. The continuum above $2\Delta_g$ is that of our model and the
pole position and residue are those found in this work for the Migdal-Eliashberg theory with the Einstein phonon.}
\label{Fig1-schematic}
\end{figure*}

In the BCS theory \cite{BCS} the amplitude mode is pinned to the edge of the
two-particle continuum and manifests itself as a resonance rather than as a
true collective mode \cite{ASchmid1968,ChubukovHiggs2023,Nosov2025}. Its
dynamics displays periodic oscillations with an amplitude which decays as
$1/\sqrt{t}$ due to Landau damping \cite{VolkovKogan1974,QReview2015}.
{Within linear response, one}
could therefore be tempted to conclude that any measured deviation of the
amplitude mode from $2\Delta$ directly probes physics beyond the standard BCS
theory. This is not so, however. The degeneracy between the energy of the
Higgs resonance and the two-particle continuum is accidental, and many
modifications of the interaction kernel which keep it instantaneous, and
remain within the weak coupling theory, should lift it. Indeed, Barankov and
Levitov have shown that giving the interaction kernel the energy dependence
moves the frequency of the Higgs mode inside the pairing gap, by an amount of
the order $O(\Delta)$ \cite{Levitov2007}. Lorenzana and Seibold have studied
the same mode in the Hubbard model using a time-dependent Gutzwiller approach,
and found that its frequency shifts below the edge of the two-particle
continuum upon increasing the interaction strength \cite{Duraki2024}. {Moreover, undamped amplitude oscillations also occur within the collisionless BCS theory with the standard, energy-independent interaction kernel after a sufficiently strong perturbation~\cite{Spivak2004,Enolski2005a,Yuzbashyan2006,Levitov2006,BarankovLevitov2006,Dzero2007,QReview2015}. Their onset at the transition from phase II to phase III can be viewed as a nonlinear counterpart of the separation of an amplitude excitation from the continuum \footnote{{At the II--III boundary, a second complex-conjugate pair of zeros of $\mathbf L^2(u)$ separates from the real axis at $u=0$. The phase-II continuum frequencies are $2\sqrt{u^2+\Delta_\infty^2}$, with $u$ measured from the chemical potential and $\Delta_\infty$ the asymptotic gap, so the oscillations begin at frequency $2\Delta_\infty$, just below the continuum of frequencies corresponding to nonlinear extension of the other normal modes. Thermal averaging can damp these oscillations for quenches from a normal initial state, whereas this dephasing is negligible for a paired initial state~\cite{YuzbashyanTsyplyatyev2009}.}}}.

Closest to the present problem is the work of Alth\"user and Uhrig
\cite{Uhrig2025,Uhrig2026}. They start from the electron-phonon Hamiltonian
with a single Einstein branch with frequency $\Omega$ and map it by a continuous unitary
transformation onto an effective electronic model, whose pairing kernel is
restricted by a cutoff on the energy transfer between the scattering
electrons. The kernel retains, in this sense, the retardation of the
phonon-mediated interaction. What it does not retain is the renormalization of
the single-particle dispersion which the same interaction generates, and which
is set aside explicitly \cite{Uhrig2025}. One is left with a static kernel
that depends on the band energies and makes the order parameter a function of
them, i.e. with the same mechanism as in Ref.~\cite{Levitov2007}. It likewise
detaches the Higgs mode from the continuum by an amount of order unity, with
or without the Coulomb interaction \cite{Uhrig2026}. In addition,
Ref.~\cite{Uhrig2026} finds that the number of discrete subgap modes grows
with the coupling: above a threshold value, each increase of the gap by
roughly $0.6\,\Omega$ adds one more, the new modes alternating between
amplitude and phase character, so that the Higgs mode becomes the lowest
member of a tower of in-gap excitations. It should be noted that this tower
depends on neither the lattice nor the position of the Fermi level.
 
The opposite conclusion is reached when the phonon dynamics is kept in full.
Murakami, Werner, Tsuji and Aoki have studied the amplitude mode in the
Holstein model by employing nonequilibrium dynamical mean-field theory with a
self-consistent Migdal impurity solver, in which retardation, mass
renormalization and the phonon self-energy all enter. These authors found the
Higgs frequency to coincide with the pair-breaking energy threshold
$2\Delta_{\rm SC}$ to within the accuracy of the simulation, even in the
strongly coupled regime beyond
BCS~\cite{Werner2016,Murakami2016damping,Tsuji2016retarded}. The additional
mode which they identify lies \emph{above} the pair-breaking continuum rather
than below it. Apparently the same conclusion is reached, at a resolution of several percent, by the real-time Keldysh simulation of Ref.~\cite{Kemper2015},
which is the closest to the present model in that the phonon is held in
equilibrium. There the order parameter oscillation frequency after a pump extrapolates to
$2\Delta_0$ in the limit of vanishing pump strength, within the
several-percent resolution of a five-period window at $T=0.4\,T_c$.
 
The two treatments start from the same electron-phonon problem and answer the
question about the energy of the amplitude mode in opposite ways. One places
the mode deep inside the gap, the other one on its edge. The two limitations
are, however, of different kinds. The effective static kernel omits an
ingredient of the problem, namely the mass renormalization which in
Migdal-Eliashberg theory accompanies the energy dependence of the pairing
vertex. The real-time simulations omit nothing of the sort, but they resolve
the position of the mode only to between several and twenty percent. Therefore one may wonder what does the standard
description of these materials give for the frequency of an amplitude mode on
the real frequency axis? In other words is the amplitude mode frequency different from the pair breaking edge $2\Delta_{\rm g}$ in the Migdal-Eliashberg theory, in which the
energy dependence of the pairing interaction is not imposed but rather follows from
the phonon dynamics, together with the mass renormalization and the
quasiparticle damping which come with it? To the best of our knowledge this
question has not been addressed yet. 
 
We address precisely this question by re-deriving nonequilibrium quasiclassical equations for
strong-coupling superconductors and then using these equations to analyze the
collective amplitude mode. Two controls are established first: our equations
are equivalent to the ones obtained by Eliashberg for a superconductor in an
external electromagnetic field~\cite{Eliashberg1972}, and in equilibrium the
mass renormalization and the order parameter reduce to the standard
Migdal-Eliashberg expressions of Refs.~\cite{Marsiglio1991,Marsiglio2020}. We
then linearize these equations about the equilibrium saddle point in the
amplitude channel, driven by an external pair field. Throughout the paper we
distinguish the order parameter $\Delta(\eps)$ from the gap edge $\Delta_g$ of
the single-particle spectrum, the latter being fixed by the root of the
equation $\mathrm{Re}\Delta(\eps)=\eps$. The two coincide in BCS theory but
not at strong coupling, where the retarded self-energy pushes them apart. It
is $2\Delta_g$, the pair breaking threshold, against which the mode frequency
must be measured; in this regard the situation is similar to the case of BCS
superconductors in the presence of weak pair breaking interactions
\cite{LiDzero2024,Kamenev2025}. Consequently, we identify the energy of the
amplitude mode as the complex frequency at which the linearized operator, an
analogue of the longitudinal pair susceptibility, becomes singular. Our main
results are summarized in Fig.~\ref{Fig1-schematic}.
 
Two independent routes lead to the shift of the Higgs mode. On the imaginary
axis the linearized Eliashberg system with a pair source at bosonic transfer
frequency yields the longitudinal pair propagator to essentially machine
precision, with no quadrature and no branch cuts. These data show that the
mass renormalization does affect the magnitude of the longitudinal pair
susceptibility, and they exclude a shift in the Higgs frequency of order
$O(1)$ or higher. They cannot, however, resolve a percent-level displacement.
The reason is that in the BCS limit the zero of the propagator sits exactly
\emph{on} its nearest branch point, which is the analytic content of the
$1/\sqrt{t}$ decay, and the continuation error grows exponentially with the
distance from the data to that coincident feature. The quantitative answer
therefore has to come from the real axis, where we assemble the linearized
Eilenberger-Keldysh problem as a dense operator acting on the retarded,
advanced and Keldysh amplitude blocks, and locate the mode from the resonance
in the resulting susceptibility, extrapolated to vanishing quasiparticle
broadening.
 
What we find is a pole below the pair-breaking threshold. In the
intermediate-coupling window $0.56\le\Delta_g/\Omega\le0.90$ its binding
energy is
\begin{equation}\label{Eq1}
  2\Delta_g-\omega_{\mathrm{H}}\;=\;0.028\,\frac{\Delta_g^{2}}{\Omega},
\end{equation}
to within $12\%$ over that range, with the coefficient itself determined to
about $20\%$ by the calibration of the extraction (Appendix~\ref{AppendixE}),
and with a damping which is zero within the numerical noise at $T\ll\Delta_g$.
The displacement is thus of order one percent of $2\Delta_g$. Its smallness is
structural rather than numerical. Retardation leaves the coincidence of the
zero with the branch point in place, and deforms only the regular part of the
inverse response, by a relative amount $\delta A/A$ of order ten percent. A
zero of square-root character moves quadratically in such a deformation, see
Eq.~\eqref{quadratic} below. A ten-percent deformation of the threshold state
therefore produces a one-percent correction to the binding energy. We
emphasize that it is the power, and not the value of the numerical
coefficient, which carries the physical content of this estimate; what
Eq.~\eqref{Eq1} adds is the coefficient of that quadratic law in the window
where the binding is resolved.
 
We are now in a position to compare Eq.~\eqref{Eq1} with the results discussed
above. The displacement itself is real, in the sense that a genuinely retarded
kernel does detach the mode, so that the static-kernel calculations are not
reporting a spurious effect. Its magnitude, however, is far smaller than what
those calculations give. For the energy-transfer kernel of
Refs.~\cite{Uhrig2025,Uhrig2026} the mode leaves the neighbourhood of the
threshold once the coupling exceeds a value of order unity, and is then bound
by a finite fraction of the gap \cite{Uhrig2026}. Equation \eqref{Eq1}, by
contrast, gives $1.2\%$ of $2\Delta_g$ at $\Delta_g/\Omega=0.9$, and in the
strong-coupling limit $\Delta_g/\Omega=\infty$ no subgap pole with a binding
above one percent is found (Appendix~\ref{AppendixF}). We note in passing that
the interval $1\lesssim\Delta_g/\Omega<\infty$ between the two is not covered
by either calculation. In Sec.~III we show where the difference comes from:
a pairing kernel which depends on the band energies makes the order parameter
itself a function of the band energy, $\Delta\to\Delta(\xi)$
\cite{Levitov2007}, and this breaks, at every $\xi$ separately, the identity
that pins the BCS mode to the threshold. A kernel which depends on the
frequency transfer, as the phonon dynamics prescribes, leaves the $\xi$-level
structure of the threshold intact, even though the identity itself no longer
holds frequency by frequency. It deforms only the regular part of the
response, and the mass renormalization which accompanies it enters at second
order.
 
As for the real-time simulations, we can make a rough comparison of
resolution. In the notation of the respective papers,
$\Delta_{\rm SC}/\omega_0\sim0.16$--$0.2$ in Ref.~\cite{Werner2016} and
$\Delta_0/\Omega\sim0.24$ in Ref.~\cite{Kemper2015}. The frequency
resolution is set by the length of the time window, $2\pi/t_{\max}$: it is
about twenty percent of $2\Delta_{\rm SC}$ for the propagation time of
Ref.~\cite{Werner2016}, and several percent for the five-period window of
Ref.~\cite{Kemper2015}. Using these ratios in Eq.~\eqref{Eq1} gives a rough estimate of order
$0.3\%$\footnote{{The frequency $\omega_0$ in Ref.~\cite{Werner2016} is the bare Holstein phonon frequency, whereas our Einstein frequency $\Omega$ is the physical, renormalized phonon frequency. For this order-of-magnitude comparison we neglect the difference between bare and renormalized frequencies.}} for the relative displacement. A displacement one
to two orders of magnitude below the resolution is invisible there, so that
the absence of a shift in those simulations is consistent with its presence
here.
 
Ordinarily a percent-level displacement would produce a percent-level effect,
but here that expectation fails. The mode detaches from a square-root branch
point, so that its residue is governed by
$\nu=\sqrt{4\Delta_g^{2}-\omega_{\rm H}^{2}}\simeq 2\sqrt{\Delta_g|E_b|}$
rather than by the binding energy $E_b$ itself. The square root converts a
displacement of one percent into a spectral weight of order twenty-five
percent. A quarter of the amplitude-channel weight leaves the
two-quasiparticle continuum and reappears as a discrete line below it,
Fig.~\ref{Fig1-schematic}. What experiment sees is therefore not the
displacement but its consequence: undamped periodic oscillations of the order parameter amplitude.
 
The rest of the paper is organized as follows. In Sec.~II we set up the
quasiclassical equations for the Migdal-Eliashberg model and discuss their
equilibrium limit. In Sec.~III we derive the linearized amplitude problem with
an external pair source, on the imaginary and on the real frequency axis.
Section~IV presents the results of our numerical solution and the mode
frequency as a function of $\Delta_g/\Omega$. In Sec.~V we discuss the
comparison of our main results with the existing literature and the
consequences for pump-probe and driven-superconductor experiments, and in
Sec.~VI we summarize our work. Technical details of the calculations are
delegated to the Appendices.

\section{Main equations}

\subsection{Model}
After integrating out the phonon degrees of freedom, we obtain the effective action for electrons interacting through virtual phonons
\beg\label{Action}
\begin{aligned}
S={}&\sum_{\sigma}\int_{\cal C}dt\int d^3\br\,
\overline{\psi}_\sigma(\br,t)
\bigl(i\partial_t-\hat H\bigr)\psi_\sigma(\br,t)\\
&-\frac12\sum_{\sigma,\sigma'}\int_{\cal C}dt\,dt'
\int d^3\br\,d^3\br'\,
\overline{\psi}_{\sigma'}(\br',t')\psi_{\sigma'}(\br',t')\\
&\qquad\times\mathcal D(\br',t';\br,t)\,
\overline{\psi}_\sigma(\br,t)\psi_\sigma(\br,t).
\end{aligned}
\en
Here $\hat H=-\nabla^2/(2m)-\mu$ is the single-particle Hamiltonian, $\mu$ is the chemical potential, and $\psi_\sigma(\br,t)$ is the fermionic field with spin projection $\sigma=\up,\dn$. Both time integrations run along the Keldysh contour $\mathcal C$. The kernel $\mathcal D(\br',t';\br,t)$ is the contour-ordered physical phonon propagator with the electron-phonon coupling vertices included. We make no assumption about the phonon dispersion at this stage.

We use the standard isotropic Migdal--Eliashberg theory, taking the physical phonon spectrum and electron--phonon matrix elements as input. These enter the theory through the Eliashberg function $\alpha^2F(\omega)$~\cite{AllenMitrovic1983,Carbotte1990,Marsiglio2008,Marsiglio2020}. We solve self-consistently for the electronic self-energies in the Migdal approximation~\cite{Migdal1958,Eliashberg1960}. The physical phonon spectrum already includes renormalization due to the electrons. We therefore do not calculate this renormalization again~\cite{BrovmanKagan1967,Geilikman1975,Giustino2017}. We keep the phonon propagator fixed when calculating the electronic response. We set the Coulomb pseudopotential $\mu^*$ to zero. We may do so because the Coulomb repulsion is instantaneous: it acts on energy scales of order $\veps_F\gg\Omega$, and on the scales below $\Omega$ which concern us here it renormalizes $\Delta_g$ without changing the kernel structure of the resulting equations (see the end of Sec.~III.A).

Three considerations motivate the choice of the Migdal-Eliashberg formulation. First, the question we address concerns the analytic structure of the amplitude susceptibility $\chi(\omega)$ on the real frequency axis, in particular the position of its pole relative to the pair-breaking branch point at $2\Delta_g$, at the percent level. As we show below, the Eliashberg equations can be formulated directly on the real axis, and the mode is then located as the frequency at which a linear operator becomes singular. Second, the Eliashberg formulation isolates retardation as the only available mechanism. {With $\alpha^2F(\omega)$ held fixed and no dynamical feedback on the physical phonon propagator, any displacement of the mode from $2\Delta_g$ within our quasiclassical formulation must originate in the frequency dependence of the electronic pairing kernel and the accompanying mass renormalization, rather than in a change of the phonon spectrum or in details of the band structure.} This exposes the microscopic origin of the effect. Third, once the phonon spectrum is reduced to a single frequency $\Omega$, as is done below, the model depends on the dimensionless coupling constant $\lambda=2\int_0^\infty d\omega\,\alpha^2F(\omega)/\omega$ and on the ratio $T/\Omega$ alone, so varying $\lambda$ is unambiguous, and $\lambda$ is the quantity in which the coupling strength of the relevant materials is conventionally reported. {We neglect corrections to the elementary electron--phonon vertex in the Migdal approximation and keep the phonon propagator fixed in the response calculation. Our calculation therefore cannot produce the additional phonon-derived collective mode generated by phonon feedback in the dynamical mean-field treatment~\cite{Werner2016}. That absence is what makes it a clean test of whether such a mode is required to displace $\omega_{\rm H}$ from the pair-breaking edge.

{For simplicity, we take physical Einstein phonons of frequency $\Omega$, as appropriate to an optical branch with negligible dispersion. This assumption concerns the physical spectrum, in contrast to the Holstein Hamiltonian, where the \emph{bare}  frequency $\omega_0$ is momentum independent, while the physical phonon spectrum is obtained by renormalization within that model. Instead, we specify the physical spectrum directly by taking}
\beg\label{alpha2Fw}
\alpha^2F(\omega)=\frac{\lambda\Omega}{2}\delta(\omega-\Omega)
\en
{and work in the Migdal limit $\Omega/\veps_F\to0$ at fixed $\lambda$, neglecting corrections to the elementary electron--phonon vertex. The restriction to Einstein phonons is only for simplicity. For an arbitrary function $\mathcal U(\omega)$, we make the replacement}
{
\beg\label{SpectralReplacement}
\frac{\lambda\Omega}{2}\mathcal U(\Omega)
\;\longrightarrow\;
\int_0^\infty d\omega\,\alpha^2F(\omega)\mathcal U(\omega)
\en
in the self-energies and response kernels before solving the equations self-consistently. All dependence on the phonon frequency, including thermal factors and shifts of the electronic frequencies, must be included in $\mathcal U$.

The quasiclassical equations we use are the standard Eilenberger-Keldysh
equations of the strong-coupling theory
\cite{Eliashberg1972,LO1975,LO1977,SereneRainer1983,Kopnin2001}. We
summarize them here to fix the conventions, in particular the separation
of the order parameter from the collision integral and the form of the
self-energies on which the linearization of Sec.~III is built, and
relegate the remaining definitions to Appendix~\ref{AppendixA}.

{We first derive the contour Dyson equations from the effective action~\eqref{Action} and evaluate the electronic self-energy in the self-consistent Migdal approximation. We define the single-particle propagators on the Keldysh contour as}
\beg\label{Gab}
G_{\alpha\alpha'}^{(jj')}(1,2)=\frac{1}{i}\langle\hat{T}_t\Psi_{\alpha}(1_j)\Psi_{\alpha'}\dg(2_{j'})\rangle.
\en
The average is taken with the action \eqref{Action}, $\Psi=(\psi_\up ~\overline{\psi}_\dn)^T$ is the Nambu spinor, $(1)=(\br_1,t_1)$, and $\hat{T}_t$ is the time-ordering operator along the Keldysh contour, whose top and bottom parts are labeled $j=1$ and $j=2$ \cite{VolkovKogan1974,Kamenev2011}. We use the standard convention that for $t_1<t_2$ an event at $t_1$ on the bottom part of the contour is later than an event at $t_2$.

Given the definition \eqref{Gab}, only three of the four functions defined on the Keldysh contour are independent, because of the relation
$\hat{G}^{(11)}+\hat{G}^{(22)}=\hat{G}^{(12)}+\hat{G}^{(21)}$.
It is therefore standard to introduce three independent functions. The retarded function is
$\hat{G}^R(1,2)=\hat{G}^{(11)}-\hat{G}^{(12)}$,
the advanced one is
$\hat{G}^A(1,2)=\hat{G}^{(11)}-\hat{G}^{(21)}$,
and the Keldysh one is $\hat{G}^K(1,2)=\hat{G}^{(11)}+\hat{G}^{(22)}$. {For the Einstein spectrum, we write the isotropically averaged interaction kernel $\mathcal D$ as $g_c^2D$, with the normalization $\nu_Fg_c^2=\lambda\Omega/2$, where $\nu_F$ is the normal-state density of states per spin at the Fermi level. Thus $g_c$ is fixed by the physical input spectrum and is not a bare Holstein coupling. The retarded, advanced and Keldysh parts of the physical phonon propagator $D$ are defined in the same way as those of $G$. In the Fourier representation they are}
\beg\label{DKRA}
\begin{aligned}
D^{R(A)}(\omega)&=\frac{2\Omega}{(\omega\pm i0)^2-\Omega^2}
\end{aligned}
\en
and $D^K(\omega)=-2\pi i[\delta(\omega+\Omega)+\delta(\omega-\Omega)]\coth(\Omega/2T)$.

Since the resulting Dyson equations contain derivatives with respect to both coordinates of $\hat{G}(1,2)$, the second step is to pass to the center-of-mass coordinates $\br=(\br_1+\br_2)/2$, $t=(t_1+t_2)/2$ and to Fourier transform with respect to the relative coordinates $\br_{12}=\br_1-\br_2$ and $t_{12}=t_1-t_2$. It is then convenient to introduce the function
\beg\label{SmallgRA}
\check{g}_{\bn\eps}(\br,t)=\frac{i}{\pi\nu_F}\int\limits_0^\infty\frac{p^2dp}{2\pi^2}\left[\begin{matrix} \hat{\tau}_3\hat{G}_{\bp\eps}^{R}(\br,t) & \hat{\tau}_3\hat{G}_{\bp\eps}^{K}(\br,t) \\ 0 & \hat{\tau}_3\hat{G}_{\bp\eps}^{A}(\br,t)
\end{matrix}\right],
\en
where $\bn=\bp/p$ and $\nu_F$ is the single-particle density of states at the Fermi level. This function satisfies the Eilenberger equation, written here in the conventions of Appendix~\ref{AppendixA}:
\beg\label{NowFinalEilenberger}
\begin{aligned}
\left[\eps\check{\tau}_3,\check{g}_{\bn\eps}\right]_{-}&+\left[\check{\Delta}_\eps(\br,t)\stackrel{\circ},\check{g}_{\bn\eps}\right]_{-}+\frac{i}{2}\left[\check{\tau}_3,\partial_t\check{g}_{\bn\eps}\right]_{+}\\&+iv_F\left(\bn{\mbox{\boldmath $\nabla$}}_\br\right)\check{g}_{\bn\eps}=\check{I}(\bn\eps;\br t).
\end{aligned}
\en
Here $\check{I}(\bn\eps;\br t)$ is the electron-phonon collision integral, and the product $A\circ B$ and the order parameter matrix
$\check{\Delta}_\eps(\br,t)$ are defined in Appendix~\ref{AppendixA}. The solution of \eqref{NowFinalEilenberger} must also satisfy the normalization condition $\check{g}\circ\check{g}=\check{1}$.

\subsection{Ground state}
We now turn to the solution of \eqref{NowFinalEilenberger} in the ground state, where $\check{g}_{\bn\eps}$ is uniform, stationary and independent of $\bn$; we denote this equilibrium value by $\check{\cal G}_\eps$. It is convenient to work directly with the self-energy parts, which amounts to re-absorbing $\check{\Delta}_\eps$ into the collision integral. We take the retarded and advanced self-energies in the form
\beg\label{SigmaREqAnsatz}
\hat{\Sigma}_\eps^{R(A)}=\eps(1-Z_\eps^{R(A)})\hat{\tau}_0+\Phi_\eps^{R(A)}\hat{\tau}_1
\en
and the retarded and advanced blocks of $\check{\cal G}_\eps$ in the corresponding form $\hat{\cal G}_\eps^{R(A)}=g_\eps^{R(A)}\hat{\tau}_3+if_\eps^{R(A)}\hat{\tau}_2$.
Here $Z_\eps^R$ and $\Phi_\eps^R$ are functions to be determined self-consistently, and $g_{\eps}^A=-(g_\eps^R)^*$, $f_{\eps}^A=-(f_\eps^R)^*$.
The phase of the equilibrium order parameter is set to zero, which is why \eqref{SigmaREqAnsatz} contains no $\hat{\tau}_2$ term. In equilibrium the Keldysh component of the self-energy is fixed by the fluctuation-dissipation theorem,
\beg\label{FDT}
\hat{\Sigma}_{\eps}^{K}=\left(\hat{\Sigma}_{\eps}^{R}
-\hat{\Sigma}_{\eps}^{A}\right)\tanh\frac{\eps}{2T}.
\en
The same theorem gives, for the propagator,
\beg\label{GKParametrization}
\hat{\cal G}_\eps^K=\left(\hat{\cal G}_\eps^R-\hat{\cal G}_\eps^A\right)t_\eps
\en
with $t_\eps\equiv\tanh({\eps}/{2T})$. Substituting \eqref{SigmaREqAnsatz} and \eqref{GKParametrization} into the retarded self-energy \eqref{SigmaREq2m} of Appendix~\ref{AppendixA}, and using the fact that the integrals of $\hat{g}^R_{\eps-\omega}D^A_\omega$ and of $\hat{g}^A_{\eps-\omega}D^R_\omega$ over $\omega$ vanish by analyticity, we obtain the system of equations
\beg\label{EliEqsSet1}
{\begin{aligned}
\eps(1-Z_\eps^R)&=-\frac{i\pi\lambda\Omega}{4}\coth\left(\frac{\Omega}{2T}\right)\sum\limits_{s=\pm}g_{\eps+s\Omega}^R\\&+\frac{\lambda\Omega}{4}\int d\eps'D_{\eps-\eps'}^R\textrm{Re}(g_{\eps'}^R)t_{\eps'}, \\
\Phi_\eps^R&=\frac{i\pi\lambda\Omega}{4}\coth\left(\frac{\Omega}{2T}\right)\sum\limits_{s=\pm}f_{\eps+s\Omega}^R\\&-\frac{\lambda\Omega}{4}\int d\eps'D_{\eps-\eps'}^R\textrm{Re}(f_{\eps'}^R)t_{\eps'}.
\end{aligned}}
\en
Here $\lambda=2\nu_Fg_c^2/\Omega$ is the coupling constant of \eqref{alpha2Fw}. It remains to express $g_\eps^R$ and $f_\eps^R$ in terms of $Z_\eps^R$ and $\Phi_\eps^R$, which is accomplished using the Eilenberger equation in equilibrium,
\beg\label{EqMEEilen}
[\eps\hat{\tau}_3-\hat{\Sigma}_\eps^R\hat{\tau}_3,\hat{\cal G}_\eps^R]=0
\en
together with the normalization condition
$(\hat{\cal G}_\eps^R)^2=\hat{1}$. With the self-energy \eqref{SigmaREqAnsatz}
the solution is
\beg\label{GRclosure}
\hat{\cal G}_\eps^R=\frac{\eps Z_\eps^R\hat{\tau}_3+i\Phi_\eps^R\hat{\tau}_2}
{\sqrt{(\eps Z_\eps^R)^2-(\Phi_\eps^R)^2}}.
\en
Furthermore, the relation $\hat{\cal G}^A_\eps=-\hat{\tau}_3(\hat{\cal G}^R_\eps)^\dagger\hat{\tau}_3$ gives $\eps Z_\eps^A=(\eps Z_\eps^R)^*$ and
$\Phi_\eps^A=(\Phi_\eps^R)^*$. In Appendix~\ref{AppendixB} we show that after substituting \eqref{GRclosure} into \eqref{EliEqsSet1} the resulting closure equations coincide with those given in Ref.~\cite{Marsiglio1991}.

The order parameter $\Delta(\eps)$ of the Introduction is the gap function $\Delta_\eps=\Phi_\eps^R/Z_\eps^R$, and the gap edge $\Delta_g$ is the root of $\textrm{Re}\Delta_\eps=\eps$, so that the pair-breaking threshold is $2\Delta_g$. It is to be distinguished from the real part of the anomalous self-energy, $\textrm{Re}\,\Phi_\eps$, which in equilibrium is what enters the matrix $\check{\Delta}_\eps$ of \eqref{NowFinalEilenberger} (Appendix~\ref{AppendixA}); the two coincide in BCS theory, where $Z_\eps=1$ and $\Phi_\eps$ is real. The result of the numerical solution of Eqs.~\eqref{EliEqsSet1} for $\Delta_\eps$ is shown in Fig.~\ref{Fig-MC}.
\begin{figure}
\includegraphics[width=0.9950\linewidth]{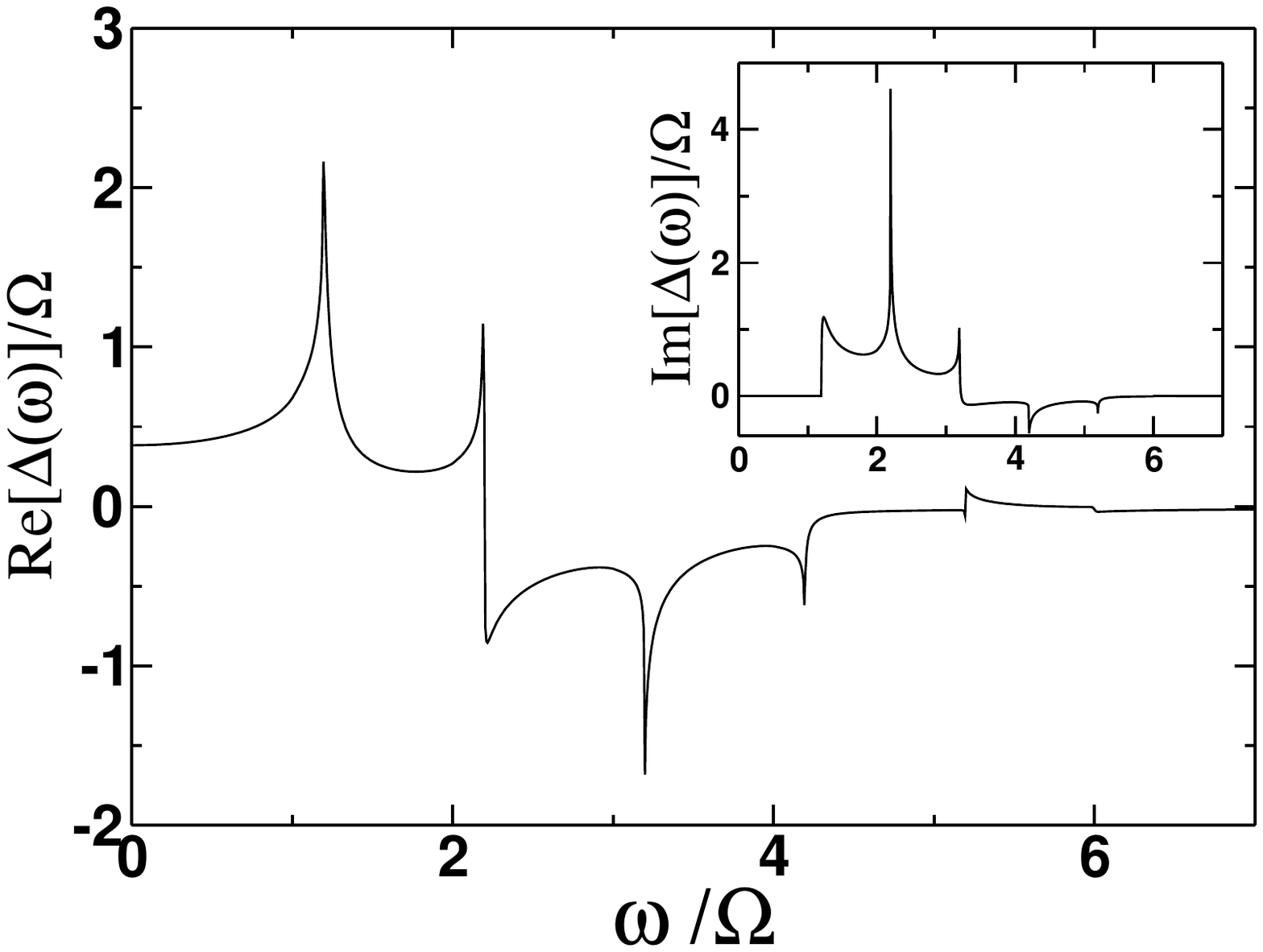}
\caption{Real (main panel) and imaginary (inset) parts of the equilibrium order parameter $\Delta(\omega)$ obtained by solving Eqs.~\eqref{EliEqsSet1} for $\lambda=0.95$ and $T=0.005\Omega$.}
\label{Fig-MC}
\end{figure}
\begin{figure}
\includegraphics[width=0.9950\linewidth]{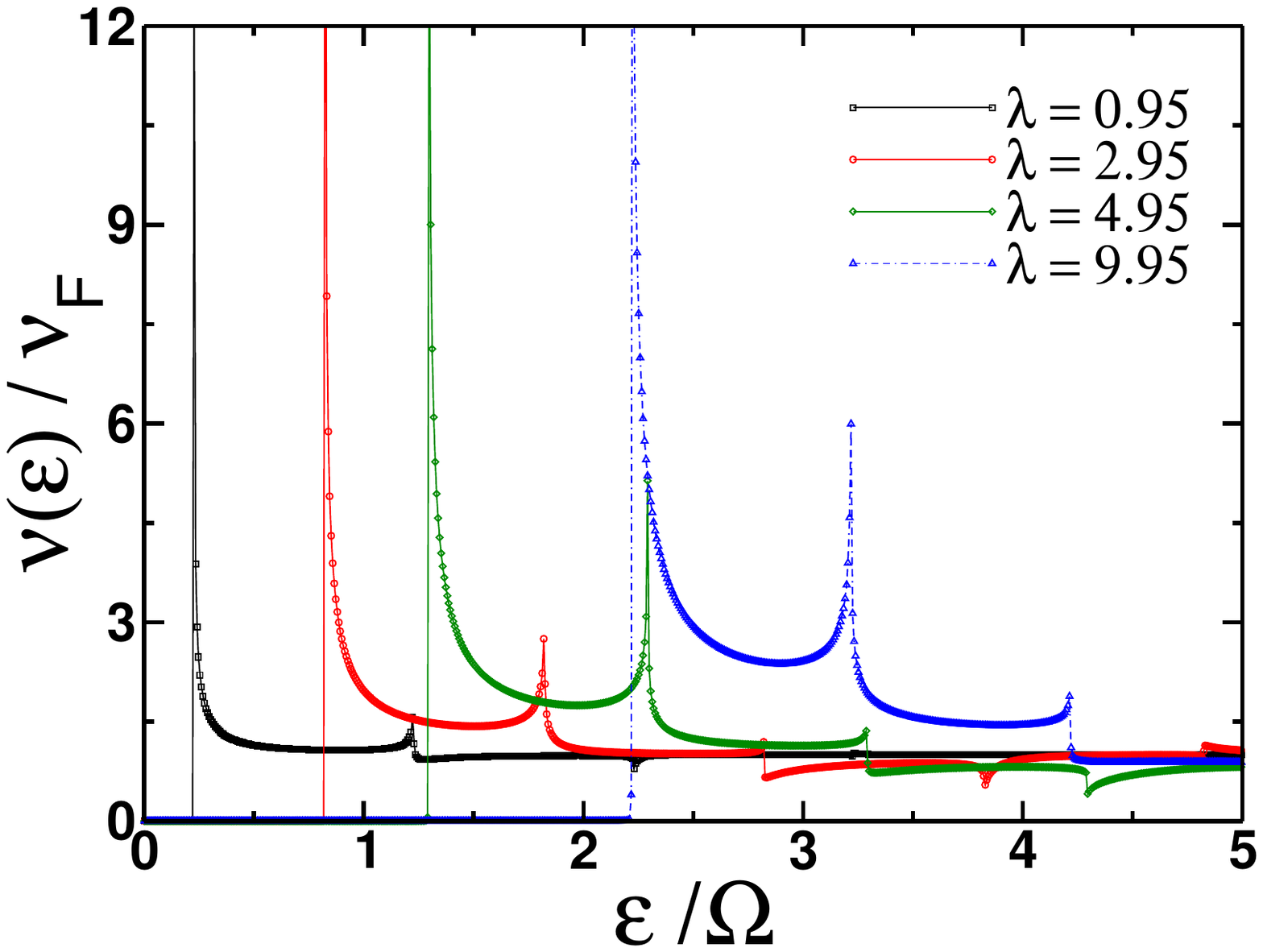}
\caption{Single-particle density of states \eqref{DOS} for several values of the dimensionless coupling constant $\lambda$ at $T=0.005\Omega$.}
\label{Fig-DOS}
\end{figure}
The same solution determines the single-particle density of states,
\beg\label{DOS}
\nu(\eps)=\nu_F\textrm{Re}\left({g_\eps^R}\right),
\en
which is shown in Fig.~\ref{Fig-DOS} for several values of $\lambda$. The spectrum has its gap edge at $\Delta_g$ and a phonon-assisted structure at $\Delta_g+\Omega$, which approaches the pair-breaking threshold $2\Delta_g$ as the coupling grows and crosses it at $\Delta_g/\Omega=1$ (Sec.~IV.A).

\section{Linear analysis}
This Section evaluates the longitudinal pair susceptibility of a superconductor subjected to a weak external perturbation. It begins by stating precisely what the BCS ``Higgs mode'' is
and what kind of object it becomes in Migdal--Eliashberg theory. 

In the BCS model
the mode condition is in fact satisfied at $\omega=2\Delta$ provided $\bq=0$ and no pair breaking is present: the inverse
propagator \eqref{BCSclosed} vanishes there, and the response
$\chi(\omega)$ diverges at the edge as $(4\Delta^2-\omega^2)^{-1/2}$. A sharp mode requires isolation, i.e. a zero of $\chi^{-1}$ must be separated from the branch point. The BCS theory fails this criterion: the zero sits exactly on the branch
point, so no pole can be split off from the two-particle continuum
\cite{ASchmid1968,Gurarie2009,ChubukovHiggs2023,Nosov2025}. Therefore, the amplitude response is a
\emph{threshold resonance}: a divergent, non-Lorentzian spectral peak pinned
at $2\Delta$, and a free periodic oscillation whose frequency is exactly
$2\Delta$ while its envelope decays algebraically,
$\delta\Delta(t)\propto\cos(2\Delta t+\pi/4)/\sqrt{2\Delta t}$
\cite{VolkovKogan1974,YuzbashyanDzero2006,QReview2015}. The frequency of the mode is sharply
defined but the mode as a discrete oscillator is not. The Anderson pseudospin
representation \cite{Anderson1958,Enolski2005,Enolski2005a,Yuzbashyan2006,QReview2015} makes the distinction physical: each
Cooper pair precesses at its own frequency $2E_\bp=2\sqrt{\eps_\bp^2+\Delta^2}$,
a continuum bounded from below by $2\Delta$ with an inverse-square-root van
Hove singularity in the density of precession frequencies at its bottom.
The BCS amplitude ``mode'' is this band edge --- a coherent superposition
dominated by the slowest oscillators --- and the $1/\sqrt{t}$ law is the
dephasing of that superposition. At long times only the pairs near the minimum
of the precession frequency, $\eps_\bp\simeq0$, remain in step, and by the
stationary-phase estimate their number shrinks as $t^{-1/2}$, with the carrier
at exactly $2\Delta$ and the phase $\pi/4$ of Refs.~\cite{YuzbashyanDzero2006,QReview2015}.
In the language of quantum-mechanical scattering it is a bound state
at exactly zero binding energy, and the algebraic decay is the standard
threshold law. In Migdal--Eliashberg theory this mode becomes isolated, as we now show. 

\subsection{Imaginary time representation}
On the imaginary axis the linear response
to a pair field is a real dense linear
system that can be solved numerically without quadrature or analytic
continuation. The response splits into a sourced channel,
the pairing vertex $\delta\Phi$, and an unsourced one, the mass renormalization
$\delta W$, and the latter is not negligible: it enters the ladder at the
ten-percent level at intermediate coupling, grows with $\lambda$, and
supplies the deformation to the pairing vertex. It
nevertheless displaces the amplitude mode by only a percent, because the unsourced
channel acts at second order in the interchannel coupling and on a zero of
square-root character. The imaginary-axis data bound the
displacement from above but do not resolve it, which is what sends the
quantitative determination to the real axis in Sec.~III.B and Sec.~IV.

On the imaginary axis the equilibrium problem is described by Eqs.~\eqref{EliEqsOpt}. 
We collect its two functions into the Nambu matrix 
$\hat g_n=g_n\hat{\tau}_3+f_n\hat{\tau}_2$,
where $g_n=\tilde\omega_n/D_n$,
$f_n=\Phi_n/D_n$, $D_n=\sqrt{\tilde\omega_n^2+\Phi_n^2}$, $\tilde\omega_n=\omega_n+\Sigma_n$ so that $g_n^2+f_n^2=1$ (the factor $i$ of \eqref{GRclosure}
is absorbed on continuation, see \eqref{ContinueGF} in Appendix~\ref{AppendixB}).
The function $\hat g_n$ is the imaginary-axis image of \eqref{GRclosure} and obeys the normalization condition
\beg\label{Normgn} 
\hat g_n^2=\hat 1, 
\en
which is the Matsubara form of the normalization condition stated
below \eqref{NowFinalEilenberger}. Writing
$\hat M_n=D_n\hat g_n$, the
equilibrium equations \eqref{EliEqsOpt} read compactly
\beg\label{eqMats}
\hat M_n=\omega_n\hat\tau_3+\pi T\sum_m\lambda_{n-m}\hat g_m,
\en
{and the interaction kernel in \eqref{eqMats} is}
{
\beg\label{GeneralMatsKernel}
\begin{aligned}
\lambda_{n-m}&=2\int_0^\infty d\nu\,
\frac{\nu\alpha^2F(\nu)}{\nu^2+(\omega_n-\omega_m)^2}\\
&\stackrel{\mathrm{Einstein}}{=}
\frac{\lambda\Omega^2}{\Omega^2+(\omega_n-\omega_m)^2}.
\end{aligned}
\en
}
{We linearize at fixed $\mathcal D$, so $\delta\mathcal D=0$. The physical phonon spectrum and the equilibrium phonon distribution remain fixed. We vary the electronic self-energies and the pair-response vertex self-consistently.}

The simplest perturbation adds to the anomalous self-energy a
$c$-number pair field of bosonic transfer frequency $\nu_l=2\pi Tl$,
\beg\label{Source}
\Phi(\tau_1,\tau_2)\to\Phi(\tau_1,\tau_2)+he^{-i\nu_l\tau_1}\delta(\tau_1-\tau_2).
\en
This amounts to the shift $\hat M_n\;\longrightarrow\;\hat M_n+h\,\hat\tau_2$ and allows us to linearize the Eilenberger equation in $h$. Since the perturbation carries the two frequencies
$\omega_m$ and $\omega_m+\nu_l$ we write $\delta\hat g_m$
for the amplitudes of the response at transfer $\nu_l$. This yields
\beg\label{dM}
\hat M_{m+l}\delta\hat g_m-\delta\hat g_m\hat M_m+\delta\hat M_m\hat g_m-\hat g_{m+l}\delta\hat M_m=0.
\en
The source preserves \eqref{Normgn}, which at transfer $\nu_l$ linearizes to
\beg\label{LinNorm}
\hat g_{m+l}\,\delta\hat g_m+\delta\hat g_m\,\hat g_m=0,
\en
and with its help the solution of \eqref{dM} for $\delta\hat{g}_m$ reads
\beg\label{Sylvester}
\begin{aligned}
&\delta\hat g_m=\frac{\delta\hat M_m-\hat g_{m+l}\,\delta\hat M_m\,\hat g_m}{D_{m+l}+D_m}.
\end{aligned}
\en

The closure equations follow from the ansatz
\beg\label{dM2comp}
\delta\hat M_m=\delta W_m\hat\tau_3+\delta\Phi_m\hat\tau_2.
\en
The variation of \eqref{eqMats}, with \eqref{dM2comp} and 
\eqref{Sylvester}, gives after a short calculation
\beg\label{MatsSystem}
\begin{aligned}
\delta W_n&=\pi T\sum_m\lambda_{n-m}
\left(P_m\,\delta W_m-Q_m\,\delta\Phi_m\right),\\
\delta\Phi_n&=h+\pi T\sum_m\lambda_{n-m}
\left(R_m\,\delta\Phi_m-Q_m\,\delta W_m\right).
\end{aligned}
\en
Here we introduced the following three kernels
\beg\label{PQR}
\begin{aligned}
P_m&=\frac{1+f_{m+l}f_m-g_{m+l}g_m}{D_{m+l}+D_m},\\
R_m&=\frac{1+g_{m+l}g_m-f_{m+l}f_m}{D_{m+l}+D_m},\\
Q_m&=\frac{g_{m+l}f_m+f_{m+l}g_m}{D_{m+l}+D_m}.
\end{aligned}
\en

Two features of \eqref{MatsSystem} matter below. First, the
kernel is $\lambda_{n-m}$ and carries \emph{no} shift by $l$: the boson
transfer between the two frequency intervals is $(\omega_n+\nu_l)-(\omega_m+\nu_l)
=\omega_n-\omega_m$. 
Second, the source enters the second of Eqs.~\eqref{MatsSystem} alone: the pair
field is $h\hat\tau_2$, whereas $\delta W$ is the $\hat\tau_3$ component of
$\delta\hat M$.  In the second of Eqs.~\eqref{MatsSystem} the $-Q_m\,\delta W_m$ term enters
with the same sign as $R_m\,\delta\Phi_m$ and contributes about ten percent of
the Matsubara ladder at every transfer frequency. The mass-renormalization response thus acts
as an additional attraction in the amplitude channel, which is the sign
required for a bound state below the threshold rather than a virtual state
above it. Setting $\delta W\equiv0$ --- as is implicit in any treatment with an
instantaneous kernel --- reduces the binding below the resolution of the
calculation, as we show below.

The longitudinal pair susceptibility at $\bq=0$ is the response of the anomalous
self-energy at the lowest Matsubara frequency,
\beg\label{chiMats}
\chi_\Phi(i\nu_l)=\frac{\delta\Phi^{(l)}_{n=0}}{h},
\en
and the amplitude mode is the zero of $\chi_\Phi^{-1}$ continued to
$\nu^2=-\omega_{\rm H}^2$. The subscript distinguishes this response of
the pairing vertex from the gap-edge response $\chi(\omega)$ of Sec.~III.B,
Eq.~\eqref{MyChi}, which shares its singularities but not its residue
(Appendix~\ref{AppendixD}); the position of the amplitude mode is the same
in both. Equations \eqref{MatsSystem} are a real,
dense $2N\times2N$ linear system in the Matsubara index at each value of coupling $\lambda$ and external frequency 
$\nu_l$ and can be solved numerically by direct factorization.
The linear system \eqref{MatsSystem} must be solved on exactly the same Matsubara
set of frequencies on which \eqref{EliEqsOpt} was converged.

At $\Delta/\Omega\to0$ (BCS weak coupling limit) the kernel becomes instantaneous, so that $Z\to1$, $\Phi_n\to\Delta$ and 
$\delta W\to0$. In this case \eqref{MatsSystem} collapses to a scalar relation whose
$T=0$ limit can be evaluated in closed form, and the vertex response
coincides with the gap-edge response of Sec.~III.B since $Z=1$ and
$\delta W=0$. With the normalization
$\chi_\Phi^{-1}(0)=1$, this is \cite{ChubukovHiggs2023,Nosov2025}
\beg\label{BCSclosed}
\chi_\Phi^{-1}(\nu)=\sqrt{\nu^2+4\Delta^2}\;
\frac{\mathrm{arcsinh}(\nu/2\Delta)}{\nu},
\en
which upon the analytic continuation yields
$\chi_\Phi^{-1}(\omega)=\sqrt{4\Delta^2-\omega^2}
{\arcsin(\omega/2\Delta)}/{\omega}$.
Written in $u=\nu^2$ this is
$\chi_\Phi^{-1}=\sqrt{u+4\Delta^2}\,\tilde K(u)$ with
$\tilde K=\mathrm{arcsinh}(\sqrt u/2\Delta)/\sqrt u$ analytic in
$|u|<4\Delta^2$ and finite and nonzero at $u=-4\Delta^2$, where
$\tilde K\to\pi/(4\Delta)$. The zero of $\chi_\Phi^{-1}$ therefore sits exactly
on its own nearest branch point, and it is a square-root zero rather
than a simple one. This is the analytic content of the statement that the BCS
amplitude mode is an edge singularity rather than a pole, and of the
$\delta\Delta(t)\propto\cos (2\Delta t+\pi/4)/\sqrt{2\Delta t}$ decay of Refs.~\cite{VolkovKogan1974,YuzbashyanDzero2006,QReview2015}.

Equations \eqref{MatsSystem} separate the two retardation effects. Setting $\delta W\equiv0$ while keeping the retarded
kernel $\lambda_{n-m}$ in place removes the mass
renormalization and leaves the retardation of the pairing vertex acting alone.
This is similar to the situation in Ref.~\cite{Uhrig2025}, where the phonon-mediated
interaction is retained only in the BCS channel and the renormalization of the
single-particle dispersion that it also generates is set aside explicitly. The
truncation affects only the response and not the equilibrium solution, so
$\Delta_{n=0}$ and the kernel are unchanged, making the comparison well defined.

Two questions must be kept apart: whether the truncated
response is pinned to the threshold exactly, and by how much the
truncation reduces the binding energy $E_b=2\Delta_g-\omega_{\rm H}$. On
the first, no exact statement is available, and none should be expected.
In BCS theory the pinning rests on an identity that holds for each state
in the band energy. The integrand of the pair
susceptibility bubble is $2\xi^2/[E(4E^2-\omega^2)]$ ($\xi$ the band
energy and $E=\sqrt{\xi^2+\Delta^2}$), and at $\omega=2\Delta$ it reduces
to the integrand of the static pair bubble, $1/2E$, so that the mode
condition at the threshold coincides with the gap equation. As a result
the inverse pair susceptibility,
$\chi_\Phi^{-1}(\omega)=(4\Delta^2-\omega^2)\int d\xi/[2E(4E^2-\omega^2)]$,
vanishes as $\sqrt{4\Delta^2-\omega^2}$ for $\omega\to2\Delta$. This
identity requires a $\xi$-independent gap and a kernel that does not weigh
different band energies. An instantaneous kernel that depends on the band
energies makes the order parameter a function of the band energy,
$\Delta\to\Delta(\xi)$, and violates the BCS identity at every $\xi$ by an
amount of order unity, which is why the mode detaches by an amount of
order unity at strong coupling in Refs.~\cite{Levitov2007,Uhrig2025}. A kernel that depends
on the frequency transfer acts differently. In the quasiclassical theory
the $\xi$ integration has been carried out and the response is an
equation in the frequency $\eps$, with the two propagators evaluated at
$\eps\pm\omega/2$. The amplitude coherence factor of
Appendix~\ref{AppendixC} is regular at the threshold and the singularity
resides in $1/(\eta_{\eps_+}+\eta_{\eps_-})$ alone, so the $\xi$-level
structure of the BCS threshold is preserved. An analogue of the BCS
identity would require that, at $\omega=2\Delta_g$, the response kernel
reduce to the gap-equation kernel at every frequency $\eps$ separately.
It does not: the expressions at $\eps\pm\Delta_g$ carry different values
of $\Phi$ and $Z$, so the two propagators are weighted unequally and no
identity holds frequency by frequency. The zero of the truncated $\chi_\Phi^{-1}$ is
therefore not protected by symmetry, and whether it lies on the branch
point is decided by the integral over $\eps$, i.e.\ quantitatively.

On the second question the real-axis calculation of
Sec.~\ref{sec:intermediate} is explicit, and its answer is a bound.
Carried through the identical grids, regularization, fit and
extrapolation, the truncated system returns a binding that is not
distinguishable from zero: writing $\delta=1-\omega_{\rm H}/2\Delta_g$ for
the binding relative to the threshold, $\delta_{\delta W=0}\lesssim0.005$ at both ends
of the coupling window, against $0.0088$ and $0.0117$ for the full system
($\delta W\neq0$). The truncated response is nevertheless not the
regularized square-root edge. The same procedure applied to a pure
threshold singularity places its apparent binding on the other side of
the threshold, negative on every grid and for every fit window. The
truncated system returns zero or a small positive value, separated
from the edge by more than the spread of either over the fit windows;
the numbers, in units of the regulator, are given in Sec.~IV.A and
Appendix~\ref{AppendixE}. The vertex channel acting alone therefore
deforms the threshold state but does not detach the mode by a resolvable
amount. The response of the mass renormalization supplies at least
$40\%$ of the binding at $\lambda=2$ and at least $55\%$ at
$\lambda=3.25$, in the sense that the truncation removes that fraction of
it, $1-\delta_{\delta W=0}/\delta\ge0.4$ and $0.55$. Since the binding is quadratic in the deformation of the
regular part of $\chi_\Phi^{-1}$, Eq.~\eqref{quadratic} below, the bound
$\delta_{\delta W=0}\lesssim0.005$ limits the vertex channel to at most
$\sqrt{0.005/0.0088}\simeq0.75$ of the total deformation at $\lambda=2$
and $\sqrt{0.005/0.0117}\simeq0.65$ at $\lambda=3.25$. The remainder, at
least a quarter to a third of the deformation, is the response of the
mass renormalization, and the square-root amplification turns the sum of
the two into the observed displacement.
\begin{figure}[t]
\includegraphics[width=0.995\linewidth]{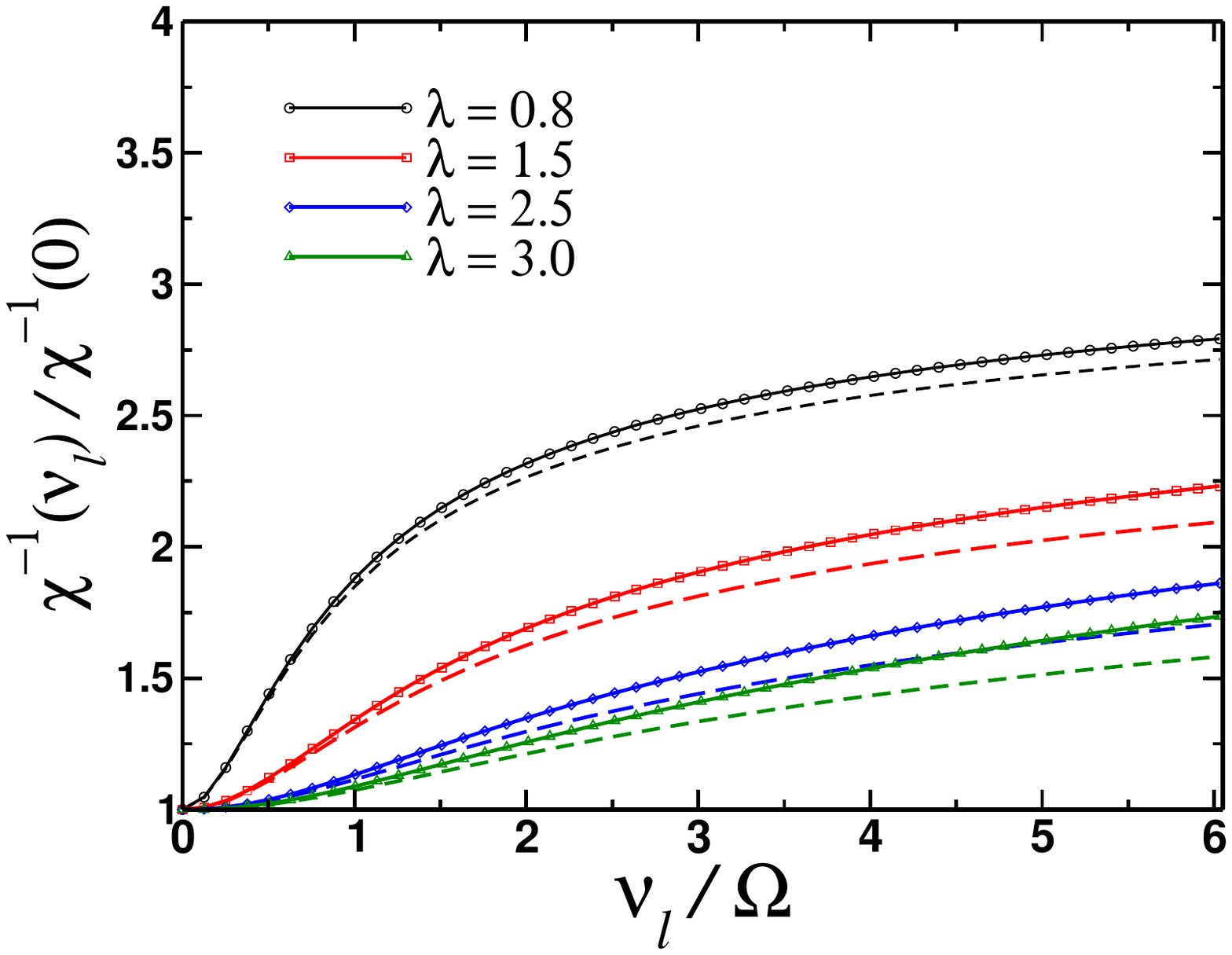}
\caption{\footnotesize Isolating the mass-renormalization response in the
longitudinal pair propagator. Normalized inverse propagator
$\chi_\Phi^{-1}(i\nu_l)/\chi_\Phi^{-1}(0)$ at four couplings. Solid: the full system
\eqref{MatsSystem}. Dashed: the same system with $\delta W\equiv0$, i.e.\
with the vertex retardation acting alone. The gap between the two curves
is the mass-renormalization response. The full system's curve rises
faster with $\nu_l$: the mass channel lowers $\chi_\Phi^{-1}$ near $\nu_l=0$
relative to its high-frequency value, which is the imaginary-axis
signature of the additional attraction in the amplitude channel discussed
in the text; removing $\delta W$ moves the mode towards the threshold.
$N=1024$, $T/\Omega=0.005$.}
\label{Fig-dWn}
\end{figure}

The zero of $\chi_\Phi^{-1}(\nu)$ cannot be located from these data, and the
obstruction is the coincidence exhibited by \eqref{BCSclosed} rather than any
deficiency of the numerics. An upper bound is nonetheless available. Continuing $\chi_\Phi^{-1}(i\nu_l)$ with
Pad\'e approximants and locating the point $u=-4\Delta_g^2(1-\delta)^2$ at
which the continuation vanishes returns a value of $\delta$. To establish when
that value can be trusted we applied the identical procedure to test functions
with a prescribed $\delta$, sampled at the same Matsubara frequencies. For example, for the test function
\beg\label{test}
\chi^{-1}_{\Phi,{\rm test}}(u)=\frac{u+\omega_c^2}{(1-\delta)^2}\,
\frac{}{}\,
\frac{\mathrm{arcsinh}(\sqrt u/2\Delta_g)}{\sqrt u\sqrt{u+4\Delta_g^2}}
\en
with $\omega_c=2\Delta_g(1-\delta)$
we find that (i) the prescribed value is recovered to better than ten percent whenever it exceeds
$0.1$; (ii) it is already low by twenty percent at $0.05$; and (iii) it is unstable once the data
carry their $10^{-4}$ uncertainty, and bears no relation to the input below
$0.02$. Applied to the computed propagator the procedure returns no stable root. A
displacement of order $0.1$ or larger would have been recovered, and was not,
so the value $\delta\ge0.1$ must be excluded, which is already well below the $0.2$--$0.5$ obtained
from static kernels~\cite{Levitov2007,Uhrig2025}. The quantitative
determination must therefore be carried out on the real axis, where the mode is
a root of the linearized operator rather than the endpoint of an analytic continuation.

The smallness of the effect has a formal explanation in the Matsubara representation. With the shorthand
$(\mathsf{K}_X)_{nm}=\pi T\lambda_{n-m}X_m$ for $X=P,Q,R$, Eqs.~\eqref{MatsSystem} read
\beg\label{nonlocal}
\begin{aligned}
&(1-\mathsf{K}_P)\delta W=-\mathsf{K}_Q\delta\Phi, \\
&(1-\mathsf{K}_R)\delta\Phi=h-\mathsf{K}_Q\delta W.
\end{aligned}
\en
Since $\delta W$ appears
linearly and without a source it may be eliminated exactly, leaving a closed
equation for the vertex response alone,
\beg\label{dPhiEff}
\Bigl[\,1-\mathsf{K}_R-\mathsf{K}_Q\bigl(1-\mathsf{K}_P\bigr)^{-1}\mathsf{K}_Q\,\Bigr]
\delta\Phi=h .
\en
The mass renormalization thus enters the pair response as the insertion
$\mathsf{K}_Q(1-\mathsf{K}_P)^{-1}\mathsf{K}_Q$, and two properties of that
insertion follow at once. It carries a plus sign, because $\mathsf{K}_Q$ occurs twice and the two minus
signs of \eqref{MatsSystem} cancel: the mass renormalization adds attraction
to the amplitude channel, so that the deformation it produces moves the zero
of $\chi_\Phi^{-1}$ onto the physical sheet below $2\Delta_g$. With the opposite
sign the same deformation would leave only a virtual state. Its contribution to $\chi_\Phi$ is $O(\mathsf{K}_Q^2)$ and never $O(\mathsf{K}_Q)$:
no term linear in the interchannel coupling exists. This does not by itself
make the contribution small --- $\mathsf{K}_Q$ is not a small operator --- but
it fixes the power with which its size enters. The insertion
$\mathsf{K}_Q(1-\mathsf{K}_P)^{-1}\mathsf{K}_Q$ is bounded by
$\|\mathsf{K}_Q\|^2\,\|(1-\mathsf{K}_P)^{-1}\|$, and with the values
computed at $\lambda=2.5$, $\|\mathsf{K}_Q\|=0.21$ and a largest eigenvalue
of $\mathsf{K}_P$ of $0.58$, this is about $0.1$: the deformation of order
ten percent seen in Fig.~\ref{Fig-dWn}. A linear term with the same
$\mathsf{K}_Q$ would have permitted a deformation several times larger. At
weak coupling the same structure supplies the scaling: $\mathsf{K}_Q$ is
generated only by the curvature of $\lambda_{n-m}$ and is
$O\bigl((\Delta_g/\Omega)^2\bigr)$, so the mass channel enters at
$O\bigl((\Delta_g/\Omega)^4\bigr)$, Section~IV.

The direction of the shift, here and for the energy-dependent kernels discussed in the Introduction, can be understood as follows. A periodic perturbation which modifies the pairing gap drives all degrees of freedom that determine the pairing gap: the Cooper pairs, the shape of the pairing function across the band, and
the mass renormalization of the quasiparticles. At $T=0$ neither Cooper pairs nor quasiparticles can be excited at energy less than $2\Delta_g$, so each
behaves as an oscillator whose natural frequency lies at or above $2\Delta_g$. It is well known that an
oscillator driven below its natural frequency follows the driving force in phase \cite{LandauLifshitz1}. Each
Cooper pair therefore responds to a periodic modulation of the gap near $2\Delta_g$ with the same
sign as the gap itself, and through the gap equation its response adds to rather than subtracts from the external periodic drive
that produced it. The larger this addition, the lower the frequency at which the
oscillation sustains itself. For example, in the BCS theory this is described by the following expression for the inverse longitudinal pair susceptibility
\beg\label{SumOverPairs}
\chi^{-1}(\omega)=\sum\limits_{\mathbf k}\frac{4\Delta_g^2-\omega^2}{2E_{\mathbf k}(4E_{\mathbf k}^2-\omega^2)}
\en
which is obtained by subtracting the bare longitudinal pair response from the self-consistency equation for the order parameter term by term \cite{Nosov2025,Kamenev2025,Kazi2026}. From \eqref{SumOverPairs} it follows that because
every pair has the same pairing gap, every term carries the same factor $4\Delta_g^2-\omega^2$:
all terms are positive below $2\Delta_g$ and vanish together at $\omega=2\Delta_g$, which is
what pins the mode to the threshold in the BCS theory.   When the gap differs from pair to pair, $\Delta_g\to \Delta_\bk$ and is largest at the Fermi surface, the factor $4\Delta_g^2-\omega^2$ in \eqref{SumOverPairs} becomes 
$4\Delta_{\mathbf k}^2-\omega^2$, with $\Delta_{\mathbf k}<\Delta_g$ away from the Fermi surface. These terms are negative at $\omega=2\Delta_g$, and to compensate for them $\omega$ must be less than $2\Delta_g$. Therefore, frequency at which 
$\chi^{-1}(\omega)$ is zero must necessarily move below the $2\Delta_g$ BCS threshold.

{In Eliashberg theory the pairing interaction depends on the frequency transfer rather than on the
band energies, and the gap varies with frequency but is the same for every pair.
By itself the retarded vertex deforms the threshold state without detaching the
mode by some amount. It is the dressing of the quasiparticles that adds a term that is
negative whatever the frequency dependence of the gap, i.e. the correction 
$\mathsf{K}_Q(1-\mathsf{K}_P)^{-1}\mathsf{K}_Q$ in Eq.~\eqref{dPhiEff}. The
oscillating gap drives the mass renormalization through matrix $\mathsf{K}_Q$. Now since every
eigenvalue of $\mathsf{K}_P$ lies below unity and the mass channel has no resonance
of its own below $2\Delta_g$, the mass renormalization follows this drive in phase,
and because it acts back on the gap through the same kernel, its back-action
reinforces the oscillation whatever the sign of $\mathsf{K}_Q$. In other words this term has  the same effect on inverse pair susceptibility as $\Delta\to\Delta_\bk$ on the sign of $4E_\bk^2-\omega^2$ near the Fermi surface in Eq. \eqref{SumOverPairs}: like the pairs with
$\Delta_\bk<\Delta_g$, this term contributes a negative amount at
$\omega=2\Delta_g$, where every term of Eq.~\eqref{SumOverPairs} vanishes, so that
$\chi_\Phi^{-1}$ changes sign below the threshold and the mode moves into the gap.
The two cases (modified BCS theory vs. Eliashberg theory) therefore may only differ in the magnitude of by how much the frequency of the mode moves below $2\Delta_g$.}

Why a deformation of this size displaces the mode by only one percent
follows from the character of the threshold. With $\delta W\equiv0$ the
amplitude response retains the inverse-square-root threshold singularity
of the BCS limit, and its zero lies within the resolution of the
calculation of the threshold, $\delta_{\delta W=0}\lesssim0.005$
Sec.~\ref{sec:intermediate}. Writing
$\chi_\Phi^{-1}(\omega)\simeq A\sqrt{2\Delta_g-\omega}+\delta A$ in the vicinity
of the threshold, with $\delta A$ collecting the deformation of the
regular part from both vertex and mass renormalization channels, a deformation displaces the zero to
\beg\label{quadratic}
2\Delta_g-\omega_{\rm H}\simeq\left(\frac{\delta A}{A}\right)^{2},
\en
which is quadratic rather than linear in the strength of the perturbation. A
deformation of order ten percent therefore displaces the mode by about
one percent, which is the magnitude obtained in Sec.~IV by solving the
linear equations on the real frequency axis. The percent-level binding is
in this sense a property of the threshold rather than of the coupling:
the channel that detaches the mode enters without a linear term and acts
upon a zero whose square-root character converts a ten-percent deformation
into a one-percent shift.
 
The Coulomb pseudopotential $\mu^*$ does not change this picture. The Coulomb
repulsion is instantaneous, so the part of the self-energy it produces does
not depend on frequency; a frequency-independent shift of the normal
self-energy is simply a shift of the chemical potential and drops out of the
mass renormalization. As a result $\mu^*$ appears only in the equation for
the pairing response $\delta\Phi$, where it reduces the pairing kernel from
$\lambda_{n-m}$ to $\lambda_{n-m}-\mu^*$, and not at all in the equation for
the mass-renormalization response $\delta W$. Eliminating $\delta W$ as
before again yields Eq.~\eqref{dPhiEff}, with the pairing kernel and one of
the two coupling factors in $\mathsf{K}_Q(1-\mathsf{K}_P)^{-1}\mathsf{K}_Q$
evaluated with $\lambda_{n-m}-\mu^*$ instead of $\lambda_{n-m}$. The
Coulomb term thus rescales the size of the mass-channel insertion but does
not change its structure: the mass channel $\mathsf{K}_Q$ still has no source of its own
and still enters at second order \eqref{dPhiEff}. Where $\mu^*$ does matter is in the equilibrium state, where at fixed $\lambda$ and $\Omega$ it lowers the gap
$\Delta_g$ and thereby, through Eq.~\eqref{Eq1}, the binding energy.

\subsection{Real time representation}
{We also keep the phonon propagators $D^{R,A,K}$ at their equilibrium values in the real-time response, so $\delta D^{R,A,K}=0$.}
On the real frequency axis the pairing amplitude is taken to vary as
$\hat{\Delta}_\eps(t)= i\hat{\tau}_2\Delta_\eps+\delta\hat{\Delta}_\eps(t)$,
with the fluctuation in the longitudinal channel, 
\beg\label{LongChannel}
\delta\hat{\Delta}_\eps(t)=i\hat{\tau}_2\delta\Delta_\eps(t).
\en
For a single Fourier component
$\delta\hat{g}_\eps(t)=\delta\hat{g}_\eps(\omega) e^{-i\omega t}$
the linearized Eq.~\eqref{NowFinalEilenberger} reads
\beg\label{ME-LinearizedEq}
\begin{aligned}
&\left[\eps\check{\tau}_3,\delta\check{g}_\eps(\omega)\right]_{-}+\delta\check{\Delta}_{\eps}(\omega)\check{\cal G}_{\eps-\frac{\omega}{2}}\\&-\check{\cal G}_{\eps+\frac{\omega}{2}}\delta\check{\Delta}_{\eps}(\omega)+\check{\Delta}_{\eps+\frac{\omega}{2}}\delta\check{g}_{\eps}(\omega)-\delta\check{g}_{\eps}(\omega)\check{\Delta}_{\eps-\frac{\omega}{2}}\\&+\frac{\omega}{2}\left[\check{\tau}_3,\delta\check{g}_\eps(\omega)\right]_{+}=\delta\check{\Pi}_{\eps}\check{\cal G}_{\eps-\frac{\omega}{2}}\\&-\check{\cal G}_{\eps+\frac{\omega}{2}}\delta\check{\Pi}_{\eps}+\check{\Pi}_{\eps+\frac{\omega}{2}}\delta\check{g}_{\eps}(\omega)-
\delta\check{g}_{\eps}(\omega)\check{\Pi}_{\eps-\frac{\omega}{2}}.
\end{aligned}
\en
The solution of this equation for the retarded and advanced components $\delta\hat{g}_\eps^{R(A)}$ must satisfy the normalization condition
\beg\label{NormdgRA}
\hat{\cal G}_{\eps+\frac{\omega}{2}}^{R(A)}\delta\hat{g}_\eps^{R(A)}+
\delta\hat{g}_\eps^{R(A)}\hat{\cal G}_{\eps-\frac{\omega}{2}}^{R(A)}=0.
\en
The normalization condition for the Keldysh part $\delta\hat{g}_\eps^{K}$ is more involved,
\beg\label{NormdgK}
\hat{\cal G}_{\eps+\frac{\omega}{2}}^{R}\delta\hat{g}_\eps^{K}+
\delta\hat{g}_\eps^{K}\hat{\cal G}_{\eps-\frac{\omega}{2}}^{A}+\hat{\cal G}_{\eps+\frac{\omega}{2}}^{K}\delta\hat{g}_\eps^{A}
+\delta\hat{g}_\eps^{R}\hat{\cal G}_{\eps-\frac{\omega}{2}}^{K}=0.
\en

For the retarded and advanced components, with the superscript $R$ suppressed, Eq.~\eqref{ME-LinearizedEq} becomes
\beg\label{ME-LinearizedEq4R}
\begin{aligned}
&\left[\eps\hat{\tau}_3,\delta\hat{g}_\eps\right]_{-}-\hat{\Sigma}_{\eps+\frac{\omega}{2}}\hat{\tau}_3\delta\hat{g}_{\eps}+\delta\hat{g}_{\eps}\hat{\Sigma}_{\eps-\frac{\omega}{2}}\hat{\tau}_3\\&+\frac{\omega}{2}\left[\hat{\tau}_3,\delta\hat{g}_\eps\right]_{+}=\delta\hat{\Sigma}_\eps\hat{\tau}_3\hat{\cal G}_{\eps-\frac{\omega}{2}}-\hat{\cal G}_{\eps+\frac{\omega}{2}}\delta\hat{\Sigma}_{\eps}\hat{\tau}_3.
\end{aligned}
\en
With the normalization condition \eqref{NormdgRA} its solution is
\beg\label{Eq4dgRA}
{\delta\hat{g}_\eps^{R(A)}=\frac{\hat{\cal G}_{\eps+\frac{\omega}{2}}^{R(A)}\delta\hat{\Sigma}_\eps^{R(A)}\hat{\tau}_3\hat{\cal G}_{\eps-\frac{\omega}{2}}^{R(A)}-\delta\hat{\Sigma}_{\eps}^{R(A)}\hat{\tau}_3}{\eta_{\eps+\frac{\omega}{2}}^{R(A)}+\eta_{\eps-\frac{\omega}{2}}^{R(A)}}.}
\en
Here $\eta_\eps^A=-(\eta_\eps^R)^*$, $\eta_\eps^R=\sqrt{(\eps Z_\eps^R)^2-(\Phi_\eps^R)^2}$ and in $\eta_\eps^R$ the branch must be chosen so that 
${\;\textrm{Im}\,\eta_\eps^R>0\;}$ always holds. The expressions for $\delta\hat{\Sigma}_\eps^{R(A)}(\omega)$ follow directly from  
\eqref{SigmaREq2m} and \eqref{SigmaAEq2} of Appendix~\ref{AppendixA}. Since $\delta \hat{g}_\eps^R$ remains analytic in the upper half-plane of complex $\eps$, the integral 
over $\eps'$ of $\delta\hat{g}_{\eps'}^RD_{\eps-\eps'}^A$ vanishes, and so does the integral of $\delta\hat{g}_{\eps'}^AD_{\eps-\eps'}^R$. Hence
\beg\label{ClosureRA}
\begin{aligned}
&\delta\hat{\Sigma}_{\eps}^{R}\hat{\tau}_3=\frac{\lambda\Omega}{8}\int{d\eps'}\left\{
\delta\hat{g}_{\eps'}^RD_{\eps-\eps'}^K+\delta\hat{g}_{\eps'}^KD_{\eps-\eps'}^R\right\}, \\
&\delta\hat{\Sigma}_{\eps}^{A}\hat{\tau}_3=\frac{\lambda\Omega}{8}\int{d\eps'}\left\{
\delta\hat{g}_{\eps'}^AD_{\eps-\eps'}^K+\delta\hat{g}_{\eps'}^KD_{\eps-\eps'}^A\right\}.
\end{aligned}
\en
The dependence of the linear corrections on $\omega$ is suppressed here for brevity.

The Keldysh component obeys an equation of the same form, but its normalization condition \eqref{NormdgK} is more involved. We therefore use the standard decomposition of the Keldysh
component into its equilibrium-like part and the anomalous function
$\delta\hat g^a$ of Larkin and Ovchinnikov \cite{LO1977,LO1986,Kopnin2001},
here written for the linear response at transfer $\omega$,
\beg\label{ansatz4gK}
\delta\hat{g}_\eps^{K}(\omega)=\delta\hat{g}_\eps^{R}(\omega)t_{\eps-\frac{\omega}{2}}-t_{\eps+\frac{\omega}{2}}\delta\hat{g}_\eps^{A}(\omega)+\delta \hat{g}_\eps^a(\omega).
\en
Inserted into \eqref{NormdgK} together with \eqref{GKParametrization}, this gives the normalization condition for $\delta \hat{g}_\eps^a(\omega)$,
\beg\label{Norm4dGK}
\hat{\cal G}_{\eps+\frac{\omega}{2}}^R\delta\hat{g}_\eps^a(\omega)+\delta\hat{g}_\eps^a(\omega)\hat{\cal G}_{\eps-\frac{\omega}{2}}^A=0.
\en
The equation for $\delta\hat{g}_\eps^a$ is
\beg\label{Eq4dga}
\begin{aligned}
&\left(\eps+\frac{\omega}{2}-\hat{\Sigma}_{\eps+\frac{\omega}{2}}^R\right)\hat{\tau}_3\delta\hat{g}_\eps^a-\delta\hat{g}_\eps^a
\left(\eps-\frac{\omega}{2}-\hat{\Sigma}_{\eps-\frac{\omega}{2}}^A\right)\hat{\tau}_3\\&+\hat{\cal R}_\eps^{R}t_{\eps-\frac{\omega}{2}}-\hat{\cal R}_\eps^{A}t_{\eps+\frac{\omega}{2}}=\hat{\cal R}_\eps^K,
\end{aligned}
\en
where the functions $\hat{\cal R}_\eps^X$ ($X=R,A,K$) are defined in Appendix~\ref{AppendixC} and the relation \eqref{FDT} has been used; the terms containing $\hat{\Sigma}_\eps^K$ cancel. Hence
\beg\label{FinalEq4dga}
\begin{aligned}
&\eta_{\eps+\frac{\omega}{2}}^R\hat{\cal G}_{\eps+\frac{\omega}{2}}^R\delta\hat{g}_\eps^a-\delta\hat{g}_\eps^a\hat{\cal G}_{\eps-\frac{\omega}{2}}^A
\eta_{\eps-\frac{\omega}{2}}^A\\&=\hat{\cal R}_\eps^K-\hat{\cal R}_\eps^{R}t_{\eps-\frac{\omega}{2}}+\hat{\cal R}_\eps^{A}t_{\eps+\frac{\omega}{2}}.
\end{aligned}
\en
With \eqref{Norm4dGK} this gives
\beg\label{dgaFin}
\delta\hat{g}_\eps^a=\frac{\hat{\cal G}_{\eps+\frac{\omega}{2}}^R\left[\hat{\cal R}_\eps^K-\hat{\cal R}_\eps^{R}t_{\eps-\frac{\omega}{2}}+\hat{\cal R}_\eps^{A}t_{\eps+\frac{\omega}{2}}\right]}{\eta_{\eps+\frac{\omega}{2}}^R+\eta_{\eps-\frac{\omega}{2}}^A}.
\en
Finally, the closing relation for $\delta\hat{\Sigma}_\eps^K(\omega)$ is
\beg\label{dsigmaKFin}
\begin{aligned}
&\delta\hat{\Sigma}_{\eps}^{K}(\omega)\hat{\tau}_3=\frac{\lambda\Omega}{8}\int{d\eps'}\left\{
\delta\hat{g}_{\eps-\eps'}^K(\omega)D_{\eps'}^K\right.\\&\left.+\left(\delta\hat{g}_{\eps-\eps'}^R(\omega)+\delta\hat{g}_{\eps-\eps'}^A(\omega)\right)\left(D_{\eps'}^R+D_{\eps'}^A\right)\right\}.
\end{aligned}
\en

These equations determine the longitudinal pair susceptibility. The unknowns are the components of the self-energy parts,
\beg\label{dSigmaRAK}
\delta\hat{\Sigma}_\eps^X\hat{\tau}_3=a_\eps^X\hat{\tau}_3+ib_\eps^X\hat{\tau}_2, 
\en
where $X=R,A,K$, collected into the columns
\beg\label{psiX}
\psi_\eps^X=\begin{pmatrix}a_\eps^X\\[2pt] b_\eps^X\end{pmatrix}.
\en
The equation for $\psi_\eps^K$ is local. With the expressions of Appendix~\ref{AppendixC},
\beg\label{ClosureK}
\begin{aligned}
\psi_\eps^{K}=\frac{i\pi\lambda\Omega}{4}\sum_{s=\pm}
&\{s(\mathbb{M}^{R}\psi^{R}-\mathbb{M}^{A}\psi^{A})\\&
-\coth({\Omega}/{2T})\mathbb{T}[\delta\hat g^{K}]\}_{\eps+s\Omega}.
\end{aligned}
\en
Here $\mathbb{M}^{R(A)}$ are the two-by-two matrices \eqref{Mmatrix} of Appendix~\ref{AppendixC} and 
$\mathbb{T}$ is the two-component trace
$\mathbb{T}[{\delta\hat{g}_\eps^K}]=\bigl(\tfrac12\mathrm{Tr}[\delta\hat g_\eps^K\hat\tau_3],\,
\tfrac1{2i}\mathrm{Tr}[\delta\hat g_\eps^K\hat\tau_2]\bigr)^{\!T}$. Given the definition \eqref{ansatz4gK} for the last term in \eqref{ClosureK} we find
\beg\label{TraceK}
\mathbb{T}\bigl[\delta\hat g_\eps^{K}\bigr]
=t_{\eps-\frac{\omega}{2}}\mathbb{M}_\eps^{R}\psi_\eps^{R}
-t_{\eps+\frac{\omega}{2}}\mathbb{M}_\eps^{A}\psi_\eps^{A}
+\mathbb{M}_\eps^{K}\zeta_\eps 
\en
and $\zeta_\eps=\psi_\eps^{K}+t_{\eps+\frac{\omega}{2}}\psi_\eps^{A}-t_{\eps-\frac{\omega}{2}}\psi_\eps^{R}$.

The equations for $\psi_{\eps}^{R(A)}$ are nonlocal, because they are determined in part by $\delta\hat{g}^a$, which, unlike $\delta\hat{g}_\eps^{R(A)}$, has no definite analytic properties. A somewhat tedious but straightforward calculation gives the two remaining closure equations
\begin{widetext}
\beg\label{ClosureRAfinal}
\begin{aligned}
\psi_\eps^{X}&=\frac{i\pi\lambda\Omega}{4}\sum_{s=\pm}
\Bigl[s\,t_{\eps+s\Omega\mp\omega/2}
-\coth\Bigl(\frac{\Omega}{2T}\Bigr)\Bigr]
\mathbb{M}_{\eps+s\Omega}^{X}\psi_{\eps+s\Omega}^{X}+\frac{\lambda\Omega}{4}\int\limits_{-\infty}^\infty d\eps'
\Bigl[v_{-}^{X}(\eps,\eps')\,\mathbb{M}_{\eps'}^{R}\psi_{\eps'}^{R}
-v_{+}^{X}(\eps,\eps')\,\mathbb{M}_{\eps'}^{A}\psi_{\eps'}^{A}\Bigr]\\
&+\frac{\lambda\Omega}{8}\int\limits_{-\infty}^\infty d\eps'\,D_{\eps-\eps'}^{X}\,
\mathbb{M}_{\eps'}^{K}\zeta_{\eps'} ,
\end{aligned}
\en
\end{widetext}
with $X=R$ (upper sign) and $X=A$ (lower sign), and $\mathbb{M}_{\eps'}^{K}$ also defined by \eqref{Mmatrix}. Equations \eqref{ClosureK} and \eqref{ClosureRAfinal} form a closed system of linear equations and are one of the main results of this paper.

The equations above are homogeneous, and to obtain the susceptibility we
must supply the real-frequency counterpart of the pair source
\eqref{Source}. On the Keldysh contour the field $h$ is a $c$-number: it
takes equal values on the forward and backward branches, i.e.\ it is a
purely classical perturbation. Upon the rotation to the
retarded--advanced--Keldysh representation a classical field enters the
retarded and advanced self-energies identically and generates no Keldysh
component,
\beg\label{SourceReal}
\delta\hat\Sigma^{R,{\rm ext}}_\eps\hat\tau_3
=\delta\hat\Sigma^{A,{\rm ext}}_\eps\hat\tau_3
= ih\,\hat\tau_2 ,
\qquad
\delta\hat\Sigma^{K,{\rm ext}}_\eps=0 ,
\en
which is also consistent with \eqref{FDT}. In equilibrium a static
classical field must drop out of the anomalous combination on which
$\delta\hat g^a$ is built, and \eqref{SourceReal} is the unique assignment
with this property. In terms of the columns \eqref{psiX} the source adds
$h\,(0,1)^T$ to the right-hand sides of \eqref{ClosureRAfinal} for both
$X=R$ and $X=A$, while \eqref{ClosureK} remains sourceless because the external source does not introduce inelastic scattering processes. As a result the closed
system becomes the driven linear problem
\beg\label{DrivenSystem}
\bigl[\mathbbm{1}-\mathbb{K}(\omega)\bigr]\Psi=\Psi_{\rm ext},
\en
where $\mathbb{K}(\omega)$ denotes the linear operator defined by the
right-hand sides of \eqref{ClosureK} and \eqref{ClosureRAfinal}, $\Psi$
collects the six functions $a^{R,A,K}_\eps$, $b^{R,A,K}_\eps$ on the
energy grid, and $\Psi_{\rm ext}$ is the source term just described. Equation
\eqref{SourceReal} is precisely the analytic continuation of
\eqref{Source}: there the same field shifts $\hat M_n\to\hat M_n+
h\hat\tau_2$ once, outside the Matsubara sum, which is the imaginary-axis
image of a classical field.

It remains to specify which scalar functional of the solution plays the
role of the response amplitude. The induced self-energy has two components,
the anomalous and the normal one, both taken from the classical
(symmetrized) combination of the retarded and advanced blocks
(see Appendix~\ref{AppendixA} for details):
\beg\label{dDltInd}
\begin{aligned}
&\delta\Phi_{\rm ind}(\eps,\omega)=\frac{1}{2}
\left[b_\eps^{R}(\omega)+b_\eps^{A}(\omega)\right],\\
&\delta W_{\rm ind}(\eps,\omega)=\frac{1}{2}
\left[a_\eps^{R}(\omega)+a_\eps^{A}(\omega)\right],
\end{aligned}
\en
where, by \eqref{SigmaREqAnsatz} and \eqref{dSigmaRAK}, $b$ is the response
of $\Phi_\eps$ and $a$ that of $\eps Z_\eps$, up to a common sign. In
contrast to the BCS case both are functions of energy, so the
susceptibility requires a projection. In Eliashberg theory the single
physical amplitude of the order parameter is the spectral gap edge
$\Delta_g$, fixed by $\textrm{Re}\Delta(\eps)=\eps$, i.e.\ by the zero of
$\eta_\eps^2=(\eps Z_\eps)^2-\Phi_\eps^2$, the branch point of the
equilibrium propagator \eqref{GRclosure}. The gap-edge response is the
shift of this zero under the perturbation. Writing
$\eps Z_\eps\to\eps Z_\eps+\delta W_{\rm ind}$,
$\Phi_\eps\to\Phi_\eps+\delta\Phi_{\rm ind}$ and $\eps=\Delta_g+\chi(\omega)h$
with $h\ll\Delta_g$, and linearizing in $h$,
yields the following expression for the pair susceptibility
\beg\label{MyChi}
\chi(\omega)=\left[\frac{\delta\Phi_{\rm ind}-\Delta_g\,\delta Z_{\rm ind}}
{hZ^R_{\eps}\left(1-\partial_\eps\Delta_\eps\right)}\right]_{\eps=\Delta_g},
\en
with $\delta Z_{\rm ind}=\delta W_{\rm ind}/\eps$ and the overall sign of
the source and projection fixed by the static stability condition
$\chi(0)>0$. Equation \eqref{MyChi} is the linearization of
$\Delta=\Phi/Z$. The response of the mass renormalization enters the gap
edge on the same footing as the response of the pairing vertex, and at the
edge, where $\eps Z_\eps=\Phi_\eps$, the two enter with equal weight,
$\Delta_\eps/\eps\to1$ up to $\textrm{Im}\Delta(\Delta_g)$, which is
exponentially small at $T\ll\Delta_g$. Retaining $\delta\Phi_{\rm ind}$ alone would misrepresent the residue of
$\chi$ by the factor $1-\delta W_{\rm ind}/\delta\Phi_{\rm ind}$, where the
ratio is about ten percent at intermediate coupling (the share of the mass
channel in the ladder, Sec.~III.A) and of order unity in the strong-coupling
limit, where the two responses are rigidly locked, Eq.~\eqref{scLock}.
At $\omega\to0$ the quantity \eqref{MyChi} is
nothing but $d\Delta_g/dh$, the shift of the equilibrium gap edge under a
static pair field, which supplies an independent check. It must be
reproduced by re-solving the equilibrium problem with a small constant
$\hat\tau_2$ source added by hand, and this check is passed only when both
terms of \eqref{MyChi} are kept.

%
The choice of projection affects only the residue of
$\chi$ and not the position of its singularity. Any generic functional of $\Psi$ produces the same pole of
$[1-\mathbb{K}(\omega)]^{-1}$, so the frequency of the amplitude mode,
extracted in Sec.~IV from the resonance in $\chi(\omega)$ at finite
quasiparticle broadening --- the regulator $\gamma$ defined in Sec.~IV --- and extrapolated to $\gamma\to0$, is projection
independent. The quantity that continues to the Matsubara propagator
$\chi_\Phi(i\nu_l)$ of \eqref{chiMats} is the bare anomalous amplitude
$b^R$ at the lowest fermionic frequency, Eq.~\eqref{MRedChi} of
Appendix~\ref{AppendixD}, and not $\chi(\omega)$ of \eqref{MyChi}: the two
share the pole but not the residue. 

The reduction to the Matsubara representation can be carried out directly on \eqref{ClosureK}, and it shows what information about the out-of-equilibrium dynamics is lost in passing from the real frequency axis to Matsubara frequencies.
At $\omega=i\nu_l$ the two thermal factors in the second term of \eqref{ClosureK} coincide,
\beg\label{MRedPeriod}
t_{\eps+{i\nu_l}/{2}}=t_{\eps-{i\nu_l}/{2}}\equiv\bar t_\eps,
\en
for all $\eps$, with $\bar t_\eps=\tanh(\eps/2T)$ for even $l$ and $\coth(\eps/2T)$ for odd $l$. Inserting \eqref{MRedPeriod} and \eqref{ansatz4gK} into \eqref{dsigmaKSimp}, the form of \eqref{dsigmaKFin} given in Appendix~\ref{AppendixC}, and using the identity
\beg\label{MRedCothId}
\pm\bigl(1-\bar t_\eps\,\bar t_{\eps\pm\Omega}\bigr)
=\coth\Bigl(\frac{\Omega}{2T}\Bigr)
\bigl(\bar t_{\eps\pm\Omega}-\bar t_\eps\bigr)
\en
one finds
\beg\label{MRedChiZero}
\delta\hat{\Sigma}_\eps^K=\bigl(\delta\hat{\Sigma}_\eps^{R}-\delta\hat{\Sigma}_\eps^{A}\bigr)\bar t_\eps.
\en
and $\delta g_\eps^a(i\nu_l)=0$. 
In the Matsubara representation the linear system therefore closes on $\psi_\eps^{R}$ and
$\psi_\eps^{A}$ alone, and Appendix~\ref{AppendixD} shows that it coincides with Eqs.~\eqref{MatsSystem}.  

{The replacement~\eqref{SpectralReplacement} applies to all the equations above. For example, for a term containing the phonon propagator it gives}
\beg\label{Replace1}
\lambda\Omega D_\Omega(\omega)\to 2\int\limits_0^\infty d\nu \alpha^2F(\nu)D_\nu(\omega).
\en
Here $D_\nu(\omega)$ denotes the propagator~\eqref{DKRA} with oscillator frequency $\nu$. We make the same replacement in the thermal factors and terms with shifted electronic frequencies. Calculations with a material-specific $\alpha^2F(\omega)$, including the effect of its spectral shape on the Higgs frequency, are left for future work.

\section{Results}
In this Section we will present the results of the numerical analysis of equations \eqref{ClosureK}--\eqref{ClosureRAfinal} which allow us to evaluate the real and imaginary part of the pair susceptibility defined by \eqref{MyChi} and then extract the frequency of the Higgs mode by fitting it to the damped harmonic 
oscillator (see Appendix~\ref{AppendixE} for details). {The numerical results at finite coupling are for the physical Einstein spectrum~\eqref{alpha2Fw}. In the second part we use this spectrum to take the universal strong-coupling limit, which is independent of the phonon spectral shape.} In this limit equations \eqref{ClosureK}--\eqref{ClosureRAfinal} can be considerably simplified.

\begin{figure}
\includegraphics[width=0.980\linewidth]{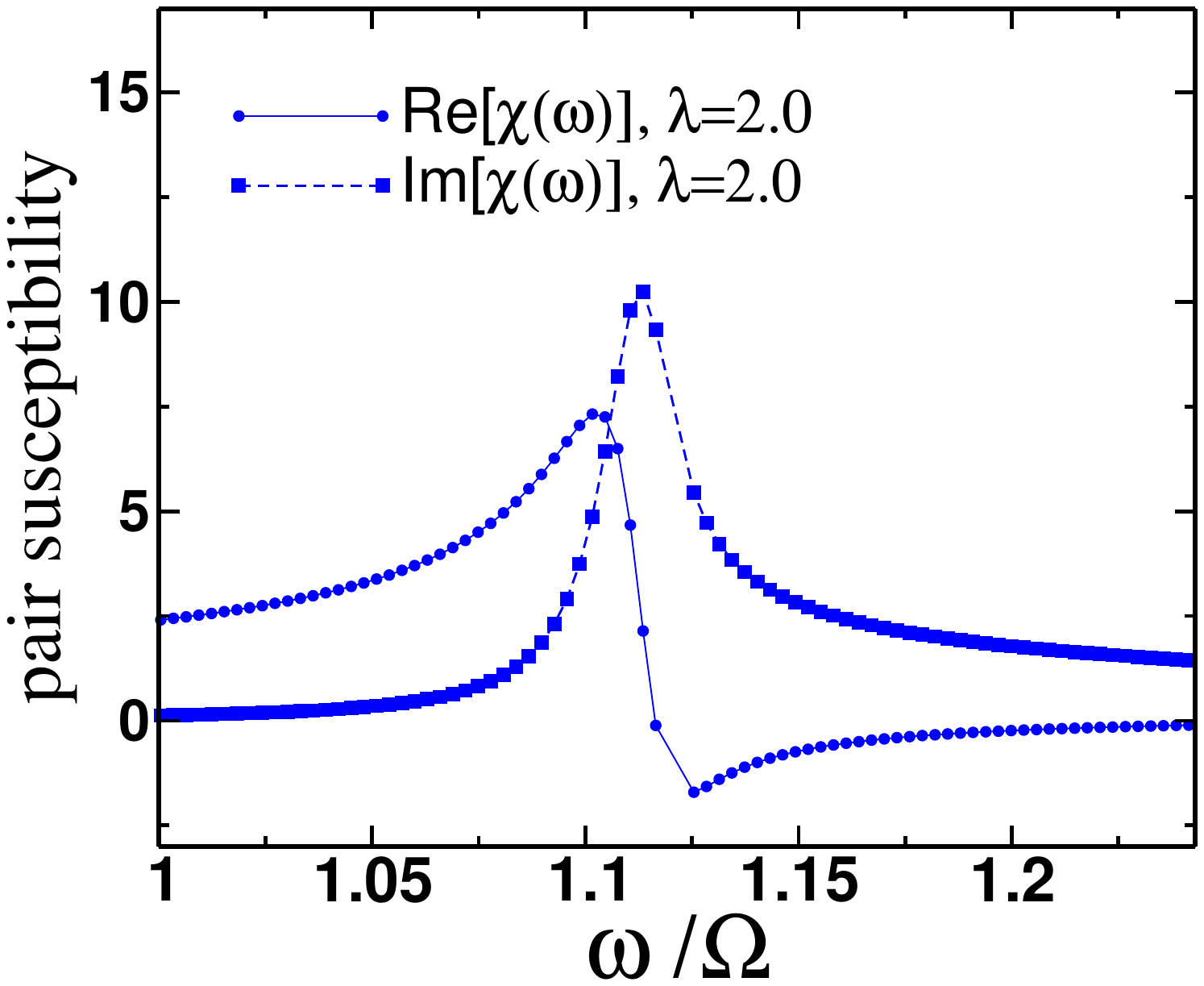}\\
\includegraphics[width=0.980\linewidth]{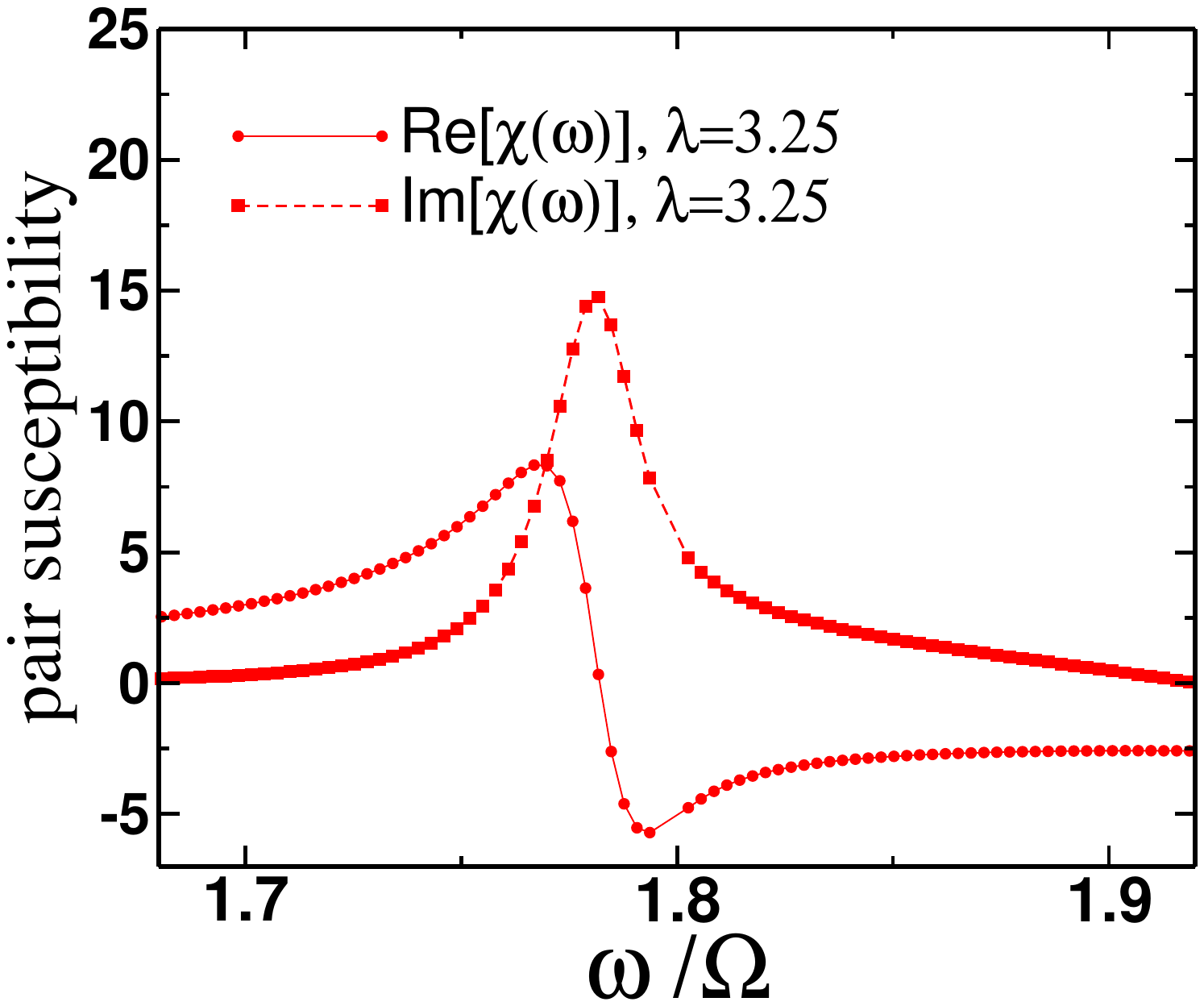}
\caption{Real and imaginary parts of the pair susceptibility $\chi(\omega)$, Eq.~\eqref{MyChi}, across the pair-breaking edge at $\lambda=2$ (top panel) and 
$\lambda=3.25$ (bottom panel) for $T/\Omega=0.005$ and $K_s=336$. The change in sign of $\textrm{Re}[\chi(\omega)]$ and the maximum of $\textrm{Im}[\chi(\omega)]$ at $\omega_{\rm H}\approx 2\Delta_g$ are the finite-$\gamma$ image of the amplitude Higgs pole.}
\label{Fig-chi}
\end{figure}

\subsection{Intermediate coupling}\label{sec:intermediate}
For a single Einstein mode the Eliashberg problem carries only two dimensionless
parameters, $\lambda$ and $T/\Omega$, so at fixed $T/\Omega$ the gap edge is a
function of the coupling alone. Varying $\lambda$ therefore traces out a single
curve, $\Delta_g/\Omega$ against $\lambda$, which is tabulated in
Table~\ref{tab:controls} and rises monotonically across the range used here.  The
quantity that controls the amplitude mode is nevertheless not $\lambda$ but the
adiabaticity ratio $\Delta_g/\Omega$, and at $T/\Omega = 0.005$ this ratio
reaches $0.96$ at $\lambda = 3.5$ (Appendix~\ref{AppendixE}) and passes
through unity slightly above that coupling.  That crossing separates two regimes in which
the numerical problem --- not the underlying physics --- is qualitatively
different.  For $\Delta_g/\Omega < 1$ the pair-breaking threshold $2\Delta_g$ lies
below the phonon-assisted threshold $\Delta_g + \Omega$, so the collective mode
resides in a clean two-particle gap and the phonon replicas at
$\pm(\Delta_g + n\Omega)\mp\omega/2$, which carry the square-root non-analyticities
of the Keldysh kernel, remain well separated on the scale of the width of the energy grid.
The gap edge stays inside the interval on which the root of
$\textrm{Re}\,\Delta(\eps)=\eps$ is bracketed.  A single discretization therefore serves the whole
of the intermediate coupling window $2.0\leq \lambda\leq3.25$.  

Once $\Delta_g$ exceeds $\Omega$ all
three properties fail simultaneously: the mode region overlaps the one-phonon
continuum, the gap edge leaves the original bracket and must be located by a
scan for the lowest crossing, and the only route into that regime is
$\lambda\to\infty$ with $\Omega\to 0$ at fixed $g = \sqrt{\lambda}\Omega$, which forces
the replica cut-off to scale as $1/\Omega$ and changes the temperature
bookkeeping.  Because the prescription itself must change, we report the two
regions separately and hold one discretization fixed across all couplings within
each, so that the extrapolation to vanishing regulator and the $\lambda$-dependence can be read
off without confounding a change in the discretization with a change in the
result (see Appendix~\ref{AppendixE}). That boundary is not merely anticipated on numerical grounds: at $\lambda = 3.5$,
the last coupling below it, the extrapolation itself loses control.
This is why we have restricted our analysis above to the intermediate-coupling $\lambda\leq 3.25$.

The pair susceptibility \eqref{MyChi} was evaluated on the real axis at
$T/\Omega=0.005$ for $2.0\leq\lambda\leq3.25$, at fifty frequencies
in the sub-threshold window $0.80\le\omega/2\Delta_g\le0.998$ on the grids
$K_s=168$ and $336$, and on a third grid $K_s=672$ over the same or a
narrower window (Appendix~\ref{AppendixE}). Every retarded energy argument of the response carries a small imaginary part, $\eps\to\eps+i\gamma$, which regularizes the inverse-square-root singularities of $\eta^R_\eps$ at the gap edge. This regulator is not a free parameter: it is tied to the grid spacing $h=\Omega/K_s$, $\gamma=3h/2$, so that $\gamma\to0$ and the continuum limit are taken together (Appendix~\ref{AppendixE}). In Fig.~\ref{Fig-chi} we plot $\textrm{Re}[\chi(\omega)]$ and $\textrm{Im}[\chi(\omega)]$ as a function of $\omega$ for $\lambda=2.0$ and $\lambda=3.25$ correspondingly. At finite
values of the regulator $\gamma$ the mode appears as a resonance: $\textrm{Re}\,\chi$
changes sign and $\textrm{Im}\,\chi$ has a maximum below the pair-breaking
edge, both superimposed on the smooth contribution of the continuum onset. We
therefore do not read $\omega_{\rm H}$ off the position of the maximum, which
is biased toward the threshold by the background, but fit the complex
$\chi(\omega)$ to a single damped oscillator plus a constant as described by 
\beg\label{AppE-fit}
\chi(\omega)\simeq C+\frac{A}{\omega_{\rm H}^{2}-\omega^{2}-i\omega\Gamma},
\en
with complex $A$ and $C$, over several intervals of frequency $\omega$. 
Two results of the fit help with the interpretation of the numerical results. First, the damping
parameter $\Gamma$ of the oscillator form \eqref{AppE-fit}, the full width
at half maximum of the resonance peak in $\textrm{Im}\chi$, tracks the
regulator, $\Gamma\simeq5$--$6\gamma$, at every coupling. Second, when the
calculation is repeated on a grid twice as fine, so that $\gamma$ is halved
exactly, the binding energy $E_b=2\Delta_g-\omega_{\rm H}$ rises by $10$--$24\%$
while $\Gamma$ halves. Extrapolating both linearly to $\gamma\to0$ gives a finite binding energy and a width that
vanishes within the fit noise, $|\Gamma_{\gamma\to0}|\lesssim0.12\,\Gamma$.
The object below $2\Delta_g$ is thus a pole on the real axis, and the
finite-$\gamma$ resonance is its regularized image. 

The interpretation rests on the intercept $[E_b]_{\gamma\to0}$ alone, since a
regularized threshold singularity also has a width proportional to $\gamma$.
Appendix~\ref{AppendixE} (Table~\ref{tab:controls}) collects the tests that
separate the two. The resolution limit of the grid falls across the window
while the binding rises, so that their ratio grows from $1.1$ to $2.4$ on
$K_s=336$ and from $2.2$ to $4.7$ on $K_s=672$. The
binding rises, not falls, when $\gamma$ is halved, at every coupling and on
both halvings, and its dependence on $\gamma$ is concave, so that the two-grid
extrapolation overestimates the intercept formed from the finest pair by
$3$--$12\%$. On the finest grid, $K_s=672$, the binding is resolved outright at
every coupling, $E_b=3.9$--$8.5\gamma$. We specifically comment on this in Appendix~\ref{AppendixE}. The fit and the position of the maximum of
$\textrm{Im}\chi$ extrapolate to the same value to $0.9$--$2.6\%$ for
$\lambda\ge2.5$ and to $10\%$ at $\lambda=2$, where the maximum lies one or
two cells from the threshold on the coarse grid. The identical procedure
applied to a threshold singularity returns a binding of the opposite sign.

The same calculation with
$\delta W\equiv0$, carried through the identical grids, regulator, fit
and extrapolation, returns a binding below the resolution of the
method at both ends of the window, $\delta_{\delta W=0}\lesssim0.005$
against $0.0088$ and $0.0117$ for the full system. The truncated
response is not the regularized edge singularity, for which the procedure
gives $E_b=-1.8\gamma$ over the full sub-threshold window and $-1.0\gamma$
over the narrowest, on every grid and at both couplings:
it gives $E_b=-0.4$ to $+0.6\gamma$ at $\lambda=2$ and $-0.1$ to
$+1.2\gamma$ at $\lambda=3.25$, separated from the edge by $1.1$--$2.8\gamma$ regulator
widths at every window: it is not the full pole either. Retardation of the
pairing vertex alone thus displaces the mode by less than the resolution
of the calculation, and the response of the mass renormalization supplies
at least $40\%$ of the binding at $\lambda=2$ and at least $55\%$ at
$\lambda=3.25$.
The values of
$1-\omega_{\rm H}/2\Delta_g$ shown in Fig.~\ref{Fig-Binding} 
are the $\gamma\to0$ extrapolated ones. 
Equation \eqref{Eq1} is an interpolation across the resolved window
$0.56\le\Delta_g/\Omega\le0.90$ and not an asymptotic law: in that window
$\Delta_g/\Omega\simeq0.28\,\lambda$, so a dependence linear in
$\Delta_g/\Omega$ cannot be told from one linear in $\lambda$, and neither
quantity is small. 

The behavior at weak coupling is fixed by the analytic
structure of the kernel. The Einstein kernel $\lambda_{n-m}$ is analytic
and even in the frequency transfer, so every correction to the BCS quantity
is a power series in $(\Delta_g/\Omega)^2$ times logarithms, as in the
familiar strong-coupling corrections to $2\Delta/T_c$
\cite{Carbotte1990}; there is no term linear in $\Delta_g/\Omega$. The
deformation of the regular part of $\chi_\Phi^{-1}$ at the branch point by the
vertex channel is therefore $O\bigl((\Delta_g/\Omega)^2\bigr)$, that by the
mass channel $O\bigl((\Delta_g/\Omega)^4\bigr)$ --- since $\mathsf K_Q$
vanishes for a constant kernel by the reflection
$\omega_m\to-\omega_{m+l}$, under which $Q_m$ is odd, and is generated
only by the curvature of $\lambda_{n-m}$ --- and by \eqref{quadratic}
\beg\label{quartic}
1-\frac{\omega_{\rm H}}{2\Delta_g}\;\xrightarrow[\ \Delta_g/\Omega\to0\ ]{}\;
c_4\,\Bigl(\frac{\Delta_g}{\Omega}\Bigr)^{4}\Bigl[1+O\bigl((\Delta_g/\Omega)^2\bigr)\Bigr],
\en
with $c_4$ depending on $\lambda$ only through
$\lambda\ln(\Omega/\Delta_g)\simeq1$. Two consequences follow. The linear
law \eqref{Eq1} must bend down toward the BCS limit below the window, so
that its extrapolation to weaker coupling is an upper bound on the
binding. At asymptotically weak coupling the vertex channel, not the
mass renormalization, dominates the deformation, the mass channel taking
over only where its $O\bigl((\Delta_g/\Omega)^2\bigr)$ suppression is no
longer effective, which the bound on the truncated binding at $\lambda=2$
above places at or just below the resolved window. The coefficient $c_4$ is computable, the
$O(\Omega^{-2})$ part of the kernel being a rank-three separable
perturbation of the BCS problem, but the resulting binding,
$\delta\lesssim10^{-3}$ at $\Delta_g/\Omega=0.3$, lies below the resolution of
the real-axis method and for this reason we do not pursue it here.

\begin{figure}
\includegraphics[width=0.9950\linewidth]{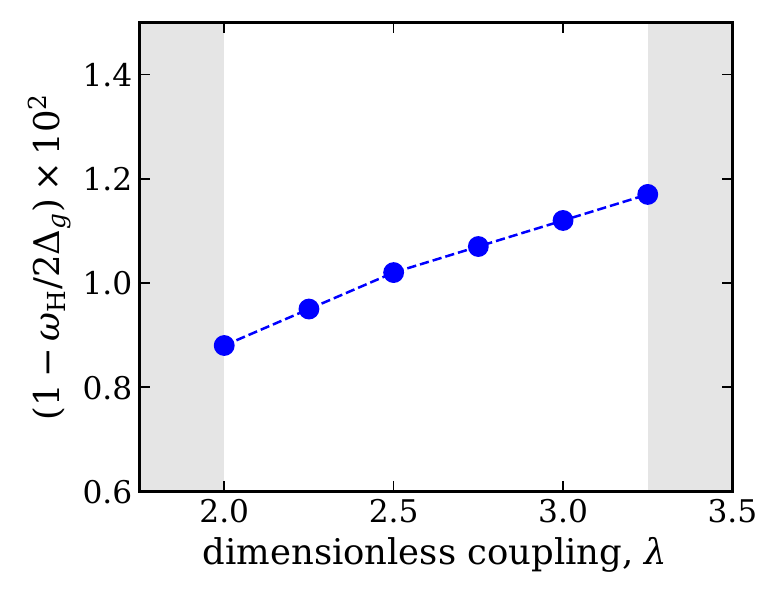}
\caption{Binding energy $1-\omega_{\rm H}/2\Delta_g$ of the amplitude mode
as a function of the coupling $\lambda$ at $T/\Omega=0.005$, extrapolated to
$\gamma\to0$ (Appendix~\ref{AppendixE}). The unshaded region,
$2\le\lambda\le3.25$ ($0.56\le\Delta_g/\Omega\le0.90$), is the window in
which the binding is resolved by the real-axis calculation. In the shaded
region to the left the binding falls below the resolution limit of the
method; in the shaded region to the right $\Delta_g$ exceeds $\Omega$, the
replica structure of the kernel enters the fit window, and the $\gamma\to0$
extrapolation loses control (Sec.~\ref{sec:intermediate}). The dashed line
connects the points and is a guide to the eye.}
\label{Fig-Binding}
\end{figure}
\begin{figure}
\includegraphics[width=0.9950\linewidth]{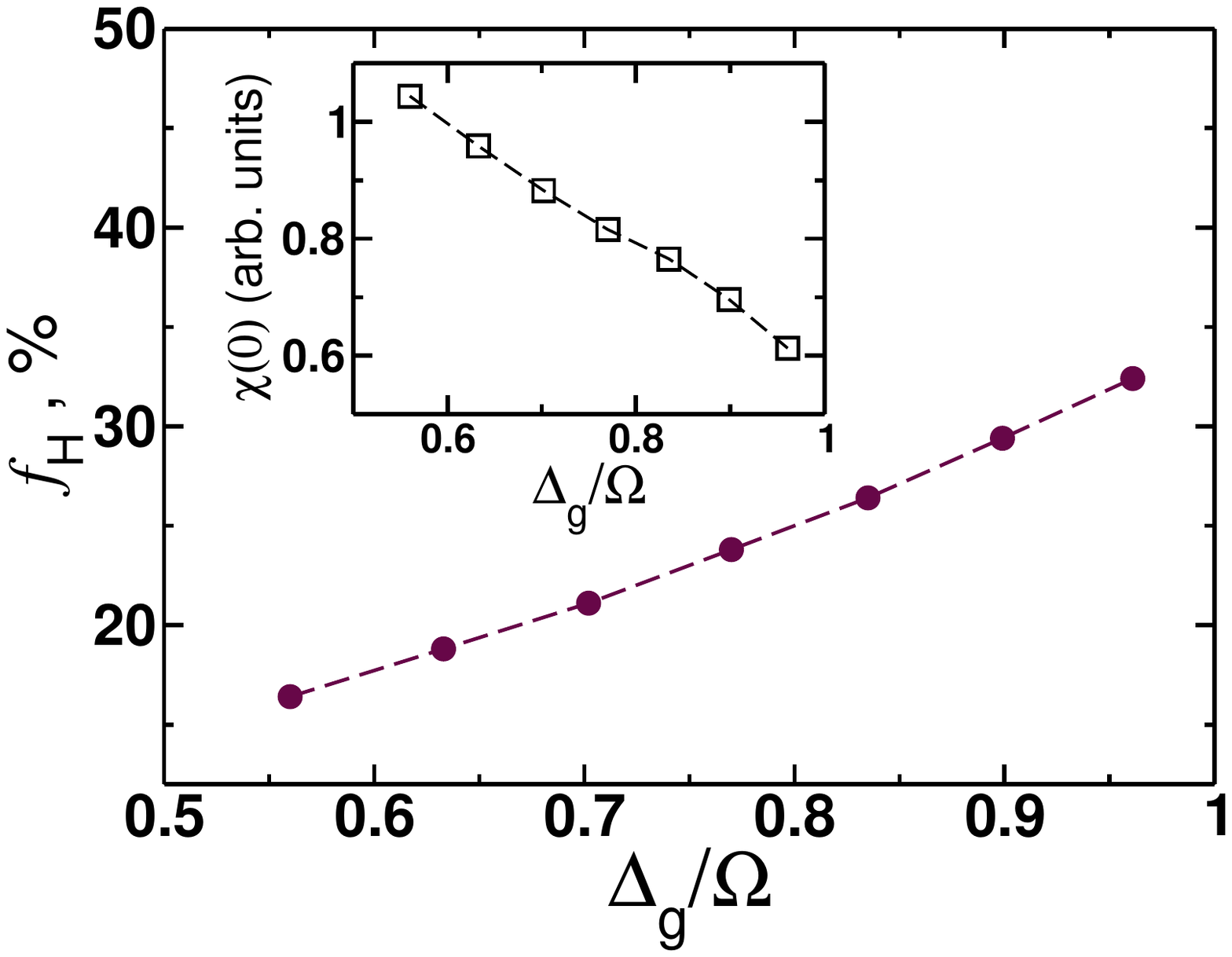}
\caption{Fraction of the amplitude-channel spectral weight carried by the
Higgs mode, $f_{\rm H}$ of Eq.~\eqref{fweight}, as a function of the dimensionless parameter $\Delta_g/\Omega$.  Since
$\chi(0)$ of Eq.~\eqref{DefineChi0} is itself the total weight, $f_{\rm H}$ is a
fraction of a conserved total and not a fit convention.  The residue ${\mathsf Z}$ and the mode
frequency $\omega_{\rm H}$ have been obtained from the sub-edge response extrapolated by taking the regulator $\gamma\to0$. The dependence of $\chi(0)$ (inset) has been computed directly
at $\omega=0$. The mode's share of the sum rule
doubles across the panel, from $16.4\%$ at $\lambda=2.00$ to $32.4\%$ at
$\lambda=3.50$; the last point, $\lambda=3.50$ ($\Delta_g/\Omega=0.96$),
lies outside the window in which the binding is resolved
(Sec.~\ref{sec:intermediate}, Appendix~\ref{AppendixE}) and is shown for
orientation only.}
\label{Fig-weight}
\end{figure}

Let us now discuss the spectral weight of the mode which may serve as a quantitative measure of the effect. The key idea here is that the shift of the Higgs mode inside the gap removes the square-root singularity of the BCS spectral weight.
In other words the smallness of the binding energy does not make the mode unobservable since the shift of the spectral weight turns out to be quite substantial. 
For our subsequent discussion let us make the following decomposition of the pair susceptibility into a contribution from 
the Higgs mode and quasiparticle continuum:
\beg\label{ChiwomDe}
\chi(\omega)=\frac{2\omega_{\rm H}{\mathsf Z}}{\omega_{\rm H}^2-\omega^2}+\frac{2}{\pi}\int\limits_{2\Delta_g}^\infty \frac{\omega'\textrm{Im}\chi(\omega')}{\omega'^2-\omega^2}d\omega'.
\en
Here and below $\chi(\omega)$ is the gap-edge response \eqref{MyChi}; its
residue, and therefore the fraction $f_{\rm H}$ defined below, refer to that
projection, whereas the position of the pole does not depend on it
(Sec.~III.B). The weight ${\mathsf Z}$ is the residue of the simple pole ${\mathsf Z}=-\left({d\chi^{-1}}/{d\omega}\right)^{-1}_{\omega=\omega_{\rm H}}$.
Near the pair-breaking threshold the natural variable is not $\omega$ but
\beg\label{svar}
s(\omega)=\sqrt{4\Delta_g^2-\omega^2},
\en
and also $\nu=\sqrt{4\Delta_g^2-\omega_{\rm H}^2}$.
The inverse susceptibility $\chi^{-1}(\omega)$ is a regular function of $s$ in a neighbourhood
of $s=0$, and the amplitude mode is a simple zero of it at $s=\nu>0$.  From \eqref{svar} and using $ds/d\omega=-\omega/s$ it then follows
\beg\label{Resnu}
{\mathsf Z}=\frac{\nu}{\omega_{\rm H}}\left[\frac{d\chi^{-1}}{ds}\right]^{-1}_{s=\nu}.
\en
We see that ${\mathsf Z}$ is proportional to $\nu$ and not to the binding energy itself. Since $\nu\simeq2\sqrt{\Delta_g|E_b|}$, 
the weight of the mode is governed by $\sqrt{|E_b|/\Delta_g}$. The enhancement comes
from the inverse-square-root edge singularity of the BCS response: the weight
contained within a window $|E_b|$ of the threshold grows as $\sqrt{|E_b|}$, so a pole
displaced by $|E_b|$ captures a fraction of the near-threshold spectral weight of order $\sqrt{|E_b|/\Delta_g}$ and not
$|E_b|/\Delta_g$. Equivalently, in the variable $s$ the
BCS spectral density is flat, and the mode removes from it a strip of width $\nu$.

Let us estimate the amount of spectral weight captured by the Higgs mode. The static pair susceptibility 
\beg\label{DefineChi0}
\chi(0)=\frac{2}{\pi}\int_0^\infty\textrm{Im}\,\chi(\omega)\,\frac{d\omega}{\omega},
\en
is itself the total spectral weight. The Higgs mode accounts for a fraction $f_{\rm H}$ of the static susceptibility --- weight
that in BCS resides entirely in the continuum, since there the zero of
$\chi^{-1}$ lies on the branch point and no discrete line exists. It then follows that the spectral weight fraction which is carried away by the Higgs mode is 
\beg\label{fweight}
f_{\rm H}=\frac{2{\mathsf Z}}{\omega_{\rm H}\,\chi(0)}\;\propto\;\sqrt{|E_b|}.
\en 
In Fig.~\ref{Fig-weight} (main panel) we show the dependence of this quantity along with $\chi(0)$ (inset) on the dimensionless ratio $\Delta_{g}/\Omega$.
At the upper end of the resolved window, $\delta=0.0117$, the estimate
$f_{\rm H}\simeq2\sqrt{2\delta}$ gives $0.31$ and the computed residue and
$\chi(0)$ give $0.29$: a displacement of one
percent leaves a discrete line carrying a fraction of $\chi(0)$ of order thirty percent. This is why a percent-level shift has consequences that
are not percent-level, and it fixes the size of the time-domain signature of
Fig.~\ref{Fig1-schematic}: the undamped component of $\delta\Delta(t)$ has
relative amplitude $\simeq f_{\rm H}$ and overtakes the decaying BCS envelope
$(\pi\Delta_g t)^{-1/2}$ after $1/\pi^2f_{\rm H}^2$ oscillation periods:
$1.4$ periods for $f_{\rm H}=0.265$ (Fig.~\ref{Fig1-schematic}) and about
four periods at the lower end of the window, $f_{\rm H}=0.16$.

\subsection{Strong coupling limit}
Beyond $\Delta_g\simeq\Omega$ the real-axis calculation of the previous
subsection loses numerical control for the reasons listed above: the replica
structure of the kernel enters the fit window, and the extrapolation of
$\omega_{\rm H}$ to vanishing regulator $\gamma$ can no
longer be carried out with the resolution it requires.
The strong-coupling limit $\lambda\to\infty$ is formal, because metals
with electron--phonon interaction beyond a certain threshold do not exist in
nature in a stable state~\cite{YuzbashyanAltshulerPatra2025,SemenokAltshulerYuzbashyan2025}. For the Einstein
phonons considered here, the stability bound is $\lambda\lesssim3.69$~\footnote{More generally, the
necessary stability condition is $\xi<1$, where $\xi$ is the stability parameter
determined by the electron--phonon contribution to the electronic specific
heat. For a fixed shape of the Eliashberg function, $\lambda=\lambda_*\xi$,
with a shape-dependent coefficient $\lambda_*$. For Einstein and Debye
phonons, $\lambda_*\simeq3.69$ and $4.72$, respectively. In experimentally
realized superconducting metals, $\lambda\lesssim4$.}.
In particular, Migdal--Eliashberg theory itself breaks down at finite coupling,
both for the
Holstein Hamiltonian~\cite{AlexandrovKabanovRay1994,AlexandrovMott1994,MillisMuellerShraiman1996I,AlexandrovMott1996,MillisMuellerShraiman1996II,BenedettiZeyher1998,Alexandrov2001,MeyerHewsonBulla2002,CaponeCiuchi2003,Esterlis2018,EsterlisKivelsonScalapino2019,ChubukovEsterlisAbanovProkofev2026,ProkofevEsterlisAbanovChubukov2026} and in the real phonon approach~\cite{Emil2022,EmilSpins,YuzbashyanKiesslingAltshuler2022}, where the physical
Eliashberg function is taken as input~\footnote{{
A first-order transition involving
lattice reconstruction, preempting complete phonon softening,
was argued for in Refs.~\cite{Emil2022,YuzbashyanAltshulerPatra2025,SemenokAltshulerYuzbashyan2025}.
Preemption of phonon softening by polaronic or bipolaronic states has also
been found in recent studies of the Holstein
model~\cite{ProkofevEsterlisAbanovChubukov2026,ChubukovEsterlisAbanovProkofev2026}.}}.

Nevertheless, this limit is relatively simple  and can
help us understand the finite-coupling
results~\cite{AllenDynes1975,Marsiglio1991}. In particular, it allows us to test
whether the small Higgs binding energy found at finite coupling becomes a
substantial fraction of the gap as $\lambda\to\infty$.

{The BCS weak-coupling limit and the strong-coupling limit are the two
universal limits of Migdal--Eliashberg theory. In both, the dependence of the
low-energy superconducting dynamics on the phonon spectrum and details of the
electron--phonon coupling is absorbed into a single energy scale. In the BCS
limit this scale is the ground-state gap, while in the
strong-coupling limit it is defined by~\cite{AllenDynes1975,Marsiglio1991,Combescot1995}
{
\beg\label{scSpectralMoment}
\omega_{\rm sc}^2
=2\int_0^\infty \,\alpha^2F(\omega)\omega d\omega
=\lambda\langle\omega^2\rangle
\en
}%
{Here $\langle\omega^2\rangle$ is the second moment with weight
$2\alpha^2F(\omega)/(\lambda\omega)$.
For Einstein phonons $\omega_{\rm sc}^2=\lambda\Omega^2$. We take the limit using this
spectrum, sending $\Omega\to0$ and $\lambda\to\infty$ at fixed
$\omega_{\rm sc}$, with $T/\Omega$ held fixed so that $T\to0$ together with
$\Omega$ (Appendix~\ref{AppendixF}). This is a formal extrapolation of the
prescribed input spectrum beyond the stability bound. The same limiting
equations follow for a general spectrum when all phonon frequencies are small
compared with $\omega_{\rm sc}$.} {This is the regime $\Delta_g/\Omega=\infty$.
To study the amplitude response in this limit, we derive the corresponding
form of Eqs.~\eqref{ClosureK}--\eqref{ClosureRAfinal}.} The details of the
analytical derivations can be found in Appendix~\ref{AppendixF}. Here we will
highlight its structure and present the result.

Similar to our discussion above it is insightful to start with the equations in the Matsubara representation. Our approach is organized around a single decomposition: the
Matsubara sum with the kernel $\lambda_{n-m}$ splits into its total weight
$\Lambda=({\pi\omega_{\rm sc}^2}/{2\Omega})\coth({\Omega}/{2T})$,
which diverges as $\Omega\to0$ at any fixed value of $T/\Omega$, and a difference operator $\mathcal{L}$ which remains finite
(Eqs.~\eqref{scSplit}--\eqref{scLzero} of Appendix~\ref{AppendixF}). Physical interpretation of the energy scale $\Lambda$ is straightforward: it is the Born scattering rate of an electron scattering off the lattice displacement treated as a static field. In the limit $\Omega\to 0$ an infinitely
slow phonon acts, at leading order, as nonmagnetic disorder of strength
$1/2\tau=\Lambda\to\infty$, and by Anderson's theorem it should drop out of every
quantity that carries physical information. On the imaginary axis our problem reduces to
the equilibrium gap equation of Allen and Dynes \cite{AllenDynes1975} with $\Delta_{n\to0}=0.755\,\omega_{\rm sc}$. On the real axis the mixed-representation equations \eqref{MSC} couple
$\Delta_\eps$ to $\Delta_{\eps\pm\Omega}$ through the phonon emission and
absorption terms, so that the solution is built in steps of $\Omega$ from
the origin to the gap edge: a grid of $\Delta_g/\Omega$ intervals whose number diverges as
$\Omega\to0$. In that limit the difference between consecutive points on the grid
becomes a derivative, $\Delta_{\eps\pm\Omega}/(\eps\pm\Omega)-\Delta_\eps/\eps
\to\pm\Omega\,\partial_\eps(\Delta_\eps/\eps)$, and the grid collapses
onto the first-order differential equation \eqref{scODE} for
$\Delta_\eps$:
\beg\label{scODE}
\begin{aligned}
&\frac{i\pi\omega_{\rm sc}^2}{2}\,g^R_\eps\,\frac{d}{d\eps}\Bigl(\frac{\Delta_\eps}{\eps}\Bigr)
=S_\Phi(\eps)-\frac{\Delta_\eps}{\eps}\bigl[\eps+iS_Z(\eps)\bigr], 
\end{aligned}
\en
where functions $S_\Phi(\eps)$ and $S_Z(\eps)$ have been defined in Appendix~\ref{AppendixF}
and $g^R_\eps={\eps}/{\sqrt{\eps^2-\Delta_\eps^2}}$.
Integrating the latter from the imaginary-axis solution we find that the gap
edge is no longer a crossing of $\textrm{Re}\,\Delta_\eps=\eps$ but a tangency,
$\Delta_\eps/\eps-1\propto(\Delta_g-\eps)^2$, located at
$\Delta_g=1.16\,\omega_{\rm sc}$, i.e.\ $2\Delta_g/T_c=12.7$ with the
Allen--Dynes $T_c=0.1827\,\omega_{\rm sc}$, in agreement with the real-axis
solution of Combescot \cite{Combescot1995}, and $\Delta_g/\Delta_{n\to0}=1.54$: at
strong coupling the spectral gap exceeds the Matsubara order parameter by half of
itself.

The tangency has a direct consequence for the single-particle spectrum.
Since $\Delta_\eps^2-\eps^2\propto(\Delta_g-\eps)^2$ at the edge,
$\sqrt{\eps^2-\Delta_\eps^2}\propto\Delta_g-\eps$ and $g^R_\eps$ has a simple pole rather
than an inverse-square-root singularity at $\eps=\Delta_g$: the density of
states carries a $\delta$-function at $E_1=\Delta_g$. 
This is the lowest member of the discrete spectrum found by
Combescot~\cite{Combescot1995} and discussed further in
Ref.~\cite{Combescot1996}. At $T=0$ in this limit the density of states is
an infinite set of $\delta$-functions at energies $E_n$, with
$E_1=1.16\,\omega_{\rm sc}$ and $2E_1/T_c=12.7$.
For Einstein phonons at large but finite $\lambda$, the discrete levels
broaden into narrow bands of width approximately
$\Omega=\omega_{\rm sc}/\sqrt{\lambda}$, even at $T=0$~\cite{Emil2022}.
Thermally excited phonons produce additional broadening at
$T>0$~\cite{Combescot1995,Combescot1996}. 

Two
consequences must be kept in mind in what follows. First, $2\Delta_g=2E_1$
is not the onset of a two-quasiparticle continuum but the lowest discrete
pair line, so the language of a pair-breaking threshold, of a branch point
and of the variable $s(\omega)$ of Sec.~\ref{sec:intermediate} does not
carry over to this limit; in particular the bare pair response diverges at
$2E_1$ as a simple pole rather than as an inverse square root. Second, an
undamped oscillation at $2E_1$ is a property of the discrete
single-particle spectrum and not evidence of a detached collective mode,
so the time-domain discriminant of Fig.~\ref{Fig1-schematic} applies to
the intermediate window and not here.

The same decomposition applies to the linear response calculation. Since the source $h$ is an
$O(1)$ addition to the $O(\Lambda)$ anomalous self-energy, the response is itself
$O(\Lambda)$, $(\delta W_n,\delta\Phi_n)=\Lambda\Psi_n$, and the kernel of
\eqref{MatsSystem} collapses at leading order onto a projector
$\mathbb P_n$, Eq.~\eqref{scProj}. The projector fixes the direction of the response,
\beg\label{scLock}
\frac{\delta W_n}{\delta\Phi_n}=-\tan\frac{\Theta_n}{2},\qquad
\Theta_n=\theta_{n+l}+\theta_n,
\en
where $g_n=\cos\theta_n$, $f_n=\sin\theta_n$, for every $n$ and every transfer $\nu_l$, and the finite parts determine its
magnitude through the closed system
\beg\label{scResp}
\Psi_n=\frac{1}{\bar\varsigma_n}\,\mathbb{P}_n\Bigl[h\,\hat{e}_2+(\mathcal{L}\Psi)_n\Bigr]
\en
and $\hat{e}_2=(0,1)^T$.
This closed system replaces \eqref{MatsSystem}--\eqref{PQR} in the
strong-coupling limit. The local weight $\Lambda$ has dropped out, as it must by
Anderson's theorem, and is replaced by the projector and by the finite part
$\bar\varsigma_n$ of the pair-propagator denominator. 
Note that unlike in the intermediate coupling regime, Eq.~\eqref{nonlocal}, at strong coupling the relation between $\delta W_n$ and $\delta \Phi_n$ is purely local with a ratio that depends on nothing but the
equilibrium solution. 
Because the direction is fixed, the mass-renormalization and
vertex components are no longer independent, and $\delta W\equiv0$ is
inconsistent with the equations at leading order. In this limit the two
channels cannot be separated, and the displacement of the mode cannot be
attributed to either of them alone. Note that equation \eqref{quadratic} does not apply in this limit in the form derived
in Sec.~III, since the gap edge is a tangency rather than a crossing and the
threshold no longer carries the square-root character on which that
argument is based. The exclusion \eqref{scBound} below is therefore a
separate statement, obtained without any assumption about the
near-threshold form of the response, and no mechanism of the kind that
produced the percent-level scale in the intermediate window is attached to
it. Through the intermediate window the mass channel enters as a second-order
correction, Eq.~\eqref{dPhiEff}; here it enters as a constraint of order
unity on the response. It is also worth mentioning that the locked system is not the truncated one with a renormalized
kernel: there $\delta W$ is absent, here it is comparable to $\delta\Phi$
and fixed by it.

The mode itself is found on the real axis by the construction of
Eq.~\eqref{scStrip}: the internal Matsubara variable is shifted by half the
external frequency, so that the two propagators entering the response are
evaluated at $z=iy\pm\omega/2$, symmetric about the real internal frequency
$y$, with $\Delta(z)$ on the two lines $\textrm{Re}\,z=\pm\omega/2$ supplied by
Eq.~\eqref{scODE}. For real $|\omega|<2\Delta_g$ both arguments
$iy\pm\omega/2$ remain inside the strip $|\textrm{Re}\,z|<\Delta_g$ in which the
propagators are analytic, so the response is analytic in $\omega$, real on
the sub-gap segment, and coincides with the Matsubara response on the
imaginary axis. We find that the two-component system
becomes equivalent to a single real scalar equation,
\beg\label{scScalar}
\begin{aligned}
{\bar\varsigma}\,a(y)&-\frac{\omega_{\rm sc}^2}{2}\dashint
\frac{[C(y,y')\,a(y')-a(y)]dy'}{(y-y')^2}\\&=h\cos\frac{\Theta(y)}{2},
\end{aligned}
\en
with the integral running over the whole real line, $C(y,y')=\cos[\Theta(y)/2-\Theta(y')/2]$ and $\Theta(y)\equiv\Theta(y;\omega)$,
$\bar\varsigma\equiv\bar\varsigma(y;\omega)$, in which the angle $\Theta$ and the finite part $\bar\varsigma$ of the
pair-propagator denominator are known functions of the equilibrium solution
(Appendix~\ref{AppendixF}). The operator on the left is real and
symmetric. Discretized on a uniform grid in $y$
the equation is almost verbatim the Matsubara system \eqref{scResp} at an effective
temperature $T_{\rm eff}=\delta y/2\pi$, so its convergence is that of a
finite-temperature Matsubara calculation, and at $\omega=0$ it reproduces the
static response $d\theta_n/dh$ of the sourced gap equation \eqref{scAD}.

Equation \eqref{scScalar} has the form $\hat L(\omega)\,a=h\cos[\Theta(y)/2]$,
where the right-hand side acts as the source and $\hat L(\omega)$ is the real
symmetric operator on the left, restricted to the sector driven by the source
($\delta\Phi$ even and $\delta W$ odd in $y$).
The response is therefore $a=h\hat L^{-1}(\omega)\cos[\Theta(y)/2]$, and it is finite for every $\omega$
at which $\hat L(\omega)$ can be inverted. A collective mode is precisely at
frequency $\omega=\omega_{\rm H}$ when the operator becomes
singular, one of its eigenvalues passes through zero, and the corresponding susceptibility
acquires a pole. Writing $\mu_j(\omega)$ for the eigenvalues of $\hat L(\omega)$,
ordered so that $\mu_0$ is the smallest, the amplitude mode is the solution of
\beg\label{scRoot}
\mu_0(\omega_{\rm H})=0,\qquad \omega_{\rm H}<2\Delta_g .
\en
It is useful to interpret $\mu_0(\omega)$ as the stiffness of the amplitude channel at
frequency $\omega$: the smaller it is, the larger the amplitude response to a
given pair field, and a collective mode is the frequency at which the channel
costs nothing to excite. The virtue of this formulation is that it replaces
every step that limited the intermediate-coupling calculation --- the regulator
$\gamma$, the lineshape fit, the extrapolation --- by a single root-finding
problem for a real function of one variable.
  
\begin{figure}
\includegraphics[width=0.995\linewidth]{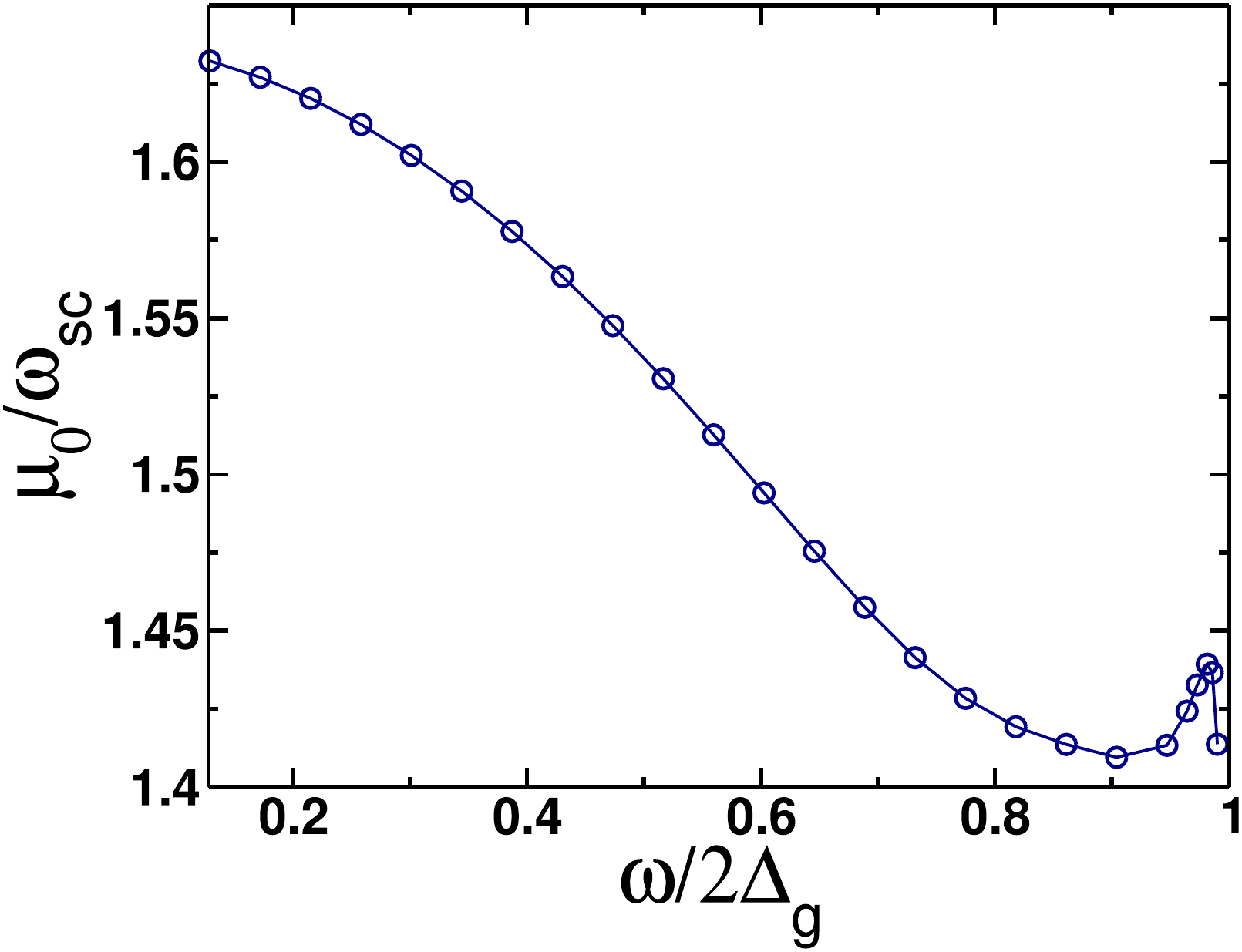}
\caption{\footnotesize Strong-coupling limit $\Omega\to0$, $\lambda\to\infty$ at
fixed $\omega_{\rm sc}=\sqrt\lambda\,\Omega$, $T=0$. Stiffness $\mu_0(\omega)$ of
the amplitude channel: the lowest eigenvalue of the operator in
\eqref{scScalar}, in the sector driven by the source. The amplitude mode would
appear as a zero of $\mu_0$; none occurs. Grid spacing $\delta
y=0.005\,\omega_{\rm sc}$, cutoff $|y|<10\,\omega_{\rm sc}$; the last point is at
$1-\omega/2\Delta_g=0.01$, where the scale $\Delta_g-\omega/2$ on which the leg
functions vary near $y=0$ equals $2.3\,\delta y$.}
\label{Fig-scEig}
\end{figure}
 
The result is shown in Fig.~\ref{Fig-scEig}. The stiffness falls from
$\mu_0(0)=1.63\,\omega_{\rm sc}$ to $1.41\,\omega_{\rm sc}$ at
$\omega=0.86\times2\Delta_g$ and then flattens, remaining between $1.41$ and
$1.44\,\omega_{\rm sc}$ up to $1-\omega/2\Delta_g=0.01$, the last point of
the scan. It shows no tendency to close, and the two grids
$\delta y=0.01$ and $0.005\,\omega_{\rm sc}$ agree to $1\%$ across the plateau.
The homogeneous problem therefore has no root on the sub-gap segment down to one
percent below $2E_1$,
\beg\label{scBound}
\mu_0(\omega)>0\quad\text{for all}\quad 0\le\omega\le0.99\times2\Delta_g\qquad(\Omega\to0),
\en
i.e.\ no collective pole of the source-coupled amplitude sector lies more than
one percent below $2E_1$ in the strong-coupling limit. This is an exclusion
and not a location. The values of $\mu_0$ depend on how the operator is discretized and normalized,
and only their sign enters~\eqref{scBound}. The last points of the scan, where
$\Delta_g-\omega/2$ spans two to five cells of the finer grid, are resolved only
to the extent that the two grids agree there, so that~\eqref{scBound} is firm a
few percent below $2E_1$ and holds at one percent within that resolution.

Whether a pole exists within the last percent, whether
the amplitude excitation in this limit is the pair line $2E_1$ itself, or
whether a collective pole lies above $2E_1$ between the pair lines
$E_n+E_m$, is not determined by this calculation. Our search above $2E_1$
requires the real-axis reduction of Appendix~\ref{AppendixF}, which we have
not carried out. The last percent below $2E_1$ is a threshold-layer
problem: near $y=0$ the leg functions vary on the scale $\Delta_g-\omega/2$,
which vanishes at $2E_1$, and both the diagonal term and the projector of
\eqref{scScalar} vary on that scale, so that it requires either a mesh
refined near $y=0$ or an analytic treatment of the layer. We do not attempt
this here. What can be stated without qualification is that the amplitude
channel 
retains a finite stiffness to within a few percent of $2E_1$, and to within one percent at the resolution of the two grids.
 
Equation \eqref{scBound} complements the results shown in Fig.~\ref{Fig-Binding}. The
binding energy grows with $\Delta_g/\Omega$ through the intermediate window and
reaches $1-\omega_{\rm H}/2\Delta_g\simeq0.012$ at its upper end,
$\Delta_g/\Omega=0.9$. The linear law \eqref{Eq1} cannot continue
indefinitely, since at $\Delta_g/\Omega=\infty$ no subgap pole with a
binding above one percent exists. The two statements refer to different
objects: in the window the binding is that of a resolved pole below a
genuine two-quasiparticle threshold, whereas at $\Delta_g/\Omega=\infty$
the single-particle spectrum is discrete and \eqref{scBound} is an
exclusion. The interval $1\lesssim\Delta_g/\Omega<\infty$ between them is not accessible to either method and is left
open. In the intermediate window the retardation of the pairing vertex, acting alone, displaces the mode by
less than the resolution of the calculation, $\delta_{\delta W=0}\lesssim0.005$. The
response of the mass renormalization supplies at least $40\%$ of the
binding at the lower end of the window and at least $55\%$ at the
upper end, and possibly all of it. Both vanish in the BCS limit $\Omega\gg\Delta_g$, where the
mode sits on the threshold. Through the intermediate window the mass
renormalization acts at second order in the interchannel coupling while in the
strong-coupling limit $\Omega\to 0$ the two responses are locked rigidly,
Eq.~\eqref{scLock}, and can no longer be separated.

\section{Discussion}

One of the main insights of our work is that the structure of the collective
mode encodes the frequency structure of the pairing interaction. Three classes
can be distinguished. An amplitude response pinned to the pair-breaking edge,
with algebraic decay, identifies an essentially instantaneous kernel. A deep
subgap mode, $\delta=O(1)$, accompanied at strong coupling by subgap
satellites, identifies a pairing kernel that depends on the band energies. In
that case the order parameter itself becomes a function of the band energy,
$\Delta\to\Delta(\xi)$ \cite{Levitov2007}, with no dynamics behind it: neither
a retarded vertex nor a mass renormalization, as in
Refs.~\cite{Levitov2007,Uhrig2025,Uhrig2026}. Finally, a single sharp pole
detached from the edge by a percent-level binding identifies a genuinely
retarded glue in the Migdal class. We have established this here for the
Einstein spectrum, in the window $0.56\le\Delta_g/\Omega\le0.90$ in which the
binding is resolved. The three classes are separated by orders of magnitude in
$\delta$ and by qualitatively different time-domain and spectral signatures,
so that the assignment does not require resolving the shift itself. Moreover,
the relation \eqref{Eq1} is invertible. Should it extend to other bosonic
spectra, which remains to be established, a measurement of the Higgs binding
energy together with the spectral gap would return the characteristic
frequency of the pairing boson, to the accuracy of the coefficient of
\eqref{Eq1}, about $20\%$.

The assignment is also protected against the principal contaminants, since
competing mechanisms that displace the mode carry their own signatures: pair
breaking shifts and damps it \cite{Kamenev2025}, and coupling to another
collective degree of freedom supplies a partner mode
\cite{LittlewoodVarma1981,Krull2016}. 
{A quantitative comparison with a particular material requires its measured or calculated Eliashberg function as input. The equations for an arbitrary spectrum follow from the replacement~\eqref{SpectralReplacement}. How phonon dispersion, spectral width, and multiple phonon branches affect the Higgs frequency at finite coupling remains to be investigated. These calculations, together with the experimental protocols that distinguish the classes, are left for future work.}
Our calculation describes spatially uniform dynamics within linear response.
Spatially uniform perturbations can nevertheless lead to the growth of spatial
inhomogeneities in superconductors and other many-body
systems~\cite{DzeroTurbulence2009,Foster2010,Foster2011,ChernBarros2019,Barresi2023,Yang2024,FanCai2024,FanGarcia2024,Grankin2026}.
For the Higgs oscillations considered here, nonlinear coupling to modes with
nonzero wave vector could produce a similar instability at finite
amplitude~\cite{Zakharov1992}. Whether this occurs, and on what time scale,
remains to be investigated.}

It is worth stating what would be required in order to displace the mode
further, since the answer bounds the retarded class from within. Equation
\eqref{dPhiEff} offers only two handles: the magnitude of the interchannel
coupling $\mathsf{K}_Q$, and the proximity of the mass-renormalization channel
to an instability of its own, measured by $(1-\mathsf{K}_P)^{-1}$. Neither is
favourable. At $\lambda=2.5$ the largest eigenvalue of $\mathsf{K}_P$ is
$0.58$, so the second factor supplies an enhancement of only $2.4$, while
$\|\mathsf{K}_Q\|=0.21$. In addition $Q_m\propto g_{m+l}f_m+f_{m+l}g_m$ is
further suppressed at precisely the low Matsubara indices that dominate the
response, where $g_n\propto\omega_n$ vanishes. Neither can the pairing source
$h$ be used. The frequency $\omega_{\rm H}$ is a root of the determinant of
the operator in \eqref{dPhiEff} and is therefore independent of $h$. A
perturbation carrying $\hat\tau_3$, such as a coherently driven phonon or a
modulated coupling constant, excites the mass-renormalization channel directly
and alters the weight of the response, but it does not move the mode. We
therefore conclude that a binding energy substantially in excess of a percent
is not generated within Migdal--Eliashberg theory at fixed $\alpha^2F(\omega)$.
The observation of a larger binding would then constitute evidence for physics
outside that class: {dynamical feedback on the phonon propagator of the kind included in}
Ref.~\cite{Werner2016}, a second band, or a pairing interaction of the kind
used in Refs.~\cite{Levitov2007,Uhrig2025}. We have established this conclusion for the Einstein spectrum in the window where the binding is resolved, and excluded a larger binding in the formal strong-coupling limit. Whether the same holds at intermediate couplings or for other phonon spectra at finite coupling remains to be established.

{At finite coupling we have taken the pairing glue to be a single physical Einstein mode. Besides the quantitative dependence on the phonon spectrum, another question is how pairing by gapless electronic fluctuations changes the result. It is natural to ask what survives when the glue is instead governed by
quantum-critical fluctuations of an electronic order parameter, where the boson
is overdamped and gapless. For example, near a two-dimensional ferromagnetic
quantum critical point the pairing kernel acquires the singular frequency
dependence $\langle\tilde I\rangle_\omega\propto\omega^{-1/3}$ and the
self-energy the non-Fermi-liquid form
$\Sigma(\omega)\propto\omega^{2/3}$~\cite{Dzero2004}. This is the case
$\gamma=1/3$ of the model with the interaction $(\bar g/|\Omega_m|)^{\gamma}$
reviewed in Ref.~\cite{AbanovChubukov2026}, and the strong-coupling limit of
Sec.~IV.B is the case $\gamma=2$ of the same model. The transition temperature
is of the order of the characteristic bosonic scale,
$T_c\sim0.06\,\omega_0$~\cite{Dzero2004}, so a critical metal sits
automatically in the strongly retarded regime. A massless boson does not by
itself fill the region below $2\Delta_g$ at $T=0$: a quasiparticle at the gap
edge has no lower state to decay into, the two-particle continuum still begins
at $2\Delta_g$, and a pole below it remains undamped. The boson does allow
quasiparticles above the edge to decay by emitting arbitrarily soft
fluctuations, and this changes the analytic character of the pair-breaking
threshold, which in Sec.~III fixed the size of the binding through
Eq.~\eqref{quadratic}. At $T>0$ thermally excited soft bosons are absorbed as
well and the pole acquires a finite width. Both effects follow from the present
formalism with the kernel replaced. The Migdal--Eliashberg treatment stays
controlled there, vertex corrections near the ferromagnetic critical point
being small in the same parameter that makes the interaction long
ranged~\cite{Dzero2004}.

\section{Conclusions}\label{sec:conclusions}
We have shown that in Migdal--Eliashberg theory the amplitude (Higgs)
excitation of a superconductor is a genuine collective mode: an isolated pole
of the pair susceptibility at a frequency $\omega_{\rm H}$ below the
pair-breaking continuum, with the small binding energy
$2\Delta_g-\omega_{\rm H}$ given by Eq.~\eqref{Eq1}. This elevates the
amplitude channel from what it is in BCS theory -- a threshold resonance
pinned exactly at $2\Delta$, without a pole -- to an undamped oscillator.
{We determine the retarded interaction kernel from the physical phonon spectrum and calculate the electronic pairing and normal self-energies self-consistently. No auxiliary collective degree of freedom is required.} The
retardation of the phonon-mediated interaction is itself the physical
deformation which lifts the threshold degeneracy. The resulting pole lies
below every branch point and, at $T\ll\Delta_g$, below any inelastic
threshold, so that the mode is well defined.

The smallness of the binding energy is not an accident of parameters but a
structural property of the theory. The two effects of retardation do not act
on the threshold state symmetrically. Acting alone, the retardation of the
pairing vertex displaces the mode by less than the resolution of the
extrapolated binding, $\delta_{\delta W=0}<0.005$. At least $40\%$ of the displacement at
the lower end of the window, and at least $55\%$ at the upper end, comes
instead from the response of the mass renormalization, and possibly all of it
(Sec.~IV, Appendix~\ref{AppendixE}). That channel carries no source of its
own, since a pair field couples to $\delta\Phi$ and not to $\delta W$. It
therefore enters the pair response only at second order in the interchannel
coupling, and it acts upon a zero of square-root character, which converts its
action into a shift of second order again. {In the formal strong-coupling limit, characterized by the single scale $\omega_{\rm sc}^2=\lambda\langle\omega^2\rangle$,} the question changes character: the two channels are rigidly
locked and cannot be separated, and the single-particle spectrum becomes
discrete. There no subgap collective pole with a binding above one percent
exists, while the fate of the amplitude excitation within the last percent of
$2E_1$, or above it, is left open.

The displacement is therefore of order one percent where it is resolved. It is
linear in $\Delta_g/\Omega$ across the window $0.56\le\Delta_g/\Omega\le0.90$,
and at $\Delta_g/\Omega=\infty$ no subgap pole bound by more than a percent
exists. The interval in between is not covered. Treatments in which the order
parameter depends on the band energy, $\Delta\to\Delta(\xi)$, are not subject
to this constraint. There the energy dependence of the pairing interaction is
not accompanied by the mass renormalization which Migdal--Eliashberg theory
ties to it, and these treatments place the mode an order of magnitude further
from the threshold. The percent-level scale is also the likely reason the
displacement has escaped self-consistent Migdal calculations at accessible
frequency resolution.

On the technical side, the kinetic equations on which this work rests are the
standard ones of the strong-coupling theory
\cite{Eliashberg1972,LO1975,LO1977,SereneRainer1983,Kopnin2001}, and real-axis
linear response of Eliashberg superconductors has already been established in
the electromagnetic channel \cite{LRZ1989,Bickers1990,Marsiglio2008}. We have
extended the existing theoretical framework to the analysis of the amplitude
channel driven by a pair source, formulated as a closed linear system on the
real frequency axis for the retarded, advanced and anomalous blocks of the
driven propagator. Two theoretical checks anchor our construction. First, the
closure equations reduce exactly to the Matsubara response when the external
frequency is continued to the bosonic axis (Appendix~\ref{AppendixD}), which
tests every kernel and contour. Second, the gap-edge projection \eqref{MyChi}
reproduces the static response of the sourced equilibrium problem at
$\omega\to0$. {The method is not restricted to the physical Einstein spectrum used in the calculations at finite coupling. An arbitrary $\alpha^2F(\omega)$ can be included using Eq.~\eqref{SpectralReplacement}. Calculations for specific materials, the dependence of the Higgs frequency on the phonon spectrum, and extensions to other response channels are left for future work.}

The experimental consequences follow from the weight rather than from the
frequency \cite{Cea2016}. Near a square-root branch point the residue scales as
$\sqrt{2\Delta_g-\omega_{\rm H}}$, so the percent-level binding leaves a pole
carrying a substantial fraction of the amplitude-channel spectral weight, in a
region that is empty at $T\ll\Delta_g$: below $2\Delta_g$ there is no
two-quasiparticle continuum to compete with it. What is hard to measure is the
\emph{position} of the line, one percent below an edge; what is not hard is
its presence. In the time domain the distinction is qualitative. The BCS
amplitude oscillation dephases across the continuum with envelope $t^{-1/2}$,
while a genuine pole oscillates undamped at $\omega_{\rm H}$, and the two
become comparable within a few periods.

Several avenues for future work follow directly. The first is the evolution of
the mode through the interval $1\lesssim\Delta_g/\Omega<\infty$ between the
resolved window and the strong-coupling limit, where the linear growth of the binding cannot continue,
together with the threshold-layer problem at $\Delta_g/\Omega=\infty$. The
second is the fate of the pole at elevated temperatures, as inelastic
thresholds descend toward it, and the driven protocols in which the crossover
from algebraic to persistent oscillations can be observed directly. {The third is the effect of phonon dispersion and spectral shape on the Higgs frequency at finite coupling. A general Eliashberg function is also needed for comparison with experiment. The fourth is the extension to a dynamical pairing glue of electronic rather than lattice origin.}

\begin{acknowledgments}
The work of MD was financially supported by the National Science
Foundation Grant No. DMR-2400484. MD thanks the Niels Bohr Institute,
where the main part of this work was completed, for its hospitality. 
Most of the numerical calculations were performed using an allocation of computing time and other resources of the Ohio Supercomputer Center~\cite{OSC1987}. The authors acknowledge the use of Anthropic's
Claude for cross-checking the analytical derivations, for independent
numerical verification of the results, and for figure preparation and
manuscript checking. All outputs of Claude were checked and validated by
the authors.
\end{acknowledgments}

\section*{Data availability} The data that support the findings of this article were generated by numerical evaluation of the analytical expressions derived herein and can be regenerated in full from the information given in the paper. For this reason the authors did not  deposit them in a public repository. The data and the codes that produce figures presented in this manuscript are available from the authors upon request. 

\begin{appendix}
\begin{widetext}
\section{Conventions and self-energies}\label{AppendixA}
The derivation of the Eilenberger--Keldysh equation
\eqref{NowFinalEilenberger} from the left and right Dyson equations of the
Nambu--Keldysh propagator \eqref{Gab} --- by subtraction of the two,
transformation to the mixed (Wigner) representation, gradient expansion
and integration over the magnitude of the momentum --- is standard
(see e.g. Refs. \cite{Eliashberg1972,LO1975,LO1977,SereneRainer1983,LO1986,Kopnin2001})
and is not repeated here. We record only the conventions in which the
main text is written: the electron--phonon self-energy and its retarded,
advanced and Keldysh components, the matrix notation in Keldysh space,
the mixed representation and the product $\circ$, the separation of the
order parameter from the collision integral, and the explicit form of
the self-energies for the Einstein spectrum \eqref{alpha2Fw}.

\paragraph{Self-energy on the contour.}
{In the self-consistent Migdal approximation, the electronic self-energy on the Keldysh contour is}
\beg\label{Self-energy}
\Sigma_{\alpha\rho}^{(jk)}(1,2)=ig_c^2(-1)^{j+k+\alpha+\rho}G_{\alpha\rho}^{(jk)}(1,2)D^{(kj)}(2,1),
\en
{where $D^{(ij)}(1,2)$ is the physical phonon propagator, $G$ is the dressed electronic propagator, and $g_c$ has the normalization specified above Eq.~\eqref{DKRA}.} $\alpha,\rho$ are
Nambu and $j,k$ Keldysh indices, and no summation is implied. Direct
substitution of \eqref{Gab} into \eqref{Self-energy} shows that
\beg\label{SelfEnergyAux}
\hat{\Sigma}^{(11)}+\hat{\Sigma}^{(22)}=-\hat{\Sigma}^{(12)}-\hat{\Sigma}^{(21)},
\en
so that, as for the propagator, only three combinations are independent.
The retarded, advanced and Keldysh self-energies are defined as
\beg\label{SigmaRA}
\hat{\Sigma}^{R}=\hat{\Sigma}^{(11)}+\hat{\Sigma}^{(12)}, \qquad
\hat{\Sigma}^{A}=\hat{\Sigma}^{(11)}+\hat{\Sigma}^{(21)}, \qquad
\hat{\Sigma}^K=\hat{\Sigma}^{(11)}+\hat{\Sigma}^{(22)}=-\hat{\Sigma}^{(12)}-\hat{\Sigma}^{(21)},
\en
the last equality by \eqref{SelfEnergyAux}. With
$\hat{\cal G}^{X}=\hat{\tau}_3\hat{G}^{X}$ ($X=R,A,K$) the propagator and
the self-energy are collected into the upper-triangular matrices in
Keldysh space
\beg\label{GSigma}
\check{\cal G}=\left(\begin{matrix} \hat{\cal G}^R & \hat{\cal G}^K \\ 0 & \hat{\cal G}^A\end{matrix}\right), \quad
\check{\Sigma}=\left(\begin{matrix} \hat{\Sigma}^R\hat{\tau}_3 & \hat{\Sigma}^K\hat{\tau}_3 \\ 0 & \hat{\Sigma}^A\hat{\tau}_3\end{matrix}\right),
\en
and $\check{\tau}_3=\hat{\gamma}_0\otimes\hat{\tau}_3$ is diagonal in
Keldysh space.

\paragraph{Mixed representation.}
We separate the centre-of-mass coordinates $\bR=(\br_1+\br_2)/2$,
$t=(t_1+t_2)/2$ from the relative ones $\br_{12}=\br_1-\br_2$,
$t_{12}=t_1-t_2$ and Fourier transform with respect to the latter,
\beg\label{Wigner}
\check{\cal G}(1,2)=\int\frac{d\eps}{2\pi}\int\frac{d^3\bp}{(2\pi)^3}\check{\cal G}_{\bp\eps}(\bR,t)e^{i(\bp\cdot\br_{12}-\eps t_{12})}.
\en
In this representation a convolution in space and time becomes the
Groenewold--Moyal product
\beg\label{GroenewoldMoyal}
\begin{aligned}
C(1,2)&=\int A(1,3)B(3,2) d3 \to C_{\bp\eps}(\bR,t)\\&=A_{\bp\eps}(\bR,t)\exp\left\{\frac{i}{2}\left(\stackrel{\leftarrow}\partial_\bR\stackrel{\rightarrow}\partial_\bp
-\stackrel{\leftarrow}\partial_t\stackrel{\rightarrow}\partial_\eps-\stackrel{\leftarrow}\partial_\bp\stackrel{\rightarrow}\partial_\bR+\stackrel{\leftarrow}\partial_\eps\stackrel{\rightarrow}\partial_t\right)\right\}B_{\bp\eps}(\bR,t)\equiv A\circ B,
\end{aligned}
\en
which is the product denoted by $\circ$ in \eqref{NowFinalEilenberger}.
For a perturbation with a single Fourier component $e^{-i\omega t}$ the
exponential can be evaluated exactly and produces the shifts
$\eps\to\eps\pm\omega/2$ of the two legs that appear throughout
Sec.~III.

\paragraph{Order parameter and collision integral.}
Following Eliashberg \cite{Eliashberg1972} and Volkov and Kogan
\cite{VolkovKogan1974}, the off-diagonal part of the self-energy that is
common to the retarded and advanced components is identified with the
order parameter,
\beg\label{DefineDelta}
\Delta(1,2)=\frac{\Sigma_{12}^{R}(1,2)+\Sigma_{12}^{A}(1,2)}{2},
\en
and the remainder with the collision integral:
\beg\label{Delta12Matrix}
\begin{aligned}
&\hat{\Sigma}^R(1,2)\hat{\tau}_3 =\left(\begin{matrix} 0 & -\Delta(1,2) \\ \Delta^*(1,2) & 0 \end{matrix}\right)+
\left(\begin{matrix} \Sigma_{11}^R & -\frac{\Sigma_{12}^R-\Sigma_{12}^A}{2} \\
\frac{\Sigma_{21}^R-\Sigma_{21}^A}{2} & -\Sigma_{22}^R
\end{matrix}\right)\equiv-\hat{\Delta}(1,2)+\hat{\Pi}^R(1,2), \\
&\hat{\Sigma}^A(1,2)\hat{\tau}_3 =\left(\begin{matrix} 0 & -\Delta(1,2) \\ \Delta^*(1,2) & 0 \end{matrix}\right)+
\left(\begin{matrix} \Sigma_{11}^A & -\frac{\Sigma_{12}^A-\Sigma_{12}^R}{2} \\
\frac{\Sigma_{21}^A-\Sigma_{21}^R}{2} & -\Sigma_{22}^A
\end{matrix}\right)\equiv-\hat{\Delta}(1,2)+\hat{\Pi}^A(1,2).
\end{aligned}
\en
The lower left entries follow from $[\Sigma^{R(A)}_{12}]^*=\Sigma^{R(A)}_{21}$,
which time reversal imposes on \eqref{Self-energy}. The collision integral of
\eqref{NowFinalEilenberger} is
\beg\label{CollisionIntegral}
\check{I}(\bp\eps;\bR t)=\left[\check{\Pi}(\bp\eps;\bR,t)\stackrel{\circ},\check{\cal G}_{\bp\eps}(\bR,t)\right], \quad
\check{\Pi}=
\left(\begin{matrix}\hat{\Pi}^R & \hat{\Pi}^K\hat{\tau}_3 \\
0 & \hat{\Pi}^A\end{matrix}\right),
\en
and $\check{\Delta}_{\eps}(\br,t)=\hat{\Delta}_{\eps}(\br,t)\otimes\hat{\gamma}_0$
is diagonal in Keldysh space; its dependence on momentum is neglected
($s$-wave pairing). Because $\check{\Delta}$ has no off-diagonal entry in
Keldysh space, the separation \eqref{Delta12Matrix} leaves the upper right
entry of $\check{\Sigma}$ in \eqref{GSigma} untouched, so that
\beg\label{PiK}
\hat{\Pi}^K=\hat{\Sigma}^K.
\en
This is also why $\hat{\Pi}^{R}$ and $\hat{\Pi}^{A}$ absorb the $\hat\tau_3$
in \eqref{Delta12Matrix} whereas $\check{\Pi}$ carries it explicitly in the
Keldysh entry. The order parameter itself is
\beg\label{SelfConsistency}
\begin{aligned}
\Delta_\eps(\br,t)&=\frac{g_c^2}{4i}\left(\frac{\pi\nu_F}{i}\right)\int\frac{d^2\bn}{4\pi}\int\limits_{-\infty}^\infty\frac{d\omega}{2\pi}[\hat{g}_{\bn\omega}^K(\br,t)]_{12}[D^R(\eps-\omega)+D^A(\eps-\omega)]\\&+\frac{i\pi\lambda\Omega}{8}\coth\left(\frac{\Omega}{2T}\right)\sum\limits_{s=\pm}
\int\frac{d^2\bn}{4\pi}\left[\hat{g}_{\bn\eps-s\Omega}^R(\br,t)+\hat{g}_{\bn\eps-s\Omega}^A(\br,t)\right]_{12},
\end{aligned}
\en
where $\bn=\bp/p$, $d^2\bn$ is the element of solid angle, and the factor
$\pi\nu_F/i$ inverts the normalisation of \eqref{SmallgRA}. The second term,
which carries the phonon population, is the one that makes $\Delta_\eps$ a
function of energy already in equilibrium.

\subsection{Self-energy parts}
We now record each component of the self-energy separately. Inverting the
definitions of $\hat{G}^{R,A,K}$ given below \eqref{Gab}, and the
corresponding ones for the phonon propagator,
\beg\label{ContourRAK}
\hat{G}^{(11)}=\hat{G}^{R}+\hat{G}^{(12)}, \qquad D^{(21)}=D^{(11)}-D^{A},
\qquad \hat{G}^{(12)}=\frac{\hat{G}^{K}+\hat{G}^{A}-\hat{G}^{R}}{2}, \qquad
D^{(11)}=\frac{D^{R}+D^{A}+D^{K}}{2},
\en
together with the relations obtained from these by $R\leftrightarrow A$ and
$(12)\leftrightarrow(21)$. From \eqref{Self-energy} and \eqref{SigmaRA},
\beg\label{SigmaREq1}
\begin{aligned}
&\hat{\Sigma}^{R}(1,2)=ig_c^2\hat{\tau}_3\left(\hat{G}^{(11)}D^{(11)}-\hat{G}^{(12)}D^{(21)}\right)\hat{\tau}_3=ig_c^2\hat{\tau}_3\left(\hat{G}^RD^{(11)}+\hat{G}^{(12)}D^{A}\right)\hat{\tau}_3=\frac{ig_c^2}{2}\hat{\tau}_3\left\{
\hat{G}^R\left(D^R+D^A+D^K\right)\right.\\&\left.+\left(\hat{G}^K+\hat{G}^A-\hat{G}^R\right)D^A\right\}\hat{\tau}_3=\frac{ig_c^2}{2}\hat{\tau}_3\left\{
\hat{G}^R\left(D^R+D^K\right)+\left(\hat{G}^K+\hat{G}^A\right)D^A\right\}\hat{\tau}_3.
\end{aligned}
\en
Here each $\hat{G}$ has an argument $(1,2)$ and each $D$ enters with an
argument $(2,1)$; the second equality uses the first two relations of
\eqref{ContourRAK} and the third equality the last two. The matrices
$\hat\tau_3$ compensate the $(-1)^{\alpha+\rho}$ of \eqref{Self-energy}. For
{the physical Einstein spectrum} \eqref{SigmaREq1} becomes
\beg\label{SigmaREq2}
\begin{aligned}
&\hat{\Sigma}_{\eps}^{R}(\br,t)\hat{\tau}_3=\frac{\lambda\Omega}{8}\int d\omega\left\{
\hat{g}_{\eps+\omega}^R(\br,t)\left(D_\omega^R+D_\omega^K\right)+\left(\hat{g}_{\eps+\omega}^K(\br,t)+\hat{g}_{\eps+\omega}^A(\br,t)\right)D_\omega^A\right\}.
\end{aligned}
\en
Here $\lambda=2g_c^2\nu_F/\Omega$, and we have used the definition \eqref{SmallgRA}. This expression can be rewritten using
\beg\label{DRAmom}
D^A(-\omega)=D^R(\omega),
\en
so that
\beg\label{SigmaREq2m}
\begin{aligned}
&\hat{\Sigma}_{\eps}^{R}(\br,t)\hat{\tau}_3=\frac{\lambda\Omega}{8}\int{d\omega}\left\{
\hat{g}_{\eps-\omega}^R(\br,t)\left(D_\omega^A+D_\omega^K\right)+\left(\hat{g}_{\eps-\omega}^K(\br,t)+\hat{g}_{\eps-\omega}^A(\br,t)\right)D_\omega^R\right\}.
\end{aligned}
\en

Similarly, for the advanced part we find
\beg\label{SigmaAEq2}
\begin{aligned}
&\hat{\Sigma}_{\eps}^{A}(\br,t)\hat{\tau}_3=\frac{\lambda\Omega}{8}\int{d\omega}\left\{
\hat{g}_{\eps-\omega}^A(\br,t)\left(D_\omega^R+D_\omega^K\right)+\left(\hat{g}_{\eps-\omega}^K(\br,t)+\hat{g}_{\eps-\omega}^R(\br,t)\right)D_\omega^A\right\}.
\end{aligned}
\en
It remains to obtain the Keldysh part of the self-energy,
\beg\label{SigmaKEq2}
\begin{aligned}
&\hat{\Sigma}_{\eps}^{K}(\br,t)\hat{\tau}_3=\frac{\lambda\Omega}{8}\int{d\omega}\left\{
\hat{g}_{\eps-\omega}^K(\br,t)D_\omega^K+\left(\hat{g}_{\eps-\omega}^R(\br,t)+\hat{g}_{\eps-\omega}^A(\br,t)\right)\left(D_\omega^R+D_\omega^A\right)\right\}.
\end{aligned}
\en

%
%
\section{Equilibrium Eliashberg equations on real axis}\label{AppendixB}
Equations \eqref{EliEqsSet1} should not be iterated in the form in which they
are written. Using
$D_\omega^R=2\Omega/[(\omega+i0)^2-\Omega^2]$ the frequency integral
splits into
\beg\label{ABsplit}
\int d\eps'D_{\eps-\eps'}^R\,\textrm{Re}(g_{\eps'}^R)\,t_{\eps'}
=\underbrace{\dashint d\eps'\,\frac{2\Omega}
{(\eps-\eps')^2-\Omega^2}\,\textrm{Re}(g_{\eps'}^R)\,t_{\eps'}}_{\rm (A)}
-\;i\pi\underbrace{\sum\limits_{s=\pm}s\,
\textrm{Re}(g_{\eps-s\Omega}^R)\,t_{\eps-s\Omega}}_{\rm (B)},
\en
and the two pieces are of entirely different character. 
Term denoted by (A) is a full kernel connecting every $\eps$ to every $\eps'$ across
the whole band. It is (A) that makes direct iteration unattractive: it must
be recomputed at every sweep, and its integrand falls off only as a power of
$\eps'$. Term (B), like the
$\coth$ term of \eqref{EliEqsSet1}, connects $\eps$ only to $\eps\mp\Omega$. 

The idea is not to compute (A) at all, but to convert it into a Matsubara
sum. It is the \emph{full} integral on the left of \eqref{ABsplit}, not (A)
itself, that has a definite analytic structure in $\eps'$, so one evaluates
that and removes (B) again afterwards. We write 
$\textrm{Re}\,g^R=\tfrac12(g^R-g^A)$ and close the $g^R$
term in the upper half plane of $\eps'$ and the $g^A$ term in the lower one.
The integral over the arcs drops out because the integrand falls as $1/\eps'^2$. The poles of
$t_{\eps'}$ at $\eps'=i\omega_m$, of residue $2T$ each, contribute the
Matsubara frequencies, the two closures supplying $m\ge0$ and $m<0$
respectively. The two half-sums join into a single sum because
$g^A(i\omega_m)=-[g^R(i|\omega_m|)]^*$ for $\omega_m<0$ is the odd
continuation of $g^R(i\omega_m)$, which $\tilde\omega_{-m-1}=-\tilde\omega_m$
and $\Phi_{-m-1}=\Phi_m$ of \eqref{EliEqsOpt} guarantee. The poles of
$D_{\eps-\eps'}^R$ at $\eps'=\eps\mp\Omega+i0$ lie in the upper half plane
and are picked up by the $g^R$ closure alone, with residue $-s$.
Hence
\beg\label{FullContour}
\int d\eps'D_{\eps-\eps'}^R\,\textrm{Re}(g_{\eps'}^R)\,t_{\eps'}
=2\pi iT\sum\limits_mg_mD^R_{\eps-i\omega_m}
-i\pi\sum\limits_{s=\pm}s\,g^R_{\eps-s\Omega}t_{\eps-s\Omega}.
\en

Subtracting (B) as prescribed by \eqref{ABsplit} replaces $g^R$ in the last
term by $g^R-\textrm{Re}\,g^R=i\,\textrm{Im}\,g^R$, and
$(-i\pi)\cdot i=\pi$, so that the result is
\beg\label{PVident}
{\rm (A)}=2\pi iT\sum\limits_m \frac{\tilde\omega_m D^R_{\eps-i\omega_m}}{\sqrt{\tilde\omega_m^2+\Phi_m^2}}
+\pi\sum\limits_{s=\pm}s\,t_{\eps-s\Omega}\,
\textrm{Im}(g_{\eps-s\Omega}^R),
\en
where $\tilde\omega_m=\omega_mZ_m$ and $\Phi_m=\Delta_mZ_m$ solve \eqref{EliEqsOpt} on the imaginary axis.
The principal-value integral is therefore \emph{not} the Matsubara sum by
itself. The leftover $\textrm{Im}\,g^R$ terms of \eqref{PVident} recombine
with (B), turning $\textrm{Re}\,g^R$ there into the full $g^R$, and the
$\coth(\Omega/2T)$ of \eqref{EliEqsSet1} then merges with the
$\tanh[(\eps\mp\Omega)/2T]$ of (B) by using
\beg\label{weights}
\frac{1}{2}\left[\coth\frac{\Omega}{2T}
+\tanh\frac{\eps-\Omega}{2T}\right]=n_B(\Omega)+n_F(\Omega-\eps),\qquad
\frac{1}{2}\left[\coth\frac{\Omega}{2T}
-\tanh\frac{\eps+\Omega}{2T}\right]=n_B(\Omega)+n_F(\Omega+\eps),
\en
(here $n_F$ and $n_B$ are the Fermi--Dirac and Bose--Einstein functions respectively).

The same steps apply to the integral of \eqref{EliEqsSet1}, which involves the anomalous function $f_\eps^R$, with one
difference that is easy to overlook: $g^R$ and $f^R$ do not continue onto the
imaginary axis in the same way. Setting $\eps=i\omega_m$ in \eqref{GRclosure}
gives $(\eps Z_\eps^R)^2-(\Phi_\eps^R)^2\to-(\tilde\omega_m^2+\Phi_m^2)$, so the
retarded branch of the square root is $i\sqrt{\tilde\omega_m^2+\Phi_m^2}$ and
\beg\label{ContinueGF}
g^R_{i\omega_m}=\frac{\tilde\omega_m}{\sqrt{\tilde\omega_m^2+\Phi_m^2}}
\equiv g_m,
\qquad
f^R_{i\omega_m}=\frac{-i\,\Phi_m}{\sqrt{\tilde\omega_m^2+\Phi_m^2}}
\equiv-i f_m.
\en
The normal amplitude continues to a real Matsubara function, the anomalous one
to $-i$ times a real function. With $\tilde\omega_{-m-1}=-\tilde\omega_m$ and
$\Phi_{-m-1}=\Phi_m$ the two half-sums then join in the anomalous channel
exactly as they do in the normal one. This single factor of $-i$ is what turns
the $i\pi T$ of the $Z$ equation into the $\pi T$ of the $\Delta$ equation of
\eqref{MSC}. It is also why the Matsubara term of the latter comes out real,
as it must: (A) is the principal value of a real integrand.

Nothing has been
approximated so far and \eqref{EliEqsSet1} becomes
\beg\label{MSC}
\begin{aligned}
\eps Z_\eps^R&=\eps+i\pi T\sum\limits_m\lambda(\eps-i\omega_m)
\frac{\omega_m}{\sqrt{\omega_m^2+\Delta_m^2}}
+\frac{i\pi\lambda\Omega}{2}\sum\limits_{s=\pm}
\left[n_B(\Omega)+n_F(\Omega+s\eps)\right]g_{\eps+s\Omega}^R,\\
\Delta_\eps^RZ_\eps^R&=\pi T\sum\limits_m\lambda(\eps-i\omega_m)
\frac{\Delta_m}{\sqrt{\omega_m^2+\Delta_m^2}}
+\frac{i\pi\lambda\Omega}{2}\sum\limits_{s=\pm}
\left[n_B(\Omega)+n_F(\Omega+s\eps)\right]f_{\eps+s\Omega}^R,
\end{aligned}
\en
{with the analytic interaction kernel}
{
\beg\label{GeneralAnalyticKernel}
\lambda(z)=2\int_0^\infty d\nu\,
\frac{\nu\alpha^2F(\nu)}{\nu^2-z^2}
\stackrel{\mathrm{Einstein}}{=}\frac{\lambda\Omega^2}{\Omega^2-z^2}.
\en
}
{Equation~\eqref{MSC} is written for the Einstein spectrum. To use a general spectrum, we apply Eq.~\eqref{SpectralReplacement} to the thermal factors and terms with shifted electronic frequencies. We compute the functions $\Delta_m$ and $Z_m$ from the Matsubara Eliashberg equations with the kernel~\eqref{GeneralMatsKernel},}
{
\beg\label{EliEqsOpt}
\begin{aligned}
\Sigma_n&=\pi T\sum\limits_m\lambda_{n-m}\frac{\omega_m+\Sigma_m}{\sqrt{(\omega_m+\Sigma_m)^2+|\Phi_m|^2}}, \\
{\Phi}_n&=\pi T\sum\limits_m\lambda_{n-m}\frac{{\Phi}_m}{\sqrt{(\omega_m+\Sigma_m)^2+|\Phi_m|^2}}.
\end{aligned}
\en
}
This is the mixed
representation of Marsiglio, Schossmann and Carbotte in Ref. \cite{Carbotte1988}. It is also worth noting that equations \eqref{MSC} match exactly the corresponding equations (4a) and (4b) of Marsiglio and Carbotte \cite{Marsiglio1991}. 

%
%
\section{Closure equations}\label{AppendixC}
In this Appendix we derive the explicit closure equations which will allow us
to compute the frequency of the amplitude Higgs mode. With
$\hat{\cal G}_\eps^{R(A)}=g_\eps^{R(A)}\hat\tau_3+if_\eps^{R(A)}\hat\tau_2$
we define the following functions:
\beg\label{ABFunctions}
\begin{aligned}
&{\cal A}^{R(A)}(\eps,\eps')=1+g_{\eps}^{R(A)}g_{\eps'}^{R(A)}+f_{\eps}^{R(A)}f_{\eps'}^{R(A)}, \quad \tilde{\cal A}^{R(A)}(\eps,\eps')=-1+g_{\eps}^{R(A)}g_{\eps'}^{R(A)}+f_{\eps}^{R(A)}f_{\eps'}^{R(A)}, \\
&{\cal A}^{K}(\eps,\eps')=1+g_{\eps}^{R}g_{\eps'}^{A}+f_{\eps}^{R}f_{\eps'}^{A}, \quad \tilde{\cal A}^{K}(\eps,\eps')=-1+g_{\eps}^{R}g_{\eps'}^A+f_{\eps}^{R}f_{\eps'}^{A}, \\
&{\cal B}^{R(A)}(\eps,\eps')=g_{\eps}^{R(A)}f_{\eps'}^{R(A)}+f_{\eps}^{R(A)}g_{\eps'}^{R(A)}, \quad {\cal B}^{K}(\eps,\eps')=g_{\eps}^{R}f_{\eps'}^{A}+f_{\eps}^{R}g_{\eps'}^{A}.
\end{aligned}
\en
 
To obtain the remaining closure equations for the functions
$a_\eps^{R(A)}$ and $b_\eps^{R(A)}$ we introduce the auxiliary functions
\beg\label{dRKRA}
\begin{aligned}
\hat{\cal R}_\eps^K(\omega)&=\delta\hat{\Sigma}_\eps^R(\omega)\hat{\tau}_3\hat{\cal G}_{\eps-\frac{\omega}{2}}^K+\delta\hat{\Sigma}_\eps^K(\omega)\hat{\tau}_3\hat{\cal G}_{\eps-\frac{\omega}{2}}^A-\hat{\cal G}_{\eps+\frac{\omega}{2}}^K\delta\hat{\Sigma}_{\eps}^A(\omega)\hat{\tau}_3-\hat{\cal G}_{\eps+\frac{\omega}{2}}^R\delta\hat{\Sigma}_{\eps}^K(\omega)\hat{\tau}_3, \\
\hat{\cal R}_\eps^{R(A)}(\omega)&=\delta\hat{\Sigma}_\eps^{R(A)}(\omega)\hat{\tau}_3\hat{\cal G}_{\eps-\frac{\omega}{2}}^{R(A)}-\hat{\cal G}_{\eps+\frac{\omega}{2}}^{R(A)}\delta\hat{\Sigma}_{\eps}^{R(A)}(\omega)\hat{\tau}_3.
\end{aligned}
\en
 
\paragraph{Local form of $\delta\hat\Sigma_\eps^K$.}
We resume the argument of the main text. In the first term of
\eqref{dsigmaKFin} the equilibrium Keldysh propagator is
$D_{\eps'}^{K}=\bigl(D_{\eps'}^{R}-D_{\eps'}^{A}\bigr)\coth(\eps'/2T)
=-2\pi i\bigl[\delta(\eps'-\Omega)-\delta(\eps'+\Omega)\bigr]
\coth(\eps'/2T)$, so that the integral collapses onto $\eps'=\pm\Omega$. In
the second term only the products $\delta\hat g^{R}D^{R}$ and
$\delta\hat g^{A}D^{A}$ survive, for the reason stated after
\eqref{dsigmaKFin}: these combinations are analytic in the upper or the lower half plane, so that the corresponding integrals vanish by Cauchy's theorem.  Shifting the integration variable to $\eps''=\eps-\eps'$
and closing each contour in the half plane where the corresponding
$\delta\hat g$ is analytic --- upward for $\delta\hat g^{R}$, downward for
$\delta\hat g^{A}$ --- one encloses the phonon poles of $D^{R(A)}$, which sit
at $\eps''=\eps+s\Omega$ ($s=\pm$) and carry residue $s$. Collecting the two
contributions,
\beg\label{dsigmaKSimp}
\delta\hat{\Sigma}_{\eps}^{K}(\omega)\hat{\tau}_3
=\frac{i\pi\lambda\Omega}{4}\sum\limits_{s=\pm}
\Bigl\{s\bigl[\delta\hat{g}_{\eps+s\Omega}^{R}(\omega)
-\delta\hat{g}_{\eps+s\Omega}^{A}(\omega)\bigr]
-\coth\Bigl(\frac{\Omega}{2T}\Bigr)\,
\delta\hat{g}_{\eps+s\Omega}^{K}(\omega)\Bigr\}.
\en
The kernel is purely local: it involves $\delta\hat g_{\eps'}^{X}$ only at
$\eps'=\eps\pm\Omega$, and not at a continuum of $\eps'$.
 
\paragraph{Reduction of the traces.}
The four functions $a_\eps^{R,A}$, $b_\eps^{R,A}$ and $a_\eps^K$, $b_\eps^K$
of \eqref{dSigmaRAK} satisfy a linear system that can be written in a single
form. Throughout, $\eps_\pm=\eps\pm\omega/2$ and $t_\eps=\tanh(\eps/2T)$. It is convenient to evaluate \eqref{ABFunctions} at the pair of arguments
$(\eps_+,\eps_-)$ and to abbreviate
\beg\label{ABsub}
{\cal A}_\eps^{X}={\cal A}^{X}(\eps_+,\eps_-),\qquad
\tilde{\cal A}_\eps^{X}=\tilde{\cal A}^{X}(\eps_+,\eps_-),\qquad
{\cal B}_\eps^{X}={\cal B}^{X}(\eps_+,\eps_-),
\en
so that the superscript $K$ carries the retarded function at $\eps_+$ together
with the advanced one at $\eps_-$. All the algebra below can then be carried
out by the single $2\times2$ matrix
\beg\label{Mmatrix}
\mathbb{M}_\eps^{X}=\frac{1}{{\cal D}_\eps^{X}}
\begin{pmatrix}
\tilde{\cal A}_\eps^{X} & -{\cal B}_\eps^{X}\\[2pt]
{\cal B}_\eps^{X} & -{\cal A}_\eps^{X}
\end{pmatrix}, \quad {\cal D}_\eps^{R(A)}=\eta_{\eps_+}^{R(A)}+\eta_{\eps_-}^{R(A)}
\en
with $X=R,A,K$ and ${\cal D}_\eps^{K}=\eta_{\eps_+}^{R}+\eta_{\eps_-}^{A}$.
 
The reason a two-component ansatz such as \eqref{dSigmaRAK} closes is the
following identity. For any $a$, $b$ and any pair of equilibrium propagators
of the form \eqref{GRclosure},
\beg\label{Bilinear}
\hat{\cal G}_{\eps_+}^{X_1}\bigl(a\hat\tau_3+ib\hat\tau_2\bigr)
\hat{\cal G}_{\eps_-}^{X_2}-\bigl(a\hat\tau_3+ib\hat\tau_2\bigr)
=\bigl(\tilde{\cal A}_\eps^{X}a-{\cal B}_\eps^{X}b\bigr)\hat\tau_3
+i\bigl({\cal B}_\eps^{X}a-{\cal A}_\eps^{X}b\bigr)\hat\tau_2 ,
\en
with $(X_1,X_2)=(R,R)$, $(A,A)$, $(R,A)$ for $X=R$, $A$, $K$: the components
along $\hat\tau_0$ and $\hat\tau_1$ cancel identically, so the right-hand side
never leaves the two-dimensional space spanned by $\hat\tau_3$ and
$i\hat\tau_2$. Writing $\mathbb{T}[\hat X_\eps]$ for the column
$\bigl(\tfrac12\mathrm{Tr}[\hat X_\eps\hat\tau_3],\,
\tfrac1{2i}\mathrm{Tr}[\hat X_\eps\hat\tau_2]\bigr)^{\!T}$, Eq.~\eqref{Eq4dgRA}
is precisely \eqref{Bilinear} divided by ${\cal D}_\eps^{R(A)}$, so that
\beg\label{TraceRA}
\mathbb{T}\bigl[\delta\hat g_\eps^{R(A)}\bigr]
=\mathbb{M}_\eps^{R(A)}\,\psi_\eps^{R(A)} .
\en
 
The anomalous piece reduces to the same bilinear. Substituting
$\hat{\cal G}_\eps^{K}=(\hat{\cal G}_\eps^{R}-\hat{\cal G}_\eps^{A})t_\eps$
from \eqref{GKParametrization} into \eqref{dRKRA}, the combination that
appears on the right-hand side of \eqref{FinalEq4dga} gives
\beg\label{Rcollapse}
\hat{\cal R}_\eps^{K}-\hat{\cal R}_\eps^{R}t_{\eps_-}
+\hat{\cal R}_\eps^{A}t_{\eps_+}
=\delta\hat{\tilde\Sigma}_\eps\hat\tau_3\hat{\cal G}_{\eps_-}^{A}
-\hat{\cal G}_{\eps_+}^{R}\delta\hat{\tilde\Sigma}_\eps\hat\tau_3,
\qquad
\delta\hat{\tilde\Sigma}_\eps\hat\tau_3
=\bigl(\delta\hat\Sigma_\eps^{K}+t_{\eps_+}\delta\hat\Sigma_\eps^{A}
-t_{\eps_-}\delta\hat\Sigma_\eps^{R}\bigr)\hat\tau_3 ,
\en
every term containing $\hat{\cal G}^{R}_{\eps_-}$, $\hat{\cal G}^{A}_{\eps_+}$
cancelling against a partner. Multiplying \eqref{Rcollapse} by
$\hat{\cal G}_{\eps_+}^{R}$ as required by \eqref{dgaFin} and using
$(\hat{\cal G}^{R})^2=\hat 1$ returns the left-hand side of \eqref{Bilinear}
with $(X_1,X_2)=(R,A)$ and with $a,b$ the components of
$\delta\hat{\tilde\Sigma}_\eps\hat\tau_3$. Hence
\beg\label{Tracea}
\mathbb{T}\bigl[\delta\hat g_\eps^{a}\bigr]=\mathbb{M}_\eps^{K}\zeta_\eps
\en
where the function $\zeta_\eps$ is defined in the main text.
Equations \eqref{TraceRA}, \eqref{Tracea} and \eqref{TraceK} are all that is
needed to convert the matrix relations of this Appendix into scalar equations.
 
\paragraph{Keldysh block.}
Taking $\mathbb{T}$ of \eqref{dsigmaKSimp} and using \eqref{TraceK} gives
equation \eqref{ClosureK} of the main text, whose two components are the
equations for $a_\eps^{K}$ and $b_\eps^{K}$. Written entrywise they read
\beg\label{Eqs4aepsKbepsK}
{\begin{aligned}
a_{\eps}^K=\sum\limits_{s=\pm}&\left\{P_{\eps+s\Omega}^{KR}a_{\eps+s\Omega}^R+P_{\eps+s\Omega}^{KA}a_{\eps+s\Omega}^A+
P_{\eps+s\Omega}^{KK}a_{\eps+s\Omega}^K+
M_{\eps+s\Omega}^{KR}b_{\eps+s\Omega}^R+M_{\eps+s\Omega}^{KA}b_{\eps+s\Omega}^A+M_{\eps+s\Omega}^{KK}b_{\eps+s\Omega}^K\right\}, \\
b_{\eps}^K=\sum\limits_{s=\pm}&\left\{\overline{M}_{\eps+s\Omega}^{KR}a_{\eps+s\Omega}^R+\overline{M}_{\eps+s\Omega}^{KA}a_{\eps+s\Omega}^A+
\overline{M}_{\eps+s\Omega}^{KK}a_{\eps+s\Omega}^K+
Q_{\eps+s\Omega}^{KR}b_{\eps+s\Omega}^R+Q_{\eps+s\Omega}^{KA}b_{\eps+s\Omega}^A+Q_{\eps+s\Omega}^{KK}b_{\eps+s\Omega}^K\right\}.
\end{aligned}}
\en
The first three of the twelve coefficient functions collect the
terms proportional to $a^{R(A)}$ in the equation for $a^{K}$:
\beg\label{Ps}
\begin{aligned}
&P_\eps^{KR}=\frac{i\pi\lambda\Omega}{4}
\left\{\frac{\tilde{\cal A}_\eps^{R}}{{\cal D}_\eps^{R}}
\left[s-t_{\eps_-}\coth\Bigl(\frac{\Omega}{2T}\Bigr)\right]
+\frac{\tilde{\cal A}_\eps^{K}}{{\cal D}_\eps^{K}}\,
t_{\eps_-}\coth\Bigl(\frac{\Omega}{2T}\Bigr)\right\},\\
&P_\eps^{KA}=-\frac{i\pi\lambda\Omega}{4}
\left\{\frac{\tilde{\cal A}_\eps^{A}}{{\cal D}_\eps^{A}}
\left[s-t_{\eps_+}\coth\Bigl(\frac{\Omega}{2T}\Bigr)\right]
+\frac{\tilde{\cal A}_\eps^{K}}{{\cal D}_\eps^{K}}\,
t_{\eps_+}\coth\Bigl(\frac{\Omega}{2T}\Bigr)\right\},
\quad P_\eps^{KK}=-\frac{i\pi\lambda\Omega}{4}\,
\frac{\tilde{\cal A}_\eps^{K}}{{\cal D}_\eps^{K}}
\coth\Bigl(\frac{\Omega}{2T}\Bigr).
\end{aligned}
\en
The remaining nine follow from \eqref{Ps} without further work:
$\tilde{\cal A}^{X}$ occupies a single entry of $\mathbb{M}^{X}$ in
\eqref{Mmatrix}, and the other entries are $-{\cal A}^{X}$ and
$\pm{\cal B}^{X}$. Hence
\beg\label{QsMs}
Q_\eps^{KY}=P_\eps^{KY}\Big|_{\tilde{\cal A}^{X}\to-{\cal A}^{X}},
\qquad
M_\eps^{KY}=P_\eps^{KY}\Big|_{\tilde{\cal A}^{X}\to-{\cal B}^{X}},
\qquad
\overline M_\eps^{KY}=-M_\eps^{KY},
\qquad Y=R,A,K,
\en
the last relation holding because the off-diagonal entries of $\mathbb{M}^{X}$
are equal and opposite. Two points about the notation of
\eqref{Eqs4aepsKbepsK} should be kept in mind. First, all twelve functions
depend on $s$ as well as on their subscript: the parameter $s$ appears
explicitly in the square brackets of \eqref{Ps}, so
$P_{\eps+s\Omega}^{KR}$ is not a function of $\eps+s\Omega$ alone. Second,
$\eps_\pm$ inside \eqref{Ps} refers to the subscript of the function itself,
i.e.\ $\eps_\pm=\eps\pm\omega/2$ becomes $\eps+s\Omega\pm\omega/2$ when the
function is evaluated at $\eps+s\Omega$.
 
\paragraph{Retarded and advanced blocks.}
These are the blocks that are not local. In \eqref{ClosureRA} the $D^{K}$ term
is again local, by the argument that led to \eqref{dsigmaKSimp}. In the second
integral we insert the ansatz \eqref{ansatz4gK}, which splits
$\delta\hat g_{\eps'}^{K}$ into three pieces, and treat each separately. Take
$X=R$ first. The piece $\delta\hat g_{\eps'}^{R}t_{\eps'-\omega/2}$ is
integrated by closing upward, where $\delta\hat g^{R}$ is analytic; the
enclosed singularities are the poles of $D_{\eps-\eps'}^{R}$ at
$\eps'=\eps+s\Omega$, with residue $s$, and the poles of
$t_{\eps'-\omega/2}$ at $\eps'=z_n$, with residue $2T$. The piece
$-t_{\eps'+\omega/2}\delta\hat g_{\eps'}^{A}$ is integrated by closing
downward, where $\delta\hat g^{A}$ is analytic and where $D^{R}$ has no poles;
only $t_{\eps'+\omega/2}$ contributes, at $\eps'=-z_n$. The third
piece is left as it stands. For $X=A$ the two closures exchange roles, the
phonon poles of $D^{A}$ now lying inside the downward contour. This is the
origin of the only asymmetry between the two equations: the local term carries
$t_{\eps+s\Omega-\omega/2}$ in the retarded equation and
$t_{\eps+s\Omega+\omega/2}$ in the advanced one. Collecting, for $X=R$
(upper sign) and $X=A$ (lower sign),
\beg\label{ClosureSigmaRA}
\begin{aligned}
\delta\hat\Sigma_{\eps}^{X}\hat\tau_3
&=\frac{i\pi\lambda\Omega}{4}\sum_{s=\pm}
\delta\hat g_{\eps+s\Omega}^{X}
\Bigl[s\,t_{\eps+s\Omega\mp\omega/2}
-\coth\Bigl(\frac{\Omega}{2T}\Bigr)\Bigr]\\
&+\frac{i\pi\lambda\Omega T}{2}\sum_{n\ge0}
\Bigl[\delta\hat g_{z_n}^{R}D_{\eps-z_n}^{X}
+\delta\hat g_{-z_n}^{A}D_{\eps+z_n}^{X}\Bigr]
+\frac{\lambda\Omega}{8}\int d\eps'\,
D_{\eps-\eps'}^{X}\,\delta\hat g_{\eps'}^{a},
\end{aligned}
\en
with $z_n=i\omega_n+\omega/2$, $\omega_n=\pi T(2n+1)$. The two Matsubara sums
still refer to $\delta\hat g^{R,A}$ off the real axis. They are returned to it
by the Cauchy representation, legitimate because $\delta\hat g^{R}$ is
analytic in the upper half plane of complex $\eps$ and $\delta\hat g^{A}$ in
the lower one:
\beg\label{UseCauchy}
\delta\hat{g}_{z_n}^{R}
=\frac{1}{2\pi i}\int\limits_{-\infty}^{\infty}
\frac{\delta\hat{g}_{\eps'}^{R}\,d\eps'}{\eps'-z_n},
\qquad
\delta\hat{g}_{-z_n}^{A}
=-\frac{1}{2\pi i}\int\limits_{-\infty}^{\infty}
\frac{\delta\hat{g}_{\eps'}^{A}\,d\eps'}{\eps'+z_n}.
\en
Introducing
\beg\label{vfun}
v_{\pm}^{X}(\eps,\eps')=T\sum_{n\ge0}\frac{D_{\eps\pm z_n}^{X}}{\eps'\pm z_n},
\qquad X=R,A,
\en
which converge as $n^{-3}$ and are computed once and for all, the sums become
integrals along the real axis and \eqref{ClosureSigmaRA} closes on itself.
Taking $\mathbb{T}$ and using \eqref{TraceRA}--\eqref{Tracea} we find equations \eqref{ClosureRAfinal} in the main text.
Together with
\eqref{ClosureK} this is a closed, homogeneous linear system for the six
functions $a_\eps^{R,A,K}$, $b_\eps^{R,A,K}$; the amplitude mode is the
complex $\omega$ at which it admits a nontrivial solution.
 
Lastly, we note that the first two terms of \eqref{ClosureRAfinal} involve $\delta\hat g^{R}$ and
$\delta\hat g^{A}$, each analytic in one half plane, and their contours have
already been closed. The last term cannot be treated the same way: through
$\mathbb{M}^{K}$ it carries both $\eta_{\eps'_+}^{R}$ and $\eta_{\eps'_-}^{A}$
in the denominator and so is analytic in neither half plane. For the
quadrature it is convenient to separate the two. With
$\tilde\eps_+=\eps_+Z_{\eps_+}^{R}$, $\Phi_+=\Phi_{\eps_+}^{R}$,
$\tilde\eps_-=\eps_-Z_{\eps_-}^{A}$, $\Phi_-=\Phi_{\eps_-}^{A}$ and
\beg\label{CompactNotations}
\rho_{+}^{R}=\tilde\eps_+^2-\Phi_+^2=\bigl(\eta_{\eps_+}^{R}\bigr)^2,
\qquad
\rho_{-}^{A}=\tilde\eps_-^2-\Phi_-^2=\bigl(\eta_{\eps_-}^{A}\bigr)^2,
\qquad
S^{K}=\tilde\eps_+\tilde\eps_-+\Phi_+\Phi_-,
\qquad
U^{K}=\tilde\eps_+\Phi_-+\tilde\eps_-\Phi_+,
\en
the entries of $\mathbb{M}_\eps^{K}$ can be rearranged so that each carries a
single inverse square root, and \eqref{Tracea} becomes
\beg\label{PQsplit}
\mathbb{M}_{\eps}^{K}\chi_{\eps}
=\frac{1}{\eta_{\eps_+}^{R}}
\begin{pmatrix}{\cal P}_\eps^{R}\\[2pt] {\cal Q}_\eps^{R}\end{pmatrix}
+\frac{1}{\eta_{\eps_-}^{A}}
\begin{pmatrix}{\cal P}_\eps^{A}\\[2pt] {\cal Q}_\eps^{A}\end{pmatrix},
\en
or, equivalently,
\beg\label{CRAz}
\delta\hat{g}_\eps^{a}
=\Bigl(\frac{{\cal P}_\eps^{R}}{\eta_{\eps_+}^{R}}
+\frac{{\cal P}_\eps^{A}}{\eta_{\eps_-}^{A}}\Bigr)\hat{\tau}_3
+\Bigl(\frac{{\cal Q}_\eps^{R}}{\eta_{\eps_+}^{R}}
+\frac{{\cal Q}_\eps^{A}}{\eta_{\eps_-}^{A}}\Bigr)i\hat{\tau}_2,
\en
where, writing $\chi_\eps=(\alpha_\eps,\beta_\eps)^{T}$,
\beg\label{DefinePRAQRA}
\begin{aligned}
&{\cal P}_{\eps}^{R}=\frac{U^{K}\beta_\eps-(\rho_{+}^{R}+S^{K})\alpha_\eps}
{\rho_{+}^{R}-\rho_{-}^{A}},
\qquad
{\cal P}_{\eps}^{A}=\frac{(S^{K}+\rho_{-}^{A})\alpha_\eps-U^{K}\beta_\eps}
{\rho_{+}^{R}-\rho_{-}^{A}},\\
&{\cal Q}_{\eps}^{R}=\frac{(S^{K}-\rho_{+}^{R})\beta_\eps-U^{K}\alpha_\eps}
{\rho_{+}^{R}-\rho_{-}^{A}},
\qquad
{\cal Q}_{\eps}^{A}=\frac{(\rho_{-}^{A}-S^{K})\beta_\eps+U^{K}\alpha_\eps}
{\rho_{+}^{R}-\rho_{-}^{A}}.
\end{aligned}
\en
The two inverse square roots are thereby exposed explicitly. Away from
$\eps=0$ the remaining factors are smooth: at a gap edge only one of
$\eta^{R}_{\eps_+}$, $\eta^{A}_{\eps_-}$ vanishes, so $\rho_+^{R}-\rho_-^{A}$
does not. At $\eps=0$ it does vanish, for every $\omega$, because
$\tilde\eps$ is odd and $\Phi$ even under $\eps\to-\eps$ up to complex
conjugation, so that $\rho^{A}_{-x}=\rho^{R}_{x}$. There
$\eta^{R}_{\eps_+}=\eta^{A}_{\eps_-}$ as well and the two poles cancel
between the terms of \eqref{PQsplit},
${\cal P}_\eps^{R}+{\cal P}_\eps^{A}=-\alpha_\eps$ and
${\cal Q}_\eps^{R}+{\cal Q}_\eps^{A}=-\beta_\eps$: the product
$\mathbb{M}_{\eps}^{K}\chi_{\eps}$ is regular there and must be evaluated
from that sum rather than term by term.

\section{Reduction to the Matsubara representation}\label{AppendixD}
In this Appendix we show that the driven system \eqref{DrivenSystem},
evaluated at the bosonic Matsubara frequencies $\omega=i\nu_l$, collapses
exactly onto the imaginary-axis system \eqref{MatsSystem}, so that the bare
anomalous amplitude of the driven real-axis system analytically continues
to the pair propagator $\chi_\Phi(i\nu_l)$ of \eqref{chiMats}. This provides a nontrivial check on the entire
real-axis construction: every kernel, prefactor and contour of
Eqs.~\eqref{ClosureK}--\eqref{ClosureRAfinal} enters the reduction.

Write
\beg\label{MRedLattice}
\bar\omega_m=\omega_m+\frac{\nu_l}{2}=\pi T(2m+1+l),
\en
so that at $\eps=i\bar\omega_m$ the two legs sit at
$\eps_\pm=i\omega_{m+l}$ and $i\omega_m$. Since $\delta\hat\Sigma^{R}$ is
analytic in the upper half plane, the retarded equation may be evaluated
at every $\bar\omega_m>0$ by redoing the contour closures of
Appendix~\ref{AppendixC} at fixed $\omega=i\nu_l$; the advanced equation
covers $\bar\omega_m<0$. Three simplifications occur:
\begin{itemize}
\item[(i)]{At $\eps=i\bar\omega_m$ the poles of
$D^{R}_{\eps-\eps'}$ at $\eps'=\eps+s\Omega$ both lie in the upper half
plane and are enclosed by the retarded closure. Their contribution
carries $\bar t_{i\bar\omega_m+s\Omega}$, and since
$\bar t_\eps\equiv t_{\eps\pm\omega/2}$, the two signs coinciding at
$\omega=i\nu_l$, this is
$t_{i\omega_m+s\Omega}=s\coth(\Omega/2T)$. The bracket
$[s\,\bar t_{\eps+s\Omega}-\coth(\Omega/2T)]$ of
\eqref{ClosureRAfinal} therefore vanishes identically, so that no local
term survives; the Matsubara system \eqref{MatsSystem} has none.}

\item[(ii)]{At $\omega=i\nu_l$ the poles of $t_{\eps'-\omega/2}$ sit at
$\eps'=i\pi T(2n+1+l)$ and are enclosed by the upward (retarded) closure
when $2n+1+l>0$; the poles of $t_{\eps'+\omega/2}$ sit at
$\eps'=i\pi T(2n+1-l)$ and are enclosed by the downward (advanced)
closure when $2n+1-l<0$. In terms of the leg index the two families
cover $m>-\tfrac{l+1}{2}$ and $m<-\tfrac{l+1}{2}$: together they cover
every fermionic leg exactly once (for odd $l$ the boundary term
$m=-\tfrac{l+1}{2}$ sits on the contour and is supplied in equal halves
by the two closures). Note that this enclosure count differs from the
real-frequency one --- as $\omega$ moves up the imaginary axis, thermal
poles cross the real axis --- so the representation \eqref{vfun}, whose
sums start at $n=0$, must \emph{not} be evaluated verbatim at
$\omega=i\nu_l$: the migrated poles must be included. Each enclosed pole
carries the residue $2T$ and the phonon factor
\beg\label{MRedDval}
D^{R}_{i(\omega_m-\omega_n)}
=-\frac{2}{\lambda\Omega}\,\lambda_{m-n},
\en
turning the prefactor $i\pi\lambda\Omega T/2$ of the Matsubara sums in
Appendix~\ref{AppendixC} into $-i\pi T\lambda_{m-n}$, with
$\lambda_{m-n}$ the kernel of \eqref{eqMats}.}

\item[(iii)]{On the
imaginary axis the equilibrium functions continue according to
\eqref{ContinueGF}, $g^{R}_{i\omega_k}=g_k$ and $f^{R}_{i\omega_k}=-if_k$,
whence $\eta^{R}_{i\omega_k}=i\omega_kZ_k/g^{R}_{i\omega_k}=iD_k$; the advanced
functions reach the same values from below,
$g^{A}_{i\omega_k}=g_k$, $f^{A}_{i\omega_k}=-if_k$ and
$\eta^{A}_{i\omega_k}=iD_k$ for $\omega_k<0$. Substitute into
\eqref{Mmatrix} with $(\eps_+,\eps_-)=(i\omega_{m+l},i\omega_m)$ and write
the components of $\psi$ as
\beg\label{MRedDict}
a_{i\bar\omega_m}=\delta W_m,
\qquad
b_{i\bar\omega_m}=-i\,\delta\Phi_m,
\en
carrying the same factor $-i$ that accompanies the anomalous function in
\eqref{ContinueGF}. The matrix products then collapse onto the kernels
\eqref{PQR}:
\beg\label{MRedRows}
-i\pi T\lambda_{m-n}\,
\mathbb{M}_{i\bar\omega_n}\psi_{i\bar\omega_n}
=\pi T\lambda_{m-n}
\begin{pmatrix}
P_n\,\delta W_n-Q_n\,\delta\Phi_n\\[2pt]
-i\bigl(R_n\,\delta\Phi_n-Q_n\,\delta W_n\bigr)
\end{pmatrix}.
\en}
\end{itemize}
Collecting these, the retarded and advanced equations at
$\eps=i\bar\omega_m$ read componentwise
\beg\label{MRedFinal}
\begin{aligned}
\delta W_m&=\pi T\sum_{n}\lambda_{m-n}
\bigl(P_n\,\delta W_n-Q_n\,\delta\Phi_n\bigr),\\
\delta\Phi_m&=ih+\pi T\sum_{n}\lambda_{m-n}
\bigl(R_n\,\delta\Phi_n-Q_n\,\delta W_n\bigr),
\end{aligned}
\en
which is the system \eqref{MatsSystem} with source $ih$: the source of
\eqref{SourceReal} enters the $b$ component as $+h$, which \eqref{MRedDict}
converts to $ih$. Since the system is linear, the
two factors of $i$ cancel in the response,
\beg\label{MRedChi}
\frac{b^{R}_{\,i\bar\omega_0}(i\nu_l)}{h}
=\frac{\delta\Phi^{(l)}_{n=0}}{h}
=\chi_\Phi(i\nu_l)
\en
of Eq.~\eqref{chiMats}, exactly and for every $l$ and every coupling.
Equation \eqref{MRedChi} equates the \emph{bare} anomalous
amplitude with the Matsubara propagator. The gap-edge projection
\eqref{MyChi} shares its singularities but not its residue, so it is
\eqref{MRedChi}, and not \eqref{MyChi}, that must be used when the
real-axis and imaginary-axis results are compared.

\section{Details of the numerical calculations}\label{AppendixE}
In this Appendix we outline the discretization used to solve the driven
system \eqref{DrivenSystem} built from Eqs.~\eqref{ClosureK}--\eqref{SourceReal}
at intermediate coupling, $\Delta_g<\Omega$, and the controls that separate
the binding energy from the resolution of the discretization. The
strong-coupling limit is treated by a different method in
Appendix~\ref{AppendixF}. {The discretization described here is for the Einstein spectrum~\eqref{alpha2Fw}. A general Eliashberg function requires the additional phonon-frequency integration in Eq.~\eqref{SpectralReplacement}.}

\emph{Equilibrium input.} The retarded functions $Z_\eps^R$ and $\Phi_\eps^R$
that enter $\mathbb{M}_\eps^{X}$ through \eqref{Mmatrix} are obtained from the
real-axis Eliashberg equations \eqref{EliEqsSet1} in the mixed
representation of Ref.~\cite{Carbotte1988}. The Matsubara sums are converged
first; the real-axis functions then follow by iterating the mixed
equations, in which those sums enter as known inhomogeneous terms,
so that no numerical analytic continuation is involved. The advanced
functions follow by complex conjugation. Every retarded argument that
enters the response carries a small imaginary part, $\eps\to\eps+i\gamma$,
which regularizes the inverse square-root singularities of
$\eta_\eps^{R}$ at the gap edge; the role of $\gamma$ is discussed below.

\emph{Gap edge.} The threshold against which the binding energy is measured is
the root $\Delta_g$ of $\mathrm{Re}\Delta(\eps)=\eps$, with
$\Delta(\eps)=\Phi^R_\eps/Z^R_\eps$ from the mixed representation, which
gives $\Delta(\eps)$ pointwise at any real $\eps$ and not only on the grid
\eqref{AppE-grid}. The interval $[10^{-5},\,3g]$, $g=\sqrt\lambda\,\Omega$,
is scanned in $200$ subintervals (a step of $0.027\,\Omega$ at
$\lambda=3.25$, about nine cells of the $K_s=336$ grid). The subinterval on which
$\mathrm{Re}\Delta(\eps)-\eps$ changes sign is passed to a Brent-type root
finder with a relative tolerance of $10^{-12}$, and the root is polished by
a secant step to a logged residual of $10^{-15}$. The crossing is regular,
$\mathrm{Im}\Delta(\eps)$
being exponentially small in $\Omega/T$ below the phonon-emission
threshold $\Delta_g+\Omega$ and the slope $1-\partial_\eps\Delta$ of order
$0.9$, so that the root is well conditioned. The real-axis functions in
the root condition are evaluated by the same routine as those entering
the operator $\mathbb K(\omega)$, with the regulator $\gamma$ at whatever
value the driver has set. The regulator moves the root by
$O(\gamma^2\partial_\eps^2\Delta)$, a relative amount of order $10^{-5}$ at
$K_s=336$, so the distinction between the root at $\gamma=0$ and at
$\gamma=3h/2$ is immaterial here.

What matters is not the absolute precision but the consistency between the
$\Delta_g$ of the equilibrium solution that enters $\mathbb K(\omega)$ and the
$2\Delta_g$ against which the fitted $\omega_{\rm H}$ is measured. Since
$\delta=1-\omega_{\rm H}/2\Delta_g$ is linear in $2\Delta_g$, a relative
mismatch $r$ between the two is an absolute error $r$ in $\delta$, i.e.\
a relative error $r/\delta\simeq100\,r$ at the percent level of binding
considered here. A mismatch of $0.25\%$ moves $\delta$ by $0.0025$, a
quarter of its value. Two facts fix the size of this systematic. First,
the binding is a property of the response relative to its own threshold.
Two runs of the equilibrium code at $\lambda=2$ have gap edges $0.19\%$
apart ($2\Delta_g=1.11885$ and $1.12097\,\Omega$),
and the responses on $K_s=336$ are shifted rigidly by the same
$0.0021\,\Omega$, with bindings relative to their own edges of
$\delta=0.0079$ and $0.0078$ on that grid, before extrapolation. A
change of the equilibrium solution at this level does not propagate to
the binding. Second, a mismatch does propagate in full: the raw output at
$\lambda=3.0$, $T/\Omega=0.005$, was at one stage normalized with the
$T/\Omega=0.02$ edge, $2\Delta_g=1.66667\,\Omega$ instead of
$1.67085\,\Omega$, a $0.25\%$ mismatch on both grids that moved the
extrapolated $\delta$ from $0.0112$ to $0.0087$; the tabulated value uses
the solution's own edge. All entries of Tables~\ref{tab:controls} and
\ref{tab:truncated} are normalized with the $2\Delta_g$ of the
equilibrium solution of the run in question, which is recorded with each
scan; the three grids of a given coupling carry the same edge to
$10^{-5}\,\Omega$.

\emph{Energy grid.} All six unknown functions $a_\eps^{X}$, $b_\eps^{X}$
($X=R,A,K$) are represented by their values on the uniform grid
\beg\label{AppE-grid}
\eps_j=-L+(j-1)h,\qquad h=\frac{\Omega}{K_s},\qquad L=N_h\Omega,
\en
$j=1,\dots,N_1$ with $N_1=1+2N_hK_s$, where $N_h$ and $K_s$ are integers.
{For the Einstein spectrum, this choice of grid uses the fact that there is only one phonon frequency:} the local
terms of \eqref{ClosureK} and \eqref{ClosureRAfinal} couple $\eps$ to
$\eps\pm\Omega$, and with \eqref{AppE-grid} this is an exact index shift
$j\to j\pm K_s$, so these terms are assembled without interpolation. For the
same reason the poles of $D^{R(A)}_{\eps-\eps'}$, Eq.~\eqref{DKRA}, always
fall on grid nodes. Between the nodes the unknowns are interpolated
linearly. The unknowns are collected into a vector of length $6N_1$; we use
$\psi^R$, $\psi^A$ and $\zeta$ as the independent blocks, $\psi^K$ being
recovered from the definition of $\zeta_\eps$ below \eqref{TraceK}. The
operator $\mathbb{K}(\omega)$ of \eqref{DrivenSystem} is then a dense complex
matrix of that dimension, assembled once and factorized by LU decomposition
for each external frequency $\omega$.

\emph{Nonlocal terms.} The last term of \eqref{ClosureRAfinal} is a
convolution with the {physical Einstein phonon propagator}, for which we use the exact
split
\beg\label{AppE-split}
\int d\eps' D^{R}_{\eps-\eps'}F(\eps')=\dashint d\eps'\,
\frac{2\Omega\,F(\eps')}{(\eps-\eps')^2-\Omega^2}
+i\pi\bigl[F(\eps+\Omega)-F(\eps-\Omega)\bigr],
\en
and the same expression with the opposite sign of the residue term for
$D^A$. The residue part is again an index shift. The principal value is
evaluated by an 8-point Gauss--Legendre rule on each cell of the grid, with
the singular contribution of the two cells adjacent to each pole subtracted
analytically. Without this subtraction the rows whose pole lies next to one
of the non-analytic points discussed next are in error by tens of percent
and do not improve with $h\to0$. The kernels $v_\pm^{X}(\eps,\eps')$ of
\eqref{vfun} are Matsubara sums which converge as $n^{-3}$; they are
truncated at $N_v\sim10^3$ terms and, since each term is a product of a
function of $\eps$ and a function of $\eps'$, the quadrature over $\eps'$ is
performed once for all rows of the matrix.

\emph{Non-analytic points.} The matrices $\mathbb{M}_{\eps'}^{X}$ are
non-smooth wherever one of the arguments $\eps'\pm\omega/2$ hits a phonon
replica of the gap edge, $\pm(\Delta_g+n\Omega)$, $n=0,1,\dots$; the
denominators ${\cal D}^{X}_{\eps'}$ carry inverse square-root cusps there.
These points are inserted as additional cell boundaries in the quadrature,
and on the cells that touch them the integration variable is mapped as
$\eps'=\eps'_0+(\eps'_1-\eps'_0)u^2$, which removes the $|\eps'-\eps'_0|^{-1/2}$
behavior and restores the accuracy of the Gauss rule. The replica ladder
is followed up to $n=N_h+1$, i.e.\ to the edge of the window. For
$\Delta_g<\Omega$ the replicas closest to the mode region,
$\pm(\Delta_g+\Omega)\mp\omega/2$, are separated from the pair-breaking
threshold by more than $\Omega-\Delta_g$, which is the property that makes a
single discretization adequate across the whole intermediate window.
Rows with $|\eps|>L-\Omega$ are not trustworthy because one of the poles in
\eqref{AppE-split} then leaves the integration range; $N_h$ is chosen so that
$L-\Omega\geq3\Delta_g$, which is satisfied with $N_h=12$ for all couplings
considered.

\emph{Regulator.} The quasiparticle broadening $\gamma$ is not a free
parameter. The quadrature is accurate either at $\gamma=0$ or when $\gamma$ is
comparable to the grid spacing $h$: intermediate values converge worse than
either, because a spectral feature narrower than one cell is not resolved by
the Gauss rule on that cell. We therefore tie the broadening to the grid,
$\gamma=3h/2=3\Omega/2K_s$. Refining the grid then reduces $\gamma$
proportionally, so the limit $\gamma\to0$ and the continuum limit $h\to0$ are
taken together. The resolution limit of a given grid is the regularized
width of the threshold in units of the threshold itself,
\beg\label{AppE-floor}
\frac{2\gamma}{2\Delta_g}=\frac{3}{2K_s\,(\Delta_g/\Omega)},
\en
which at $K_s=336$ is $0.0080$ at $\lambda=2$ and $0.0050$ at $\lambda=3.25$,
and half of that at $K_s=672$ (Table~\ref{tab:controls}); it falls with
coupling because $\Delta_g/\Omega$ grows.

\emph{Extraction of $\omega_{\rm H}$.} The susceptibility \eqref{MyChi} is
computed at $50$ frequencies uniformly spaced in $\omega/2\Delta_g$ over
the sub-threshold window and fitted by a single damped oscillator on a
smooth background, Eq.~\eqref{AppE-fit} of the main text. The fit is
repeated over five windows whose lower ends are $\omega/2\Delta_g=0.80$,
$0.85$, $0.90$, $0.92$ and $0.95$ (or the start of the scan, where that
lies higher). On
$K_s=168$ and $336$ the scans cover $0.800\le\omega/2\Delta_g\le0.998$,
the last point $0.38$ to $1.2$ cells below the threshold; on $K_s=672$
they cover the same range at $\lambda=2$ and $3.25$ and
$0.90\le\omega/2\Delta_g\le0.999$ at $2.25\le\lambda\le3.0$ (35 points at
$\lambda=2.25$), the last point $0.85$ to $2.4$ cells below the threshold.
Intercepts that combine two grids are always formed from fits over the
same window in $\omega/2\Delta_g$ on both. The spread of
$\omega_{\rm H}$ over the windows is taken as the fitting uncertainty; it is
$5$--$14\%$ of $E_b$ on $K_s=168$ and about $1\%$ on $K_s=336$ and $672$, so
that the window dependence of every extrapolated value quoted below comes
from the coarsest grid. At finite $\gamma$ the width $\Gamma$ is dominated
by the regulator, $\Gamma\simeq5$--$6\gamma$. Each coupling is computed on
three grids, $K_s=168$, $336$ and $672$, so that $\gamma$ is halved twice,
and the binding energy $E_b=2\Delta_g-\omega_{\rm H}$ is extrapolated
linearly in $\gamma$ from the two coarser grids,
\beg\label{AppE-Rich}
[E_b]_{\gamma\to0}=2\,E_b(K_s=336)-E_b(K_s=168),
\en
which is the rule behind Table~\ref{tab:controls} and Eq.~\eqref{Eq1}.
The binding rises by $10$--$24\%$ from $K_s=168$ to $336$ and by
$-1$ to $+8\%$ from $336$ to $672$; the dependence on $\gamma$ is concave at
every coupling, so that \eqref{AppE-Rich} overestimates the intercept
formed from the finest pair, $2E_b(672)-E_b(336)$, by $3$--$12\%$
(Table~\ref{tab:controls}, sixth column, and item (ii) below). We keep
\eqref{AppE-Rich} for the tabulated result because it is defined identically
at every coupling and because the difference lies inside the calibration
systematic of control (v); the finest pair is reported alongside it and used
in controls (i), (ii) and (v). The same extrapolation applied to $\Gamma$ gives
zero within the fit noise at every coupling and on both pairs of grids,
which is the basis for the statement that the mode is undamped below the
threshold at $T\ll\Delta_g$. The procedure loses control at
$\Delta_g/\Omega\simeq0.96$ ($\lambda=3.5$ at $T/\Omega=0.005$, four grids
$K_s=168$, $252$, $336$, $672$). There the dependence on $\gamma$ changes
from concave to convex, the binding rising by $5\%$ from $K_s=168$ to $336$
but by $14\%$ from $336$ to $672$, the intercepts from different pairs of
grids spread by $20\%$ ($\delta_{\gamma\to0}=0.0094$ to $0.0114$) against
at most $11\%$ elsewhere, and the fitted width on the finest grid rises to
$6.8\gamma$. We therefore quote $\lambda\le3.25$ as the resolved range.

\emph{Weak coupling.} The resolution limit \eqref{AppE-floor} grows toward
weak coupling while the binding energy shrinks; at $K_s=336$ the two cross
at $\lambda\simeq2$, which is why the window is quoted as beginning there.
Resolving $\lambda=1.5$ would require $K_s\simeq640$ ($N_1\simeq1.5\times10^4$)
and $\lambda=1$ would require $K_s\simeq1800$ ($N_1\simeq4.4\times10^4$), i.e.\
dense systems of $9\times10^4$ and $2.6\times10^5$ unknowns against
$9.7\times10^4$ at $K_s=672$. The factorization cost grows as the cube of
that dimension, so $\lambda=1$ is two orders of magnitude beyond the
production runs, not a factor of two. The $\gamma=0$ branch would remove
the constraint but is not usable in the present form. At $\gamma=0$ the
Keldysh kernel acquires inverse square-root singularities at the replica
points, which the panel quadrature treats correctly but the nodal
representation of the unknowns does not, so the result becomes controlled
by the commensuration of those points with the grid. Consistently, at
$\lambda=0.5$ we find no resonance: $\mathrm{Re}\chi$ is quadratic in
$\omega$ to within $10^{-4}$ over the sub-threshold window and only the
bound $1-\omega_{\rm H}/2\Delta_g<0.065$ can be stated, compatible with the
BCS result $\omega_{\rm H}=2\Delta$ but not a test of it.

\emph{Controls.} On the two coarser grids the binding is smaller than the
fitted linewidth, $E_b/\Gamma=0.14$--$0.33$ on $K_s=168$ and
$0.32$--$0.79$ on $K_s=336$, and the
extrapolated value \eqref{AppE-Rich} rests on the difference between them;
on $K_s=672$ the binding exceeds the linewidth for $\lambda\ge2.5$ and is
$0.76$ and $0.97$ of it at $\lambda=2$ and $2.25$. The resolution limit
\eqref{AppE-floor} is
itself of order one percent at the grids used, so it must be shown that
$\delta$ is not this limit restated. The width does not decide this: for a
regularized threshold singularity
$\chi\propto(2\Delta_g-2i\gamma-\omega)^{-1/2}$ the fitted $\Gamma$ is also
proportional to $\gamma$ and also extrapolates to zero. The discriminating
quantity is the intercept $[E_b]_{\gamma\to0}$ alone, and five independent
tests bear on it (Tables~\ref{tab:controls} and \ref{tab:truncated}).
\begin{itemize}
\item[(i)]{\emph{The resolution limit and the binding move in opposite
directions.} The limit \eqref{AppE-floor} falls by $38\%$ across the window
because $\Delta_g/\Omega$ grows, while $\delta_{\gamma\to0}$ rises by
$33\%$. Their ratio grows monotonically from $1.1$ to $2.4$ on $K_s=336$
and from $2.2$ to $4.7$ on $K_s=672$. No monotone function of the
resolution can produce this. At fixed $\gamma$ the binding in units of the
regulator doubles across the window, $E_b/\gamma=1.94\to3.95$ on $K_s=336$
and $3.92\to8.50$ on $K_s=672$ (Table~\ref{tab:controls}), whereas a
regulator-set offset would be independent of $\lambda$.}

\item[(ii)]{\emph{The sign and form of the $\gamma$ dependence.} The
binding \emph{rises} as $\gamma$ is halved, at every coupling and on both
halvings: $E_b=0.0075$, $0.0087$, $0.0088\,\Omega$ at $\lambda=2$ and
$0.0143$, $0.0177$, $0.0190\,\Omega$ at $\lambda=3.25$ on $K_s=168$, $336$,
$672$, with the four other couplings in between. The regulator pulls the
resonance toward the threshold, which is the opposite of what an offset
generated by $\gamma$ would do. Measured in units of the regulator the
binding doubles at each halving, from $E_b/\gamma=0.8$ to $1.9$ to $3.9$
at $\lambda=2$ and from $1.6$ to $4.0$ to $8.5$ at $\lambda=3.25$. The
binding therefore stays fixed while the regulator is halved, as it must for a
pole at fixed $E_{b,0}$ and cannot for a regularized edge, whose apparent
binding is proportional to $\gamma$ (control v). On the finest grid the
binding is thus resolved outright at every coupling, $E_b=3.9$--$8.5\gamma$,
with a window spread of about $1\%$.

The dependence on $\gamma$ is concave: the
slope $dE_b/d\gamma$ falls in magnitude between the pairs $(168,336)$ and
$(336,672)$ at every coupling ($-0.77$ to $-0.57$ at $\lambda=3.25$, $-0.27$
to $-0.02$ at $\lambda=2$), so that the intercept from the finest pair
lies $3$--$11\%$ below the value of \eqref{AppE-Rich}, the difference
decreasing with coupling above $\lambda=2.25$ (Table~\ref{tab:controls}).
The same rule applied to
the width gives $2\Gamma_{336}-\Gamma_{168}=-0.0031\pm0.0001\,\Omega$ at all
six couplings, $-12\%$ of $\Gamma_{336}$, and
$2\Gamma_{672}-\Gamma_{336}=-4\%$ to $-11\%$ of $\Gamma_{336}$, consistent
with zero; the residual $0.003$ in units of $2\Delta_g$ at $\lambda=2$ is
the systematic of the extrapolation and is used as such in item (iv).}

\item[(iii)]{\emph{Independence of the extraction.} The position of the
maximum of $\mathrm{Im}\chi$, which is biased toward the threshold by the
continuum onset at finite $\gamma$, agrees with the oscillator fit after
the same extrapolation \eqref{AppE-Rich} to $0.9$--$2.6\%$ for $\lambda\ge2.5$,
to $5.3\%$ at $\lambda=2.25$ and to $10\%$ at $\lambda=2$, where the maximum
lies one or two cells from
the threshold on the coarse grid (Table~\ref{tab:controls}). On $K_s=672$
the maximum lies $4$ to $12$ cells below the threshold but the scan step
there is $1.2$--$3.3\gamma$, which quantizes its position too coarsely for
the same comparison; on that grid the fit alone is used.}

\item[(iv)]{\emph{The truncated system.} The system \eqref{DrivenSystem}
with $\delta W\equiv0$ --- same grids, regulator, $\Delta_g$, fit and
windows, the only difference in the operator being which columns are
retained --- gives the values collected in Table~\ref{tab:truncated}. The
truncated scans differ from those of the full system in one respect: they
consist of $80$ frequencies at a step of $1.025h$ across the threshold,
$0.89\le\omega/2\Delta_g\le1.11$ at $\lambda=2$ and $0.93$--$1.07$ at
$\lambda=3.25$, of which the $40$ sub-threshold points enter the fit; the
edge of control (v) is fitted at these same points. Every one of the
truncated values lies below the resolution limit \eqref{AppE-floor} of its
own grid, by factors of $1.7$ to $5.3$. The
coarse-grid ranges straddle zero, the spread over fit windows is an order
of magnitude larger than for the full system, and the two-point intercepts
\eqref{AppE-Rich}, combined over the window extremes, span $0.001$--$0.008$
at $\lambda=2$ and $0.003$--$0.006$ at $\lambda=3.25$ --- ranges of the same
size as the systematic of item (ii). The truncated binding is therefore
not distinguishable from zero, and we quote it as a bound,
\beg\label{AppE-dWbound}
\delta_{\delta W=0}\lesssim0.005,
\en
the resolution of the procedure established in item (v).

What the same
numbers do establish is that the truncated response is not the regularized
square-root edge of item (v). Sampled at the same points and fitted over
the same five windows, that edge returns $E_b=-1.8\gamma$ for the full
sub-threshold window and $-1.0\gamma$ for the narrowest, on every grid and
at both couplings. The truncated system returns $E_b=-0.4$ to
$+0.6\gamma$ at $\lambda=2$ and $-0.1$ to $+1.2\gamma$ at $\lambda=3.25$,
moving in the opposite direction as the window is narrowed. The separation
between the two is $1.1$--$1.7\gamma$ on $K_s=168$ and
$1.6$--$2.8\gamma$ on $K_s=336$, and it persists at every window.
Against the full
values $0.0088$ and $0.0117$ the truncation removes at least $40\%$ and
$55\%$ of the binding respectively; against the finest-pair values
$0.0079$ and $0.0113$, at least $36\%$ and $55\%$. Consistently, the maximum of
$|\chi|$ of the truncated system lies $1.0$--$1.7\gamma$ below the threshold
on every grid, at the second scan point, i.e.\ at a fixed distance
in units of the regulator. In the full system that distance doubles
at each halving of $\gamma$ (item (ii)), as it must for a pole at
fixed $E_{b,0}$. The exponent of
the sub-threshold decay of $|\chi|$ does not discriminate between a pole of
this size and an edge at this resolution and is not used.}

\item[(v)]{\emph{Calibration on functions with a known answer.} The
identical fit and extrapolation, at the frequencies of each production run
and the same pairs of $\gamma$, were applied to a regularized
threshold singularity,
$\chi_{\rm edge}=C+A[(2\Delta_g-2i\gamma)^2-\omega^2]^{-1/2}$, and to the
displaced-pole test function \eqref{test} continued to the real axis with
the same leg broadening $\omega\to\omega+2i\gamma$ and a prescribed
$\delta$. The edge returns a \emph{negative} apparent binding at every
coupling: $\delta_{\gamma\to0}=-0.002$ to $-0.004$ on the pair $(168,336)$
and $-0.003$ on the pair $(336,672)$, with single-grid values $-0.003$ to
$-0.008$, i.e.\ $E_b=-1.0$ to $-2.5\gamma$ depending on coupling, grid and
window, and $-1.7$ to $-3.8\gamma$ on $K_s=672$. The fitted width,
$\Gamma\simeq5$--$7\gamma$, is indistinguishable from that of the data. The
procedure therefore places an edge above the threshold, opposite in sign to
what is measured, and cannot manufacture a positive binding.

On the pair $(168,336)$ a prescribed $\delta=0.02$ is returned as
$0.019$--$0.020$ and $\delta=0.01$ as $0.009$--$0.012$, over the six
couplings and the fit windows: to within $5\%$ and $\pm20\%$ respectively,
with no systematic sign. The single-grid values overestimate $0.01$ by
$6$--$10\%$, and the extrapolation scatters about the input. On the pair
$(336,672)$ a prescribed $\delta=0.01$ is returned as $0.0097$--$0.0101$ at
both ends of the window, and the measured values themselves, $0.0079$ at
$\lambda=2$ and $0.0113$ at $\lambda=3.25$, as $0.0082$--$0.0085$ and
$0.0115$, i.e.\ to within $8\%$. A prescribed $\delta=0.005$ is returned as $0.005$--$0.0075$
on the coarse pair and $0.006$--$0.0066$ on the fine pair, overestimated at
every coupling and window.

The resolution of the pipeline is therefore about $0.005$ in $\delta$:
bindings at or below it are overestimated after extrapolation and cannot be
quoted as numbers, and bindings of $0.01$ are recovered to $\pm20\%$ on the
coarse pair and to better than $10\%$ on the fine pair. The measured values
of the full system lie in the latter regime, so that the coefficient of
\eqref{Eq1} carries a systematic uncertainty of that order, comparable to
the $\gamma$ corrections themselves. The truncated values of item (iv) lie
in the former, which is why they are quoted as a bound and not as a number.
Fit windows that scale with the regulator,
$2\Delta_g-\omega<30\gamma$, reduce the extrapolated edge to
$|\delta|\le0.0005$ at all couplings but do not improve the recovery of
$\delta\lesssim0.005$, which is limited by the regulator and not by the
window.}
\end{itemize}

The numerical systematics of the binding --- the extrapolation in $\gamma$, the
fit model, the width of the energy window and the quadrature --- are controlled
only to within about ten percent each, and we do not quote the binding to
better than the $20\%$ stated for the coefficient of Eq.~\eqref{Eq1}. None of our
conclusions requires more: they rest on the sign of the binding, on its being of
order one percent rather than of order unity, on the spectral weight that a
binding of this size carries, and on comparisons made with identical
discretizations, such as that with the truncated system of Table~II.
Two of them are stated rather than defended. At $\lambda=2$ the full
binding on $K_s=336$ alone, $\delta=0.0078$, sits at the resolution limit
$0.0080$ of that grid, and it is the $K_s=672$ run, on which the same
binding is twice its resolution limit, that establishes the lower end of
the window. The couplings $\lambda\ge2.25$ lie increasingly above their
limits on every grid and carry the result. The
convergence in the window half-width $N_h$ has been tested only up to
$N_h=10$: at $\lambda=2.04$ going from $N_h=5$ to $10$ moved $\delta$ by
$7.6\%$ and $\chi$ pointwise by up to $14\%$. The production runs use
$N_h=12$, i.e.\ $L/\Delta_g$ between $21$ and $13$, and the residual
dependence on $N_h$ at that value has not been measured.

Tables~\ref{tab:controls} and \ref{tab:truncated} quote $E_b/\gamma$ rather
than $E_b$ because the regulator is the resolution of the calculation, and
the ratio states directly how far a resonance sits from the threshold in
units of the finest feature the grid can represent. At any finite $\gamma$
the fit \eqref{AppE-fit} returns a resonance position whether or not a pole
exists, since the regulator rounds the inverse-square-root threshold
singularity into a bump of its own; control (v) calibrates that bump, and
item (ii) gives the test that separates the two. For a pole the position is
a property of the $\gamma=0$ problem, so halving $\gamma$ leaves $E_b$
nearly fixed and doubles $E_b/\gamma$; for the edge $\gamma$ is the only
energy scale that has been added, so halving it halves $E_b$ and leaves
$E_b/\gamma$ fixed. The two hypotheses thus predict opposite behavior of the
same tabulated number under grid refinement, and a value of $E_b/\gamma$
below unity is not distinguishable from the threshold whatever its value in
units of $\Omega$. This is what separates the three columns: the edge at
$-1$ to $-1.8$, the truncated system of Table~\ref{tab:truncated} at
$0$--$1.2$, and the full system of Table~\ref{tab:controls} at
$1.9$--$4.0$ on $K_s=336$ and $3.9$--$8.5$ on $K_s=672$, doubling at each
halving. On a single grid $E_b$ in units of $\Omega$ conveys none of this.

\begin{table}[t]
\caption{\footnotesize Robustness of the binding energy against the
resolution of the real-axis discretization, $T/\Omega=0.005$. Resolution
limit: Eq.~\eqref{AppE-floor} at $K_s=336$ (half of it at $K_s=672$).
$\delta_{\gamma\to0}$: from the oscillator fit and the rule
\eqref{AppE-Rich} on the grids $K_s=168$ and $336$, the values used in
Eq.~\eqref{Eq1} and Fig.~\ref{Fig-Binding}; $\delta^{\max}_{\gamma\to0}$:
the same extrapolation applied to the position of the maximum of
$\mathrm{Im}\chi$; $\delta^{(672)}_{\gamma\to0}$: the same rule on the
grids $K_s=336$ and $672$. $E_b/\gamma$: binding in units of the regulator
on $K_s=336$ and on $K_s=672$.}
\label{tab:controls}
\begin{ruledtabular}
\begin{tabular}{cccccccc}
$\lambda$ & $\Delta_g/\Omega$ & resolution limit & $\delta_{\gamma\to0}$ &
$\delta^{\max}_{\gamma\to0}$ & $\delta^{(672)}_{\gamma\to0}$ &
$E_b/\gamma$ (336) & $E_b/\gamma$ (672) \\
\hline
2.00 & 0.560 & 0.0080 & 0.0088 & 0.0079 & 0.0079 & 1.94 & 3.92 \\
2.25 & 0.633 & 0.0071 & 0.0095 & 0.0090 & 0.0085 & 2.45 & 4.87 \\
2.50 & 0.702 & 0.0064 & 0.0102 & 0.0101 & 0.0093 & 2.92 & 5.85 \\
2.75 & 0.770 & 0.0058 & 0.0107 & 0.0106 & 0.0100 & 3.36 & 6.81 \\
3.00 & 0.835 & 0.0053 & 0.0112 & 0.0111 & 0.0107 & 3.72 & 7.72 \\
3.25 & 0.899 & 0.0050 & 0.0117 & 0.0114 & 0.0113 & 3.96 & 8.50 \\
\end{tabular}
\end{ruledtabular}
\end{table}

\begin{table}[t]
\caption{\footnotesize The truncated system $\delta W\equiv0$ through the
identical pipeline, $T/\Omega=0.005$. Ranges are the spread over fit
windows. Resolution limit: Eq.~\eqref{AppE-floor} for the grid in
question. Edge: the apparent binding returned by the same procedure for
the regularized threshold singularity of control (v) at the same points,
over the same fit windows (full window to narrowest).
Full: $\delta_{\gamma\to0}$ of the full system from Table~\ref{tab:controls}.
The two-point intercept \eqref{AppE-Rich} over the window extremes is
$0.001$--$0.008$ at $\lambda=2$ and $0.003$--$0.006$ at $\lambda=3.25$.}
\label{tab:truncated}
\begin{ruledtabular}
\begin{tabular}{ccccccc}
$\lambda$ & $K_s$ & resolution limit & $\delta_{\delta W=0}$ & $E_b/\gamma$ &
edge $E_b/\gamma$ & full \\
\hline
2.00 & 168 & 0.016  & $-0.0030$ -- $0.0011$ & $-0.37$ -- $0.14$ & $-1.8$ -- $-1.0$ & 0.0088 \\
2.00 & 336 & 0.0080 & $0.0012$ -- $0.0025$  & $0.30$ -- $0.6$  & $-1.8$ -- $-1.0$ & 0.0088 \\
3.25 & 168 & 0.010  & $-0.0004$ -- $0.0020$ & $-0.08$ -- $0.40$ & $-1.8$ -- $-1.0$ & 0.0117 \\
3.25 & 336 & 0.0050 & $0.0025$ -- $0.0030$  & $1.02$ -- $1.23$  & $-1.8$ -- $-1.0$ & 0.0117 \\
\end{tabular}
\end{ruledtabular}
\end{table}

\section{Strong coupling limit}\label{AppendixF}
The intermediate-coupling results of Sec.~IV cover $\Delta_g/\Omega\lesssim1$.
{
The opposite regime is the formal universal strong-coupling limit of isotropic Eliashberg theory~\cite{AllenDynes1975,Marsiglio1991,Combescot1995}. For a general physical phonon spectrum the single energy scale in this limit is}
{
\beg\label{scScale}
\omega_{\rm sc}=\left[2\int_0^\infty d\omega\,
\omega\alpha^2F(\omega)\right]^{1/2}
\stackrel{\mathrm{Einstein}}{=}\sqrt{\lambda}\,\Omega.
\en
}
{We take all phonon frequencies small compared with this fixed scale. At nonzero frequency transfer on the scale $\omega_{\rm sc}$, Eq.~\eqref{GeneralMatsKernel} then tends to $\omega_{\rm sc}^2/(\omega_n-\omega_m)^2$, independently of the spectral shape. The singular local contribution is separated before taking the limit and cancels from the equations below. For the derivation we use Einstein phonons, taking $\Omega\to0$ and $\lambda\to\infty$ at fixed $\omega_{\rm sc}$.} {In this limit} $T_c\to0.1827\,\omega_{\rm sc}$ \cite{AllenDynes1975} and every energy
scales with $\omega_{\rm sc}$. The limit is taken at $T\ll\omega_{\rm sc}$, so that
$T\to0$ together with $\Omega\to0$. The ratio $T/\Omega$, which decides whether the
slow phonon is thermally populated, may be held at any fixed value. It does not
appear in the leading-order equations; it sets only the width of the layer around
the gap edge in which the expansion in $\Omega$ fails, and that width vanishes in
the limit. Taken literally, the real-axis equations
\eqref{ClosureK}--\eqref{SourceReal} do not become simpler in this limit. The
phonon replicas $\eps\pm n\Omega$ that organize them become dense, and
the number of replicas inside any fixed energy window grows as $1/\Omega$. They do,
however, acquire a rigid analytic structure (every local term grows as
$1/\Omega$ at fixed $T/\Omega$, every retarded kernel stays finite), and this is what allows the limit to be taken in closed form. We do so in three places: (i) on the
imaginary axis, where the result is a closed system that replaces \eqref{eqMats},
\eqref{MatsSystem} and \eqref{PQR}; (ii) in the equilibrium problem on the real axis,
where the replica ladder of \eqref{MSC} collapses onto a differential equation;
and (iii) in \eqref{ClosureK}--\eqref{SourceReal}, which carry the sub-threshold
susceptibility.

\subsection{Matsubara axis}
The single identity behind everything below is the decomposition of the
Matsubara sum with the kernel $\lambda_{n-m}$ into its diagonal weight and a
difference operator,
\beg\label{scSplit}
\pi T\sum_m\lambda_{n-m}F_m=\Lambda F_n+\mathcal{L}[F]_n,
\qquad
\mathcal{L}[F]_n\equiv\pi T\sum_{m\neq n}\lambda_{n-m}\left(F_m-F_n\right),
\en
valid for any $F_m$, with
{
\beg\label{scLambda}
\begin{aligned}
\Lambda&\equiv\pi T\sum_m\lambda_{n-m}
=\pi\int_0^\infty d\nu\,\alpha^2F(\nu)\coth\frac{\nu}{2T}\\
&\stackrel{\mathrm{Einstein}}{=}\frac{\pi\lambda\Omega}{2}\coth\frac{\Omega}{2T}
=\frac{\pi\omega_{\rm sc}^2}{2\Omega}\,\bigl[2n_B(\Omega)+1\bigr]\;\xrightarrow[\;\Omega\to0\;]{}\;\infty .
\end{aligned}
\en
}
$\Lambda$ is independent of $n$ and has a simple interpretation. An infinitely slow
phonon acts at leading order as potential disorder, and $\Lambda$ is the Born
scattering rate off it, $\tau^{-1}=2\Lambda\to\infty$. The factor
$2n_B+1$ is the mean-square displacement of the oscillator, zero-point plus
thermal: for $T\ll\Omega$ it equals $1$ and $\Lambda=\pi\omega_{\rm sc}^2/2\Omega$,
for $T\gg\Omega$ it equals $2T/\Omega$ and $\Lambda=\pi\omega_{\rm sc}^2T/\Omega^2$
is the classical thermal disorder. Either way $\Lambda$ diverges, and either way
it drops out of every physical quantity below.
The operator $\mathcal{L}$ carries all the dynamics. For $m\neq n$ the kernel tends to
$\lambda_{n-m}\to\omega_{\rm sc}^2/(\omega_n-\omega_m)^2$, so $\mathcal{L}$ has a
finite, $\Omega$-independent limit. At $T\to0$ it is given by
\beg\label{scLzero}
\mathcal{L}[F(\omega)]=\frac{\omega_{\rm sc}^2}{2}\,\dashint d\omega'\,
\frac{F(\omega')-F(\omega)}{(\omega-\omega')^2},
\en
which converges at $\omega'=\omega$ for any differentiable $F$ and at infinity for
any bounded $F$.

\paragraph{Equilibrium.}
Applying \eqref{scSplit} to \eqref{EliEqsOpt}, whose right-hand sides are
$\pi T\sum_m\lambda_{n-m}g_m$ and $\pi T\sum_m\lambda_{n-m}f_m$, we find
\beg\label{scEq1}
\Sigma_n=\Lambda g_n+(\mathcal{L}g)_n,\qquad
\Phi_n=\Lambda f_n+(\mathcal{L}f)_n .
\en
Hence $\tilde\omega_n=\omega_n+\Sigma_n$ and $\Phi_n$ both diverge as $\Lambda$,
and so does $D_n$:
\beg\label{scD}
\begin{aligned}
D_n&=\sqrt{\tilde\omega_n^2+\Phi_n^2}=\Lambda+\varsigma_n+O(\Lambda^{-1}),\\
\varsigma_n&=g_n\bigl[\omega_n+(\mathcal{L}g)_n\bigr]+f_n(\mathcal{L}f)_n .
\end{aligned}
\en
The definitions $g_n=\tilde\omega_n/D_n$, $f_n=\Phi_n/D_n$ are satisfied
identically at order $\Lambda$; at order $\Lambda^0$ they require
$f_n[\omega_n+(\mathcal{L}g)_n]=g_n(\mathcal{L}f)_n$, i.e.
\beg\label{scGap}
\begin{aligned}
&\Delta_n\bigl[\omega_n+(\mathcal{L}g)_n\bigr]=\omega_n(\mathcal{L}f)_n, \quad \varsigma_n=\frac{\omega_n+(\mathcal{L}g)_n}{g_n}=\frac{(\mathcal{L}f)_n}{f_n},
\end{aligned}
\en
with $\Delta_n=\omega_n f_n/g_n=\Phi_n/Z_n$. Equation \eqref{scGap} is the
Eliashberg gap equation in the limit of Allen and Dynes \cite{AllenDynes1975}:
\beg\label{scAD}
\Delta_n=\pi T\sum_{m\neq n}\frac{\omega_{\rm sc}^2}{(\omega_n-\omega_m)^{2}}\,
\frac{\Delta_m-\Delta_n\omega_m/\omega_n}{\sqrt{\omega_m^2+\Delta_m^2}}.
\en
The divergent
parts $\Lambda g_n$, $\Lambda f_n$ of $\Sigma_n$ and $\Phi_n$ cancel exactly in
$\Delta_n$, which is Anderson's theorem for the static part of the phonon field,
while $Z_n=\Lambda/\sqrt{\omega_n^2+\Delta_n^2}+O(1)$ diverges and the
quasiparticle residue vanishes. The quantities that remain finite are the ratios $g_n$, $f_n$,
$\Delta_n$ and the finite part $\varsigma_n$ of $D_n$. At $T\to0$ we find
$\Delta_{n\to0}\simeq0.755\,\omega_{\rm sc}$
from \eqref{scGap} with \eqref{scLzero}.

\paragraph{Linear response.}
We write \eqref{MatsSystem} as $\psi_n=h\,\hat{e}_2+\pi T\sum_m\lambda_{n-m}\mathbb{K}_m\psi_m$
with $\psi_m=(\delta W_m,\delta\Phi_m)^T$, $\hat{e}_2=(0,1)^T$ and
\beg\label{KmOm0}
\mathbb{K}_m=\bigl(\begin{smallmatrix}P_m&-Q_m\\-Q_m&R_m\end{smallmatrix}\bigr).
\en
By \eqref{PQR} and \eqref{scD}, $D_{m+l}+D_m=2\Lambda+2\bar\varsigma_m$ with
$\bar\varsigma_m=\tfrac12(\varsigma_{m+l}+\varsigma_m)$, so that
\beg\label{scProj}
\begin{aligned}
\Lambda\,\mathbb{K}_m&=\Bigl(1-\frac{\bar\varsigma_m}{\Lambda}\Bigr)\mathbb{P}_m
+O(\Lambda^{-2}),\\
\mathbb{P}_m&=\frac12\begin{pmatrix}1-c_m&-B_m\\-B_m&1+c_m\end{pmatrix},
\end{aligned}
\en
where $c_m=g_{m+l}g_m-f_{m+l}f_m$ and $B_m=g_{m+l}f_m+f_{m+l}g_m$. Given the normalization condition
$g_n^2+f_n^2=1$, it follows that  $c_m^2+B_m^2=1$ and $\mathbb{P}_m^2=\mathbb{P}_m$: at
leading order the kernel of \eqref{MatsSystem} is a rank-one projector. Writing $g_m=\cos\theta_m$, $f_m=\sin\theta_m$ and
$\Theta_m=\theta_{m+l}+\theta_m$, one has $c_m=\cos\Theta_m$, $B_m=\sin\Theta_m$
and $\mathbb{P}_m=v_mv_m^T$ with $v_m=(-\sin\tfrac{\Theta_m}{2},\cos\tfrac{\Theta_m}{2})^T$.

Inserting \eqref{scSplit} and \eqref{scProj} into \eqref{MatsSystem} gives
\beg\label{scSing}
(1-\mathbb{P}_n)\psi_n=h\,\hat{e}_2-\frac{\bar\varsigma_n}{\Lambda}\mathbb{P}_n\psi_n
+\frac1\Lambda\bigl(\mathcal{L}[\mathbb{P}\psi]\bigr)_n+O(\Lambda^{-2}),
\en
in which the operator on the left is singular. The response is therefore
$O(\Lambda)$ and is fixed in two steps. At order $\Lambda$, $\psi_n$ must lie in
the null space of $1-\mathbb{P}_n$, $\psi_n\propto v_n$; this is the locking
\eqref{scLock}. At order $\Lambda^0$,
multiplying \eqref{scSing} by $\mathbb{P}_n$ and using $\mathbb{P}_n\psi_n=\psi_n+O(1)$
gives, with $\psi_n=\Lambda\Psi_n$, equation \eqref{scResp} in the main text.
We have verified numerically that the solution of the
full system \eqref{MatsSystem} converges to that of \eqref{scResp} as $\Omega\to0$
at fixed $\omega_{\rm sc}$ and $T$: at $T=0.05\,\omega_{\rm sc}$ the relative
deviation of $\delta\Phi_n$ is $1.5\%$, $0.4\%$ and $0.1\%$ for
$\Omega/\omega_{\rm sc}=0.025$, $0.0125$ and $0.00625$, i.e. it falls as
$\Lambda^{-1}$, and the ratio $\delta W_n/\delta\Phi_n$ approaches $-\tan(\Theta_n/2)$
at the same rate.

\subsection{Real axis: equilibrium}
The mixed representation \eqref{MSC} of Appendix~\ref{AppendixB} is the natural
starting point. For $|\eps|\gg T$ the Fermi factors become steps,
$n_F(\Omega\mp\eps)\to\theta(\pm\eps-\Omega)$, while $n_B(\Omega)$ is fixed by
$T/\Omega$ and is kept. For $\eps>\Omega$ the two replicas then carry the weights
$n_B+1$ (emission, $s=-$) and $n_B$ (absorption, $s=+$):
\beg\label{scMSC}
\begin{aligned}
\eps Z_\eps^R&=\eps+iS_Z(\eps)+i\Lambda_-\,g^R_{\eps-\Omega}+i\Lambda_+\,g^R_{\eps+\Omega},\\
\Delta_\eps^RZ_\eps^R&=S_\Phi(\eps)+i\Lambda_-\,f^R_{\eps-\Omega}+i\Lambda_+\,f^R_{\eps+\Omega},
\end{aligned}
\en
with
\beg\label{scLpm}
\Lambda_-=\frac{\pi\lambda\Omega}{2}\bigl[n_B(\Omega)+1\bigr],\quad
\Lambda_+=\frac{\pi\lambda\Omega}{2}\,n_B(\Omega),\qquad
\Lambda_-+\Lambda_+=\Lambda,\quad
\Lambda_--\Lambda_+=\frac{\pi\lambda\Omega}{2}=\frac{\pi\omega_{\rm sc}^2}{2\Omega},
\en
so that the sum of the two replica weights is the static weight \eqref{scLambda}
and their difference, the emission--absorption asymmetry, is
$T$-independent. Here $S_Z(\eps)=\pi T\sum_m\lambda(\eps-i\omega_m)g_m$ and
$S_\Phi(\eps)=\pi T\sum_m\lambda(\eps-i\omega_m)f_m$ are the Matsubara sums of
\eqref{MSC}. As $\Omega\to0$, $\lambda(\eps-i\omega_m)\to-\omega_{\rm sc}^2/(\eps-i\omega_m)^2$
for $\eps\neq0$, and both sums have finite limits,
\beg\label{scS}
S_Z(\eps)=-\omega_{\rm sc}^2\,\pi T\sum_m\frac{g_m}{(\eps-i\omega_m)^2},
\quad
S_\Phi(\eps)=-\omega_{\rm sc}^2\,\pi T\sum_m\frac{f_m}{(\eps-i\omega_m)^2},
\en
smooth functions of $\eps$ built from the imaginary-axis solution alone. The
replica terms carry the divergent coefficients $i\Lambda_\pm$; together they make
$\eps Z^R_\eps\simeq i\Lambda g^R_\eps$ and $\Phi^R_\eps\simeq i\Lambda f^R_\eps$,
so that $\eta^R_\eps=\sqrt{(\eps Z^R_\eps)^2-(\Phi^R_\eps)^2}\to i\Lambda$, a
constant, by virtue of $(g^R)^2-(f^R)^2=1$. In the ratio
$\Delta_\eps=\Phi_\eps/Z_\eps$ the divergence cancels once more. Using
$f^R_{\eps'}=\Delta_{\eps'}g^R_{\eps'}/\eps'$, which is exact, and eliminating $Z$
between the two equations \eqref{scMSC},
\beg\label{scLadder}
\begin{aligned}
&i\Lambda_-\,g^R_{\eps-\Omega}\Bigl[\frac{\Delta_\eps}{\eps}-\frac{\Delta_{\eps-\Omega}}{\eps-\Omega}\Bigr]
+i\Lambda_+\,g^R_{\eps+\Omega}\Bigl[\frac{\Delta_\eps}{\eps}-\frac{\Delta_{\eps+\Omega}}{\eps+\Omega}\Bigr]\\
&=S_\Phi(\eps)-\frac{\Delta_\eps}{\eps}\bigl[\eps+iS_Z(\eps)\bigr].
\end{aligned}
\en
The two brackets are $\pm\Omega\,\partial_\eps(\Delta_\eps/\eps)+O(\Omega^2)$, so
the first-order terms combine into
$i\Omega(\Lambda_--\Lambda_+)\,g^R_\eps\,\partial_\eps(\Delta_\eps/\eps)$, in which
the Bose factors cancel identically and $\Omega(\Lambda_--\Lambda_+)=\pi\omega_{\rm sc}^2/2$
stays finite. The second-order terms are
$O(\Omega^2\Lambda)=O\bigl(\omega_{\rm sc}^2\,\Omega\coth(\Omega/2T)\bigr)$ and
vanish as $\Omega\to0$ at fixed $T/\Omega$ (as $\omega_{\rm sc}^2T$ when
$T\gg\Omega$). Relative to the first-order term they are of order
$\Omega\coth(\Omega/2T)\,\partial_\eps^2(\Delta_\eps/\eps)/\partial_\eps(\Delta_\eps/\eps)$;
near the gap edge, where $\partial_\eps(\Delta_\eps/\eps)\to0$ while the
second derivative stays finite (see below), this ratio grows as
$\Omega\coth(\Omega/2T)/(\Delta_g-\eps)$, so that the expansion holds
outside a layer of width $\sim\Omega\coth(\Omega/2T)\simeq\max(\Omega,2T)$
around the edge. At fixed $T/\Omega$ this layer shrinks to zero; taken
instead at fixed $T$, it has the finite width $\sim2T$ and contains the
thermal broadening of the discrete levels of Ref.~\cite{Combescot1996}.
Emission and absorption shift the energy argument in opposite directions: their
sum is the static field, which cancels, and their difference survives as the
derivative. The replica ladder of \eqref{MSC}, whose length
$\Delta_g/\Omega$ diverges, therefore collapses to a first-order differential
equation in $\eps$, \eqref{scODE} given in the main text.
The square root in $g^R_\eps$ is the continuation of $g_m$ from the imaginary axis, cf.
\eqref{ContinueGF}: inside the gap, $\sqrt{\eps^2-\Delta_\eps^2}=i\sqrt{\Delta_\eps^2-\eps^2}$
with the principal root, so that $g^R_\eps$ is purely imaginary and the density of
states vanishes. Equation \eqref{scODE} holds for $\eps\gg\Omega\to0$, i.e.
everywhere except at $\eps=0$; its initial condition is the continuity of
$\Delta$ across the boundary layer $|\eps|\lesssim\Omega$,
$\Delta_{\eps\to0^+}=\Delta_{n\to0}$, and the same equation with
$\eps\to z=\eps+i\omega'$ continues $\Delta$ into the complex plane along
horizontal lines, starting from the Matsubara data $\Delta(i\omega')$. The
equation is the strong-coupling image of \eqref{MSC}: the emission--absorption
asymmetry of the two replica coefficients has become the derivative term, and
the analog of the $\varsigma_n$ of \eqref{scD} is the finite part of
$\eta^R_\eps/i$,
\beg\label{scVarsigma}
\begin{aligned}
\eta^R(z)&=i\bigl[\Lambda+\varsigma(z)\bigr],\qquad \Re z>0,\\
\varsigma(z)&=\left[{-iz+S_Z(z)-\tfrac{\pi\omega_{\rm sc}^2}{2}\,\partial_z g^R(z)}\right]\left[g^R(z)\right]^{-1}
=g^R(z)\bigl[S_Z(z)-iz\bigr]+if^R(z)S_\Phi(z),
\end{aligned}
\en
the two forms being equivalent on the solution of \eqref{scODE}. Like the
derivative term, $\varsigma$ is built from $\Lambda_--\Lambda_+$ alone; it
reduces to $\varsigma_n$ of \eqref{scGap} on the imaginary axis.

Three properties of \eqref{scODE} are worth recording. First, on the sub-gap
segment all its coefficients are real ($S_\Phi$ real, $iS_Z$ real, $g^R$ and the
prefactor both imaginary), so $\Delta_\eps$ is real for $|\eps|<\Delta_g$ and
there is no sub-gap spectral weight. Second, for $|\eps|\gg\Delta_g$ the equation
has the particular solution $\Delta_\eps\simeq-\omega_{\rm sc}^2C/(2\eps^2)$,
$C=\int d\omega\,f(i\omega)$, which matches the Matsubara tail, and a homogeneous
solution $\Delta_\eps/\eps\propto\eps^{2i/\pi}e^{i\eps^2/\pi\omega_{\rm sc}^2}$ of
unit modulus along the real axis and decaying as $e^{-2\eps\omega'/\pi\omega_{\rm sc}^2}$
along $\Im z=\omega'>0$: the continuation into the upper half plane is stable, and
on the real axis $\Re\Delta_\eps=\eps$ acquires the further crossings at high
energy that we encountered at finite $\Omega$. Third, the coefficient of the
derivative, $g^R_\eps=\eps/\eta$, diverges where $\eta\to0$, so that
$d(\Delta_\eps/\eps)/d\eps\to0$ there: in the limit $\Omega\to0$ the gap edge is
not a crossing but a tangency, $\Re\Delta_\eps-\eps$ touching zero with zero
slope. The finite-$\Omega$ replica structure $\Delta_g+n\Omega$ has collapsed
onto the edge, and the continuation of \eqref{scODE} beyond it requires matching
through that layer, which we do not attempt here. For the sub-threshold response,
what is below the edge is all that is needed. Integrating \eqref{scODE} from the
imaginary-axis solution of \eqref{scGap} we find the tangency at
$\Delta_g\simeq1.16\,\omega_{\rm sc}$, i.e. $2\Delta_g/T_c\simeq12.7$
and $\Delta_g/\Delta_{n\to0}\simeq1.54$: at strong coupling the spectral gap
exceeds the Matsubara order parameter by half of itself. As discussed in
Sec.~IV.B, the tangency makes $g^R_\eps$ a simple pole at $\eps=\Delta_g$,
so that $E_1=\Delta_g$ is the lowest of the discrete levels of the
strong-coupling spectrum~\darkred{\definecolor{strongcouplingcitecolor}{rgb}{0.6,0,0}\hypersetup{citecolor=strongcouplingcitecolor}\cite{Combescot1995,Combescot1996}}.

\subsection{Real axis: response and the sub-threshold susceptibility}
The same two ingredients --- a divergent local weight and finite retarded kernels
--- govern the response equations, and they reduce \eqref{ClosureK}--\eqref{SourceReal}
to the same singular structure as \eqref{scSing}. For $|\eps_\mp|\gg T$,
$t_{\eps+s\Omega\mp\omega/2}\to{\rm sgn}\,\eps_\mp$, and with
$\coth(\Omega/2T)=2n_B+1$ the brackets of \eqref{ClosureRAfinal} take, for
$\eps_\mp>\Omega$, the values $-2(n_B+1)$ at $s=-$ and $-2n_B$ at $s=+$: the local
terms of \eqref{ClosureK} and \eqref{ClosureRAfinal} connect $\eps$ to both
replicas $\eps\mp\Omega$ with the coefficients $-i\Lambda_\mp$ of \eqref{scLpm}.
By $\eta^{R,A}_\eps\to i\Lambda$ the matrices \eqref{Mmatrix} become
$\mathbb{M}^X_\eps\to\mathbb{N}^X_\eps/2i\Lambda$ with
$\mathbb{N}^X_\eps=\bigl(\begin{smallmatrix}\tilde{\cal A}^X_\eps&-{\cal B}^X_\eps\\
{\cal B}^X_\eps&-{\cal A}^X_\eps\end{smallmatrix}\bigr)$, so that each local operator
is of order one, the sum of the two replicas reproduces the local term and their
difference produces a derivative,
\beg\label{scShift}
-\frac{1}{2\Lambda}\Bigl\{\Lambda_-[\mathbb{N}^X\psi^X]_{\eps-\Omega}
+\Lambda_+[\mathbb{N}^X\psi^X]_{\eps+\Omega}\Bigr\}
=-\frac12\mathbb{N}^X_\eps\psi^X_\eps
+\frac{\pi\omega_{\rm sc}^2}{4\Lambda}\,\partial_\eps\bigl[\mathbb{N}^X\psi^X\bigr]_\eps
+O\bigl(\omega_{\rm sc}^2\,\Omega\coth\tfrac{\Omega}{2T}\bigr),
\en
in which the derivative term is $O(1)$ because $\psi^X=O(\Lambda)$ and, like the
derivative in \eqref{scLadder}, carries no $T/\Omega$; the
neglected terms are bounded outside the edge layer by the same estimate as
for \eqref{scLadder}. The retarded kernels stay
finite: $\tfrac{\lambda\Omega}{4}v^X_\pm(\eps,\eps')\to
\tfrac{\omega_{\rm sc}^2}{2}T\sum_{n\ge0}(\eps\pm z_n)^{-2}(\eps'\pm z_n)^{-1}$
by \eqref{vfun}, and $\tfrac{\lambda\Omega}{8}D^X_{\eps-\eps'}\to
\tfrac{\omega_{\rm sc}^2}{4(\eps-\eps'\pm i0)^2}$. Since the self-energy responses are
$O(\Lambda)$, the leading order of \eqref{ClosureRAfinal} is homogeneous and
algebraic,
\beg\label{scAlign}
\Bigl[1+\tfrac12\mathbb{N}^X_\eps\Bigr]\psi^X_\eps=0,\qquad
\det\Bigl[1+\tfrac12\mathbb{N}^X_\eps\Bigr]=\tfrac14\Bigl[1-\bigl(g_{\eps_+}g_{\eps_-}+f_{\eps_+}f_{\eps_-}\bigr)^2
+\bigl(g_{\eps_+}f_{\eps_-}+f_{\eps_+}g_{\eps_-}\bigr)^2\Bigr]=0,
\en
for $X=R,A$, the determinant vanishing by $(g^X)^2-(f^X)^2=1$ on each leg. Thus
$\psi^R_\eps\propto({\cal B}^R_\eps,\,1+g_{\eps_+}g_{\eps_-}+f_{\eps_+}f_{\eps_-})^T$,
which is the real-axis image of the locking \eqref{scLock} under the dictionary
\eqref{MRedDict}, and likewise for $A$. The Keldysh block \eqref{ClosureK} has
the same structure: at leading order
$[1+\tfrac12\mathbb{N}^K_\eps]\psi^K_\eps=-\tfrac12\bigl[t_{\eps_-}(\mathbb{N}^R_\eps-\mathbb{N}^K_\eps)\psi^R_\eps
-t_{\eps_+}(\mathbb{N}^A_\eps-\mathbb{N}^K_\eps)\psi^A_\eps\bigr]$, and
$1+\tfrac12\mathbb{N}^K_\eps$, built from the mixed pair $(R,A)$, is singular for
the same reason. The three scalar amplitudes are then determined at the next
order by projecting the $O(1)$ terms --- the derivative terms, the retarded
kernels, the source \eqref{SourceReal} and the $\varsigma$-corrections to the
projectors --- onto the null vectors. This is the
strong-coupling form of \eqref{ClosureK}--\eqref{SourceReal}: three coupled
first-order integro-differential equations on the real axis in place of the
six-component driven system \eqref{DrivenSystem}. We do not write them out. For
the quantity of interest here, the position of the pole below $2\Delta_g$,
the same information is obtained more directly from \eqref{scResp}, as we now
show; the real-axis reduction is needed only for $\omega>2\Delta_g$, where the
pair spectrum consists of the lines $E_n+E_m$ and of any collective modes
between them.

At $T\to0$ the propagator components are one analytic function on the plane cut
along $|\Re z|\ge\Delta_g$: with the branch of \eqref{scODE},
\beg\label{scGF}
\mathcal{G}(z)=\frac{-iz}{\sqrt{\Delta(z)^2-z^2}},\quad
\mathcal{F}(z)=\frac{\Delta(z)}{\sqrt{\Delta(z)^2-z^2}},\quad
\mathcal{G}^2+\mathcal{F}^2=1,
\en
one has $\mathcal{G}(i\omega_m)=g_m$, $\mathcal{F}(i\omega_m)=f_m$ (cf.
\eqref{ContinueGF}), $\mathcal{G}=g^R$ and $\mathcal{F}=if^R$ just above the
real axis, and no singularity on the gap segment, since there is no sub-gap
spectral weight. Now take \eqref{scResp} at $T\to0$, where the Matsubara index
becomes a continuous variable, and shift it by half the transfer, so that the two
legs sit at $i(y\mp\nu/2)$ and the internal variable $y$ is symmetric between
them. The operator $\mathcal{L}$ acts on $y$ alone and does not see the shift:
the phonon transfer between internal points is $i(y-y')$ whatever the external
frequency, cf. the remark after \eqref{PQR}. The only $\nu$-dependence is
through the propagators ${\cal G}(iy\pm\omega/2)$ and ${\cal F}(iy\pm\omega/2)$, and the continuation $i\nu\to\omega$ is the
replacement $i(y\pm\nu/2)\to iy\pm\omega/2$ in them:
\beg\label{scStrip}
\Psi(y;\omega)=\frac{2}{\varsigma(iy+\omega/2)+\varsigma(iy-\omega/2)}\,\mathbb{P}(y;\omega)
\Bigl[h\,e_2+(\mathcal{L}\Psi)(y;\omega)\Bigr],
\en
with $\mathbb{P}$ built as in \eqref{scProj} from $\mathcal{G}_\pm=\mathcal{G}(iy\pm\omega/2)$,
$\mathcal{F}_\pm=\mathcal{F}(iy\pm\omega/2)$, which keeps $\mathbb{P}^2=\mathbb{P}$
since $\mathcal{G}^2+\mathcal{F}^2=1$ holds identically,  and $\varsigma(-z)=\varsigma(z)$. For real $|\omega|<2\Delta_g$
every leg argument satisfies $|\Re(iy\pm\omega/2)|<\Delta_g$ for all $y$ and stays
inside the strip of analyticity; the right-hand side of \eqref{scStrip} is
therefore analytic in $\omega$ on $|\Re\,\omega|<2\Delta_g$, coincides with
\eqref{scResp} on the imaginary axis, and by uniqueness is the retarded response
there. Without the shift, a leg at $iy+\omega$ would sit at $\Re z=\omega$ and
leave the strip as soon as $\omega>\Delta_g$; this is why the direct continuation of \eqref{MatsSystem} fails
while the shifted one does not. On the sub-gap segment $\chi(\omega)$ is real,
and the amplitude mode is the real $\omega_{\rm H}<2\Delta_g$ at which the
homogeneous version of \eqref{scStrip} has a nontrivial solution --- a root of a
real function of one real variable, not the end point of an analytic
continuation. The pole is real by construction, $\Gamma\equiv0$ below the
threshold, which is the statement that was reached at intermediate coupling
only by the $\gamma\to0$ extrapolation of Sec.~IV. The equilibrium input,
$\Delta(z)$ on the two lines $\Re z=\pm\omega/2$, is supplied by \eqref{scODE}
integrated from the imaginary axis. The only discretization is then the quadrature in
$y$, which must resolve the scale $\Delta_g-\omega/2$ near $y=0$
on which the leg functions vary as $\omega\to2\Delta_g$: the leg at
$iy+\omega/2$ approaches the pole of \eqref{scGF} at $z=\Delta_g$, so that
$\mathcal{G}\propto1/[iy-(\Delta_g-\omega/2)]$. (The square-root scale
$\sqrt{\Delta_g(2\Delta_g-\omega)}$ familiar from BCS theory refers to the
band-energy variable $\xi$, which has been integrated out here.) Equation \eqref{scStrip}
thus determines whether a sub-gap collective pole survives in the
strong-coupling limit, with no
$\Omega$ in the calculation. It is the form in which the question of
Sec.~IV --- whether the binding survives past $\Delta_g/\Omega=1$ --- is
addressed at $\Delta_g/\Omega=\infty$.

\end{widetext}
\end{appendix}

\bibliography{eliedge}

\end{document}